\documentclass[numberedappendix]{emulateapj}
\slugcomment{The Astrophysical Journal, 891:26 (38pp), accepted 2020 January 11} 
\usepackage{natbib}

\usepackage{amsmath}
\usepackage{xcolor}
\usepackage{soul}
\sethlcolor{green}

\newcommand{\myemail}{t.r.almeyda@soton.ac.uk}

\usepackage{array}
\usepackage{ragged2e}
\newcolumntype{L}[1]{>{\RaggedRight\vspace{0pt}}p{#1}}
\newcolumntype{C}[1]{>{\centering\vspace{0pt}}p{#1}}

\shorttitle{Modeling Dusty Torus in AGN}
\shortauthors{Almeyda et al.}

\begin{document}

\title{Modeling the infrared reverberation response of the circumnuclear dusty torus in AGNs: an investigation of torus response functions}

\author{Triana Almeyda$^{1,2}$}
\author{Andrew Robinson$^2$}
\author{Michael Richmond$^2$}

\author{Robert Nikutta $^{3}$}
\author{Bryanne McDonough $^{4,2}$}

\affil{$^1$Department of Physics and Astronomy, University of Southampton, Southampton, UK; \myemail}
\affil{$^2$School of Physics and Astronomy, Rochester Institute of Technology, Rochester, NY 14623, USA}
\affil{$^3$NSF's National Optical-Infrared Astronomy Research Laboratory, 950 N. Cherry Ave., Tucson, AZ 85719, USA}
\affil{$^4$Department of Astronomy, Boston University, Boston, MA 02215, USA}

\begin{abstract}

The size and structure of the dusty circumnuclear torus in active galactic nuclei 
(AGN) can be investigated by analyzing the temporal response of the torus's infrared (IR)
dust emission to variations in the AGN ultraviolet/optical luminosity. This method,
reverberation mapping, is applicable over a wide redshift range, but the IR response is
sensitive to several poorly constrained variables relating to the dust distribution and
its illumination, complicating the interpretation of measured reverberation lags. We have
used an enhanced version of our torus reverberation mapping code (TORMAC) to conduct a
comprehensive exploration of the torus response functions at selected wavelengths, for the
standard interstellar medium grain composition. The shapes of the response functions vary widely over the
parameter range covered by our models, with the largest variations occurring at shorter
wavelengths ($\leq 4.5\,\mu$m). The reverberation lag, quantified as the response-weighted
delay (RWD), is most affected by the radial depth of the torus, the steepness of the
radial cloud distribution, the degree of anisotropy of the AGN radiation field, and the
volume filling factor. Nevertheless, we find that the RWD provides a reasonably robust
estimate, to within a factor of $\sim 3$, of the luminosity-weighted torus radius,
confirming the basic assumption underlying reverberation mapping. However, overall, the
models predict radii at $2.2\,\mu$m that are a typically factor of $\sim 2$ larger than those
derived from K-band reverberation mapping. This is likely an indication that the innermost
region of the torus is populated by clouds dominated by large graphite grains.

\end{abstract}

 ---


\section{Introduction}\label{intro}

A key element of the unified model of active galactic nuclei (AGN) is a toroidal distribution of dusty molecular gas (the ``torus''), which surrounds the supermassive black hole (SMBH) and its accretion disk \citep{Krolik:1988aa, Antonucci:1993aa, Urry:1995aa}. This structure causes heavy obscuration from soft X-ray to near-infrared (NIR) wavelengths, with the result that the observed spectrum (i.e., whether that of a broad-line/ type 1 or narrow-line/ type 2 AGN) is determined by the orientation of the torus with respect to the observer's line of sight \cite[e.g.,][]{Antonucci:1993aa}. The dust is heated by the absorbed ultraviolet (UV)/optical radiation of the accretion disk and re-emits in the IR; thus, the torus is a major contributor of mid-IR radiation in most AGN \citep{Telesco:1984aa, Sanders:1989aa}. 

In the years since it was first formulated, much evidence has emerged suggesting that the simple unified model based on a dusty circumnuclear torus does not by itself provide a complete framework for explaining AGN phenomenology. In particular, there may be other sources of obscuration, such as galactic-scale dust lanes \citep[e.g.,][]{Prieto:2014aa}, and there are situations where a torus-like structure may not be present, such as in low-luminosity AGN (``true'' type 2s; e.g., \citealp{Merloni:2014aa}) or in high-luminosity AGN fueled by accretion in major mergers. Nevertheless, there is strong observational evidence that a compact, clumpy, axisymmetric obscuring structure containing hot dust is present in most AGN (see the recent \citealp{Netzer:2015aa} review for a discussion of these issues). This dusty axisymmetric obscuring structure, which, for brevity and convention, we will refer to as the ``torus'', is therefore important to our picture of AGN but our limited knowledge of its size and structure, as well as how these properties vary with luminosity, hinders our ability to understand the obscured AGN population and hence the cosmic evolution of SMBHs.

The inner radius of  the  ``torus'', is set by the dust sublimation radius, $R_{\rm d} \lesssim1$\,pc, which in turn is set by the dust sublimation temperature, $T_{\rm sub}$, \citep{Barvainis:1987aa}.
As this corresponds to angular scales $\sim$milli-arcseconds, even in the closest AGN, the innermost regions of the torus are too small to be directly imaged by any existing single aperture telescope. Constraints on the size and structure of the torus have been obtained in recent years using near and mid-IR interferometry \citep[e.g.,][]{ Kishimoto:2011aa, Burtscher:2013aa}, but only in bright, nearby AGN.

Recent ALMA (Atacama Large Millimeter/submillimeter Array) observations have revealed disk-like structures with scales $\sim 10$\,pc in several AGN, including one of the closest, NGC 1068, which appear to be the sub-millimeter counterparts of the torus \citep{Gallimore:2016aa, Garcia-Burillo:2016aa, Imanishi:2016aa, Imanishi:2018aa, Alonso-Herrero:2018aa, Combes:2019aa}, but this is far beyond the wavelength ($\sim 30\, \mu$m) at which the torus dust emission peaks. 

An alternative approach, which is not fundamentally distance limited, is ``reverberation mapping''. This technique has been extensively applied to the broad emission line region in AGN (e.g., \citealp{Blandford:1982aa, Peterson:1993aa, Peterson:2014aa, Shen:2016aa} and references therein) but can also be used to infer the size and structure of the torus from the temporal response of its IR emission to changes in the AGN optical luminosity \citep{Clavel:1989aa, Barvainis:1992aa}. The delay (``reverberation lag''), due to light-travel times, between the optical variations and the IR response can be directly measured from the light curves and provides an estimate of the size of the torus. 

The dust reverberation radius has now been measured by NIR (K-band) reverberation mapping for about 20 AGN (e.g., \citealp{Nelson:1996aa,Oknyanskij:2001aa,Minezaki:2004aa,Suganuma:2006aa,Lira:2011aa,Koshida:2014aa,Pozo-Nunez:2014aa,Lira:2015aa,Mandal:2018aa}). 
Some of the authors of this paper are members of a collaboration that recently conducted a 2.5 year monitoring campaign, during which 12 broad-line AGN were observed using the \textit{Spitzer Space Telescope} at 3.6 and 4.5 $\mu$m, supported by ground-based optical observations. The first results from this campaign were reported in \cite{Vazquez:2015aa} and \cite{Vazquez:2015ab}. 

The size of the torus can be inferred from the measured reverberation lag alone, but the IR response is the convolution of the AGN UV/optical light curve with a transfer function that contains information about the geometry and structure of the torus. In principle, therefore, by analyzing the light curves, it is possible to constrain structural properties such as the radial depth and inclination of the torus, as well as measuring the reverberation radius. However, since many parameters influence the response, we need models to properly interpret the observed lags and extract the torus structural information.
We have developed a computer code for this purpose, which simulates the IR reverberation response of a 3D ensemble of dust clouds at selected wavelengths, given an input optical light curve (\citealp{Almeyda:2017aa}, hereafter A17; see also \citealp{Almeyda:2017ab}). 

The code, Torus Reverberation Mapping Code (TORMAC), is based on the response mapping code of Robinson and collaborators (e.g., \citealp{Robinson:1990aa, Perez:1992aa}) and adopts a geometry similar to that used in the CLUMPY torus model described in \cite{Nenkova:2008ab}. In A17, we described the code and presented simulated dust emission response functions at 3.6, 4.5, and 30 $\mu$m to explore the effects of certain geometrical and structural properties, dust cloud orientation, and anisotropy of the illuminating radiation field. We also briefly discussed the effects of cloud shadowing and presented some example synthetic light curves using the observed optical light curve of the Seyfert 1 galaxy NGC 6418 as the input.  

In this paper, we introduce the effects of cloud occultation into the code, which represents the extinction of the IR emission from a given cloud by intervening clouds along the line of sight to the observer. However, our main purpose is to present the results of a comprehensive investigation of the torus response as a function of the parameters that control the overall structure of the torus, its cloud population, and its illumination by the AGN radiation field, including both local and global optical depth effects. For this purpose, we have computed a large set of torus response functions, i.e., numerical approximations to the transfer function, at selected wavelengths, which we characterize in terms of  a response-weighted delay (RWD; the first moment of the transfer function) 
and a ratio that measures the relative power that emerges in the core of the response function (``core-to-total area ratio," CTAR).

The RWD is essentially the reverberation lag, which can be determined observationally from time series analysis. An important motivation for this study is to investigate the relationship between the RWD and  the characteristic size of the torus for emission at a given wavelength, as represented by the luminosity-weighted radius (LWR), in order to establish whether the former does indeed provide a robust measure of the latter.

This paper is organized as follows. First, the computer model is summarized and the implementation of cloud occultation is outlined in Section~\ref{sec:model}. In Section~\ref{sec:rBBRF}, we discuss the basic properties of the response function of a disk using simple models in which the dust clouds are isotropically emitting blackbodies. We also define
quantities that characterize the torus size (LWR) and its response (RWD, CTAR).

In Section~\ref{sec:DustRF}, we show models in which the  dust cloud emission is computed by interpolation in a grid of dust radiative transfer models, beginning with a globally optically thin torus (i.e., containing optically thick clouds, but the volume filling factor of the torus is negligible), and successively explore the effects of anisotropic illumination of the torus, cloud shadowing, and cloud occultation. In Section~\ref{sec:combo}, we present the response functions for the more realistic case of a globally optically thick (GOT) torus, which incorporates cloud shadowing and cloud occultation, with the volume filling factor governing their importance.
In Section~\ref{sec:dis}, we discuss the implications of our results and the limitations of the current treatment. Lastly, in Section~\ref{sec:conc}, we conclude with a summary of our findings and an outline of future work. Additional figures showing response functions and the variations of the descriptive quantities with various model parameters can be found in  Appendix~\ref{app:figs}.


\section{Review of Model: TORMAC}\label{sec:model}
TORMAC was introduced in detail in A17. Here we will summarize the code's main features and new capabilities. 

TORMAC sets up a 3D ensemble of dust clouds randomly distributed within a disk structure 
defined by the ratio of the outer to the inner radius, $Y$, an angular width, $\sigma$, and the inclination to the observer's line of sight, $i$ (the torus is face-on when $i = 0^{\circ}$). We also assume that the polar axes of the torus and the accretion disk are aligned.
In the radial coordinate, $r$ (measuring the radial distance from the central source), the clouds are arranged following a power-law distribution with index $p$. In polar angle, $\theta$, there are two choices for the cloud distribution. In one, the torus surface has a ``sharp edge''  defined by $\sigma$ and the clouds are distributed uniformly in $\theta$ within the range $(90^{\circ} - \sigma) \leq \theta \leq (90^{\circ} + \sigma)$ (see Figure 1 in A17). 
In the other, ``fuzzy torus'' case, the clouds follow a Gaussian distribution in angle $\beta$ (which is measured from the equatorial plane and is the complement of the polar angle, i.e., $\beta = 90^{\circ}-\theta$). Finally, the clouds are distributed uniformly in the azimuthal angle, $\phi$.

The dust clouds are either heated directly by the UV/optical continuum emitted from the accretion disk or, if shadowed (see below), indirectly by the diffuse radiation field produced by the directly illuminated clouds. (Diffuse radiation heating of directly heated clouds generally has only a small effect compared to direct heating and is not included, \citealp{Nenkova:2008aa}.)
The illuminating AGN radiation field may be anisotropic, due to ``edge darkening" of the accretion disk. To model this, the AGN luminosity is assumed to have the following polar angle dependence: 
\begin{equation}\label{Ltheta}
L(\theta)=[s+(1-s)(1/3)\cos\theta(1+2\cos\theta)]L_{\rm AGN}
\end{equation}
\citep[see][]{Netzer:1987aa}, where $L_{\rm AGN}$ is the isotropic bolometric luminosity of the AGN and the parameter $s$ determines the degree of anisotropy. The inner radius of the torus is taken to be the dust sublimation radius, and for $s < 1$, this radius will be a function of both the dust sublimation temperature, $T_{\rm sub}$, and $\theta$. For the standard Galactic grain mixture adopted in the models presented in this paper, the sublimation radius is \citep{Nenkova:2008ab}
\begin{equation}\label{Rdtheta}
R_{\rm d}(\theta)\simeq 0.4\left(\frac{L(\theta)}{10^{45} \,\rm erg\, s^{-1}}\right)^{1/2}\left(\frac{1500\,\rm K}{T_{\rm sub}}\right)^{2.6} \rm pc.
\end{equation}

For an isotropically illuminated torus, $s=1$, Equation~\ref{Ltheta} becomes $L(\theta)=L_{\rm AGN}$ and, thus, the inner radius, $R_{\rm d, iso}$, given by Equation~\ref{Rdtheta} is no longer dependent on $\theta$. 

The emitted flux of a cloud at a particular observer time is determined by interpolation in a grid of synthetic cloud spectra that was created for the CLUMPY torus model described by \cite{Nenkova:2008aa,Nenkova:2008ab}. 
The spectral energy distribution (SED) of the illuminating AGN continuum is represented by a broken power law as defined by Equation~13 in \cite{Nenkova:2008aa}. The spectrum of any given cloud within the model grid depends on the surface temperature of its illuminated side,  $T_{\rm cl}$, the orientation angle, $\alpha$ (the angle between the directions of the cloud and the observer from the central source), and its optical depth, $\tau_{V}$. The spectrum depends on $\alpha$ because clouds heated directly by the AGN emit anisotropically, since the side facing the central source has a higher temperature than the non-illuminated side. This effect is analogous to lunar phases, where $\alpha$ determines the fraction of the illuminated surface the observer sees. We henceforth refer to this as the ``cloud orientation" effect.
The surface temperature of the directly heated clouds is determined from the incident AGN continuum flux at the position of the cloud as defined by $r$ and $\alpha$, at the retarded time corresponding to the current observer time. As grain heating and cooling occur effectively instantaneously \citep{Nenkova:2008aa, vanVelzen:2016aa}, we assume that individual clouds respond on timescales that are much less than the light-crossing time of the inner dust-free cavity of the torus, $\tau_{\rm d}=2R_{\rm d, iso}/c$ (see also \citealp{Ichikawa:2017aa}).   

The emission from a given cloud is subject to two ``global'' opacity effects. The first is cloud shadowing, whereby an outer cloud at some radius, $r$, has its line of sight to the central AGN continuum source blocked by one or more inner clouds at smaller $r$, closer to the central source. Clouds that are shadowed in this way are heated indirectly by the diffuse radiation emitted by surrounding directly illuminated clouds  (see A17). The second is cloud occultation, whereby the emitted spectrum of a cloud may be attenuated by clouds intervening along the line of sight to the observer. The implementation of this effect in TORMAC is described in Section~\ref{sec:model_clocc}. The probability that any given cloud is either shadowed or occulted is determined by the volume filling factor, which is specified as a free parameter and is given by  $\Phi = NV_{\rm cl}/V_{\rm tor}$, where $V_{\rm cl}$ and $V_{\rm tor}$ are the volumes of an individual cloud and the torus, respectively, and $N$ is the total number of clouds. The radius of an individual cloud is then $R_{\rm cl} = [3\Phi V_{\rm tor}/4\pi N]^{1/3}$, assuming that they are spherical.

As the torus inner radius is taken to be the dust sublimation radius, the surface temperature of clouds near the inner radius will typically exceed $T_{\rm sub}$ as the AGN luminosity varies. However, optically thick dust clouds have a strong internal temperature gradient, and only dust grains close to the illuminated surface will have temperatures $\gtrsim T_{\rm sub}$. These clouds will be gradually eroded rather than being instantaneously destroyed. Therefore, in the models presented here we have adopted a simplified treatment of dust sublimation, in which the cloud surface temperature is constrained to $T_{\rm cl} = T_{\rm sub}$, where $T_{\rm sub} = 1500$\,K, even when the AGN continuum luminosity is such that the computed cloud surface temperature exceeds the sublimation temperature. 
A consequence of this assumption is that the torus response tends to ``saturate'' at shorter wavelengths, since the emission from clouds at temperature $T_{\rm cl} = T_{\rm sub}$ is constant. 
How this saturation effect modifies the torus response is discussed in Section~\ref{sec:dis_sat}.

In summary, TORMAC sets up a 3D ensemble of clouds defined by input parameters $Y$, $\sigma$, $i$, $p$, $L_{\textrm{AGN}}$, $s$, and $\Phi$. In the models presented here, all clouds in each model have the same size (determined by $\Phi$) and V-band optical depth, $\tau_{\rm V}$. The dust grain composition is a standard Galactic ISM mix of 53\% silicates and 47\% graphites assuming the Mathis-Rumpl-Nordsieck (\citealp{Mathis:1977aa}) grain size distribution. 
Given an input AGN optical light curve, the observed cloud emission is determined by taking into account the AGN illumination (which may be anisotropic), light-travel delays within the torus, cloud shadowing, and cloud occultation effects. The torus luminosity at selected IR wavelengths is computed at each observer time step by summing over all clouds in order to produce an IR light curve for each wavelength.

The TORMAC model parameters are summarized in Table~\ref{tormac_param}, which also lists the values of those parameters used in the simulations presented in this paper. 
The models discussed in Sections~\ref{sec:DustRF} and~\ref{sec:combo} explore specific axes of the multi-dimensional parameter space rather than sampling all possible combinations.
Therefore, we define a ``standard" set of parameter values, which are the values of the parameters that are held fixed when other parameters are varied. These are indicated in bold type in the table.

\begin{table}
\begin{center}
\caption{{Torus Parameters}} 

\label{tormac_param}

\renewcommand{\arraystretch}{1.6}

\scriptsize{
\begin{tabular}{ccc}
\hline
\hline
Parameter & Values & Explanation\\ 
 \hline
 \hline
$Y$ & 2, 5, {\bf 10}, 20, 50 & radial depth, $R_{\rm o}/R_{\rm d}$  \\

\hline
$\sigma$ & 15, 30, {\bf 45}, 60, 75 & angular width (deg)  \\
\hline
$p$ & $-2$, {\bf 0}, 2, 4 & radial cloud distribution  \\
&  & power-law index, $\propto r^p$  \\

\hline
$i$ & {\bf 0}, 15, 30, {\bf 45}, 60, 75, {\bf 90} & inclination (deg) \\

\hline
$s$ & 0.01, 0.05, {\bf 0.1}, 0.5, 0.99  & degree of anisotropy  \\
\hline
$\Phi$ & 0.001, 0.0032,  & volume filling factor  \\
& {\bf 0.01}, 0.032, 0.1 & \\
\hline
$\tau_{\rm V}$ & 5, 10, 20 {\bf 40}, 100 & V-band optical \\
& & depth of cloud  \\
 
\hline
\end{tabular}}

\end{center}

\begin{center}
\scriptsize{{\bf Note:} Boldface type indicates ``standard'' model parameters.} 
\end{center}

\end{table}

	\subsection{Cloud Occultation}\label{sec:model_clocc}

Dust clouds start to become optically thick to emission at NIR and shorter MIR wavelengths when the V-band optical depth $\tau_{V} > 10$ (e.g., $\tau_{3.6\,\mu\rm m}\gtrsim 1$ for $\tau_V\gtrsim 25$). Therefore, the IR emission observed from a given cloud can be attenuated due to extinction by any intervening clouds along the line of sight to the observer. The amount of attenuation depends on the number of clouds along the line of sight and the cloud's optical depth. 

The number of clouds along the line of sight between an emitting cloud and the observer is determined by the cloud's position in the torus, $\alpha$, and the radial distribution of clouds (i.e., input parameter $p$). Clouds on the side of the torus farthest from the observer are most likely to be occulted.
For any given cloud, the number of occulting clouds is determined by calculating whether foreground clouds (i.e., those with a shorter path length through the torus, toward the observer) intersect with the cloud of interest on the plane perpendicular to the line of sight. 

Assuming spherical clouds with size $R_{\rm cl}$, two clouds intersect when the projected area of intersection $A_{\rm occ} > 0$. Since all clouds in TORMAC currently have the same size, then $$A_{\rm occ}=2R_{\rm cl}^{2}\arccos\left(\frac{d_{\rm sep}}{2R_{\rm cl}}\right)-\frac{1}{2}d_{\rm sep}\sqrt{4R_{\rm cl}^2-d_{\rm sep}^2},$$ where $d_{\rm sep}$ is the distance between the centers of both clouds. For simplicity, in the current version of TORMAC we consider a cloud to be completely occulted when 50\% or more of the cloud's projected surface area is blocked by an intervening cloud, or clouds (i.e., when $A_{\rm occ}\geq0.5\pi R_{\rm cl}^2$).

Once the number of occulting clouds, $N_{\rm cl\_occ},$ is determined, we then use a discrete version of Equation 5 in \cite{Nenkova:2008aa} to determine the attenuated cloud flux, $F_{\rm cl\_att,\lambda}= F_{\rm cl,\lambda}e^{-t_{\lambda}},$ where  $t_{\lambda}=N_{\rm cl\_occ}(1-e^{-\tau_{\lambda}})$ and $F_{\rm cl,\lambda}$ is the cloud's emitted flux. 
As noted by \cite{Nenkova:2008aa}, the attenuation becomes independent of wavelength when $\tau_{\lambda} >> 1,$ such that $t_{\lambda} = N_{\rm cl\_occ}$; and becomes analogous to a uniform medium in which clouds are the ``absorbing particles" when $\tau_{\lambda} << 1$, such that $t_{\lambda} = N_{\rm {cl\_occ}} \tau_{\lambda}.$ 
The cloud optical depth at a particular wavelength is $\tau_{\lambda}=\tau_{\rm V}(C_{\rm ext,\lambda}/C_{\rm ext,V}),$ where $C_{\rm ext,\lambda}$ and $C_{\rm ext,V}$ are extinction coefficients calculated using the extinction curves of \cite{Weingartner:2001aa} and \cite{Li:2001aa}.


\section{Reverberation Response of a Disk: General Features}\label{sec:rBBRF}

The response of the IR emission to the driving time variability of the AGN UV/optical continuum can be expressed as the convolution of the UV/optical light curve with a transfer function, whose form depends on the geometrical and structural properties of the torus, 
\begin{equation}\label{eq:transeq}
L(t)=\int_{-\infty}^{\infty} \! \Psi(\tau')L_{c}(t-\tau') d\tau',
\end{equation}
where $\tau'$ is an arbitrary delay,  $\Psi(\tau')$ is the transfer function, $L(t)$ is the IR light curve, and $L_{c}(t-\tau')$ is the continuum light curve at an earlier time \citep{Peterson:1993aa}. 
The transfer function is considered to be the response to a delta function input continuum pulse.

For the TORMAC simulations presented in this paper, the input AGN light curve is a square wave pulse with a finite amplitude and a much shorter duration than the light travel time to the torus inner radius, which gives a numerical approximation to a delta function. The time dependent torus specific luminosity at a given wavelength, $L_\lambda(t)$, produced by each simulation is then a close numerical approximation to the transfer function that we, hereafter, refer to as the response function.

As outlined in Section~\ref{sec:model}, the torus is modeled as a disk of constant angular width. In this section, our aim is to characterize the basic shape of the RF and how it is affected by the geometrical parameters that define the dust distribution's structure ($Y$, $\sigma$, $i$) and the radial distribution of clouds ($p$). In order to facilitate comparison with simple analytical transfer functions and isolate the effects of the geometrical parameters, the simulations presented in this section are based on the simplest case of directly heated clouds represented as perfect isotropically-emitting blackbodies. Quantities that characterize the torus RFs will also be introduced, and their dependence on the geometrical parameters of the dust distribution will be explored.
The wavelength-dependent effects associated with dust radiative transfer, cloud orientation, anisotropic illumination of the torus, cloud shadowing, and cloud occultation will be explored in Sections~\ref{sec:DustRF} and~\ref{sec:combo}.

The response functions presented in this paper were constructed by averaging the specific torus luminosity, $L_\lambda(t)$, over 5 simulation runs, each including 50,000 clouds. These numbers were chosen to minimize statistical noise, while maintaining reasonably short run times. The averaged luminosity is normalized so that its amplitude varies in the range $0 \rightarrow 1$. Thus, the normalized response function (RF) is defined as ${\rm RF } =\frac{L_\lambda(t)-L_{\lambda,0}}{L_{\lambda, \rm max} - L_{\lambda,0}}$, where $L_{\lambda,\rm max}$ is the maximum luminosity reached in the response and $L_{\lambda,0}$ is the torus luminosity in its initial steady state, i.e., prior to the onset of  the continuum pulse.

 The time delay is normalized to the light-crossing time of the torus, i.e. $\tau = ct/2R_{\rm o}$. 
In these units, the light-crossing time of the  cavity within $2R_{\rm d, iso}$ is $\tau_{\rm d} = R_{\rm d, iso}/R_{o} = 1/Y$.

\subsection{Characteristic Descriptive Quantities of the Reverberation Response}\label{sec:anaDP}

It is generally assumed that the lag between the optical and IR light curves, as determined by reverberation mapping, is a good measure of the ``size'' of the torus. 
One of the aims of this work is to investigate how well this assumption holds, given that most of the parameters that influence the torus response are poorly constrained.  
For this purpose, it is useful to introduce quantities that characterize the effective size of the torus, the characteristic delay associated with its reverberation response, and the general shape of its RF. 

The effective radius of the torus can be characterized by the dimensionless luminosity-weighted radius (LWR), 

\begin{equation}\label{eqn:LWR}
LWR=\frac{\int_{V} r\varepsilon(r) dV}{(2R_{o})\int_{V} \varepsilon(r) dV}
\end{equation}
where $\varepsilon (r)$ is the volume emissivity distribution.

The number density of clouds is $n(r)\propto r^{p-2},$ and for spherical blackbody clouds with the same radius, $R_{\rm cl}$, the luminosity of an individual cloud is $L_{\rm cl} = \pi R_{\rm cl}^2 L_{\rm AGN}/(4\pi r^2)$. 
Therefore,  $\varepsilon (r) = nL_{\rm cl} \propto r^{p-4},$ and for a spherical shell (or, indeed, the sharp-edged torus geometry described above),

\begin{equation}\label{eqn:LWR_ss}
LWR = \frac{R_{d}}{2R_{o}}\times \left\{ \begin{array}{ll}
 \frac{Y-1}{\ln Y} & p=1 \\
\frac{Y\ln Y}{(Y-1)} &    p=0 \\
\frac{p-1}{p}\left(\frac{Y^{p}-1}{Y^{p-1}-1}\right) &    p\neq 1,0.
\end{array}
\right. 
\end{equation}

The characteristic delay (or lag) of the transfer function may be characterized by the dimensionless response-weighted delay (RWD),

\begin{equation}\label{eqn:RWD}
RWD= \frac{\int_{0}^{1} \tau\Psi(\tau)d\tau}{\int_{0}^{1} \Psi(\tau)d\tau}.
\end{equation}

Using the transfer function given by \cite{Robinson:1990aa} and  \cite{Perez:1992aa} for blackbody clouds within a spherical shell, the dimensionless RWD is identical to the dimensionless LWR as given by Equation 5. Thus, in this particular case (a spherical shell containing isotropically-emitting blackbody clouds), $\rm RWD=\rm LWR$ for all values of $Y$ and $p$ \citep{Robinson:1990aa}. In physical units, the time delay that corresponds to the RWD is $t_{\rm RW} = 2R_{\rm o} \rm{RWD}/c = 2YR_d\rm{RWD}/c$. This is essentially equivalent to the lag, $t_{\rm lag}$, that is recovered from reverberation mapping, i.e., $t_{\rm lag} \approx t_{\rm RW}$. On the other hand, the quantity we seek to determine by measuring the reverberation lag is the characteristic radius of the torus, which in physical units is $r_{\rm LW}={\rm LWR} \times R_{\rm o}$. That is, the reverberation lag is assumed to be the light-travel time corresponding to the characteristic radius,  $t_{\rm lag}\approx  t_{\rm RW}\approx r_{\rm LW}/c$.

Lastly, the overall shape of the torus transfer function can be simply characterized by the ratio of the core-to-total luminosity of the response, where the ``core'' of the transfer function is bounded by the inner cavity light crossing time,  $\tau \leq 1/Y$ (see the blue shaded region in Figure~\ref{bbdisk2} for an example). 
This quantity, the ratio of core area to the total area of the transfer function (``core-to-total area ratio", CTAR) is defined by
\begin{equation}\label{eqn:CTAR}
CTAR=  \frac{\int_{0}^{\frac{1}{Y}} \Psi(\tau)d\tau}{\int_{0}^{1} \Psi(\tau)d\tau} 
\end{equation}

The CTAR for a spherical shell containing blackbody clouds is

\begin{equation}\label{eqn:CTAR_ss}
CTAR=  \left\{ \begin{array}{ll}
 \frac{Y-1}{Y\ln Y} & p=1 \\
\frac{\ln Y}{(Y-1)} &    p=2 \\
\frac{p-1}{p-2}\left(\frac{Y^{p-2}-1}{Y^{p-1}-1}\right) &    p\neq 1,2.
\end{array}
\right. 
\end{equation}

\begin{figure}[t!]
\centering
	\includegraphics[scale=0.49]{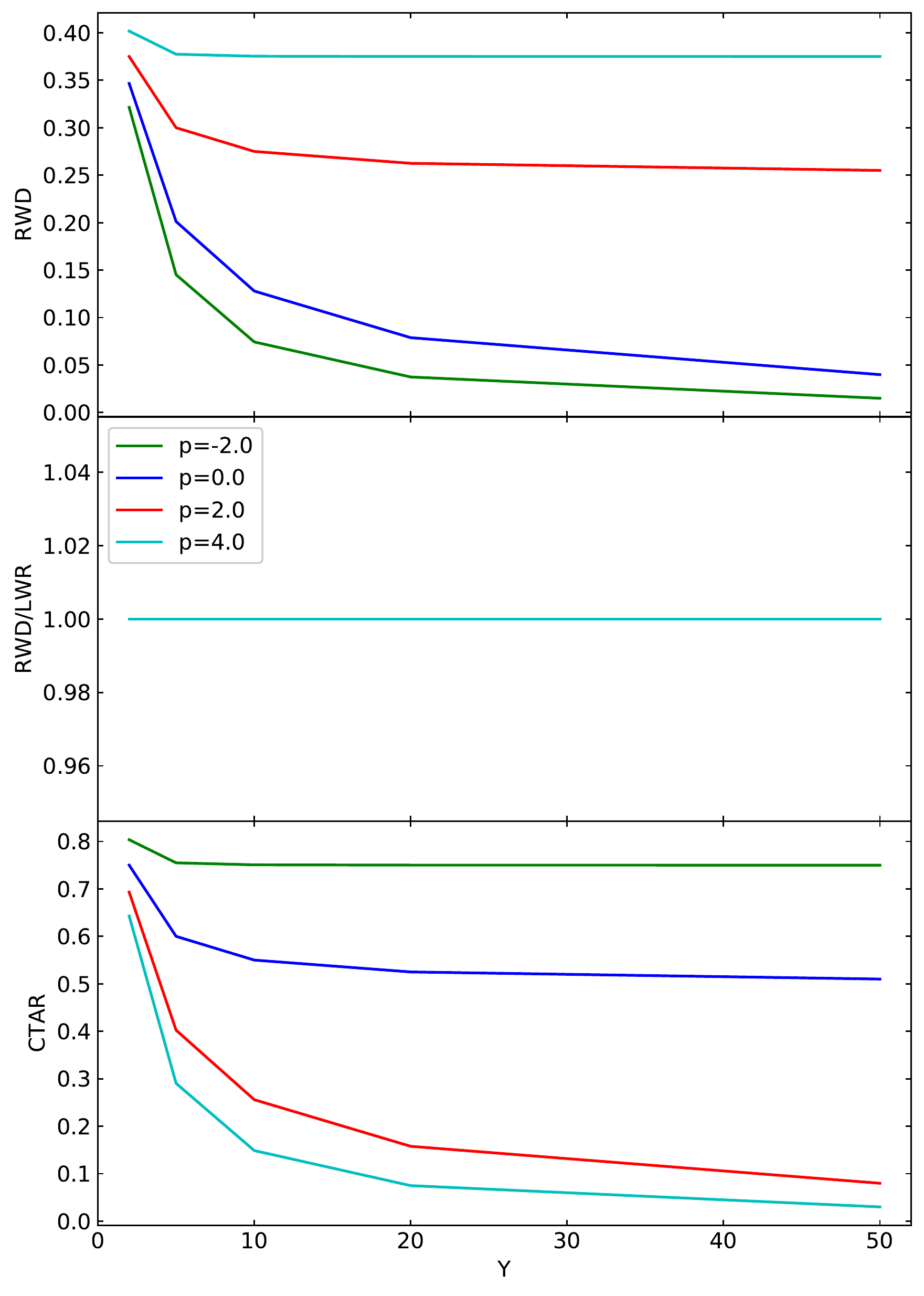}
\caption[Analytical DPs for a spherical shell of BB clouds versus $Y$]{\label{anaBBsphereDP} \textbf{Spherical shell of blackbody clouds versus $Y$:} Analytic descriptive quantities RWD (top panel), RWD/LWR (middle panel), and CTAR (bottom panel) for a spherical shell filled with isotropically emitting blackbody clouds for $Y =2-50$. The different colors show different volume emissivity distributions,  $\varepsilon(r)\propto r^{\eta}\propto r^{p-4}$, for $p = -2$ (green), 0 (blue), 2 (red), and 4 (cyan).}
\end{figure}

Henceforth, we will refer to the LWR, RWD, and CTAR collectively as  ``descriptive quantities'' of the models.

Figure~\ref{anaBBsphereDP} shows how the descriptive quantities defined above (Equations~\ref{eqn:LWR_ss} and~\ref{eqn:CTAR_ss}), for a spherical shell filled with blackbody clouds, vary with the radial depth parameter $Y$, for several values of the index $p$ of the radial cloud distribution.
The RWD (top panel) decreases as  $Y$ increases, but the amount by which it decreases is strongly dependent on the radial distribution of clouds; for $p\leq 0$ (i.e., most clouds reside closer to $R_{\rm d}$) the RWD decreases much more steeply than for $p >0$. For example, RWD is nearly independent of $Y$ when $p=4$ (cyan line), corresponding to $\varepsilon (r) \sim$ constant.

\begin{figure*}[!ht]
\begin{center}
	\includegraphics[scale=0.6]{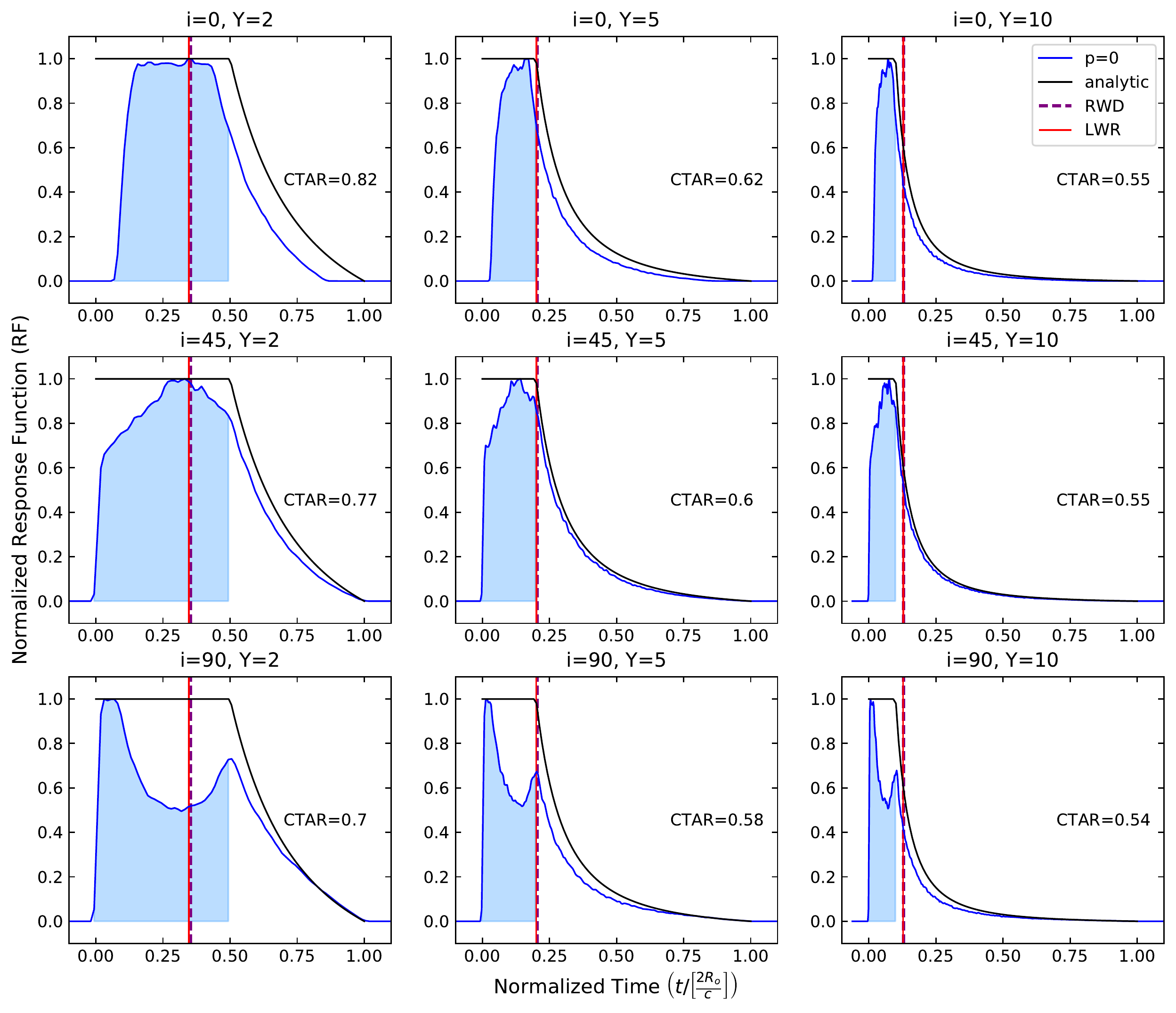}
\end{center}
\caption[RFs for BB simulations of a ``standard" disk torus case ($\sigma$=45$^\circ$)]{\label{bbdisk2} \textbf{Torus of blackbody clouds versus $Y$:} TORMAC RFs for a torus with $\sigma=45^{\circ}$ filled with isotropically emitting blackbody clouds for $p=0$ (blue), where each row represents a different inclination from top ($i$=0$^{\circ}$) to bottom ($i$=90$^{\circ}$). Each column represents a different $Y$ value increasing from left ($Y=2$) to right ($Y=10$). The black line represents the analytical transfer function for a spherical shell with $p=0$ for comparison. The descriptive quantities for each RF are shown in each panel with RWD as the dashed purple line, LWR as the solid red line, and the CTAR value printed. The shaded blue area represents the ``core" area of the response function used to calculate CTAR.}
\end{figure*}

As already noted, the ratio RWD/LWR = 1 (middle panel). Thus, for this simple model, the lag recovered from reverberation mapping (i.e., RWD) is a direct measure of the LWR regardless of $Y$ or $p$. The bottom panel shows that, as for the RWD, CTAR decreases as $Y$ increases, as would be expected since the ``core width''  of the transfer function is $1/Y$ by definition. However, CTAR shows the opposite behavior with $p$, decreasing more rapidly with increasing $p$, and hardly changing for $p=-2$ (green line). Again, this is because the clouds are highly concentrated close to the inner radius ($n(r)\propto r^{-4}$ for $p=-2$), so that the response is dominated by the ``core'' at all values of $Y$.

The RWD and CTAR are defined in terms of the transfer function in Equations~\ref{eqn:RWD} and~\ref{eqn:CTAR}. In order to calculate these quantities for our TORMAC models, we replace $ \Psi(\tau)$ in those equations with the numerical RFs obtained from the simulations. Similarly, the numerical volume emissivity is used in Equation~\ref{eqn:LWR}.

\begin{figure*}[!h]
\begin{center}
	\includegraphics[scale=0.6]{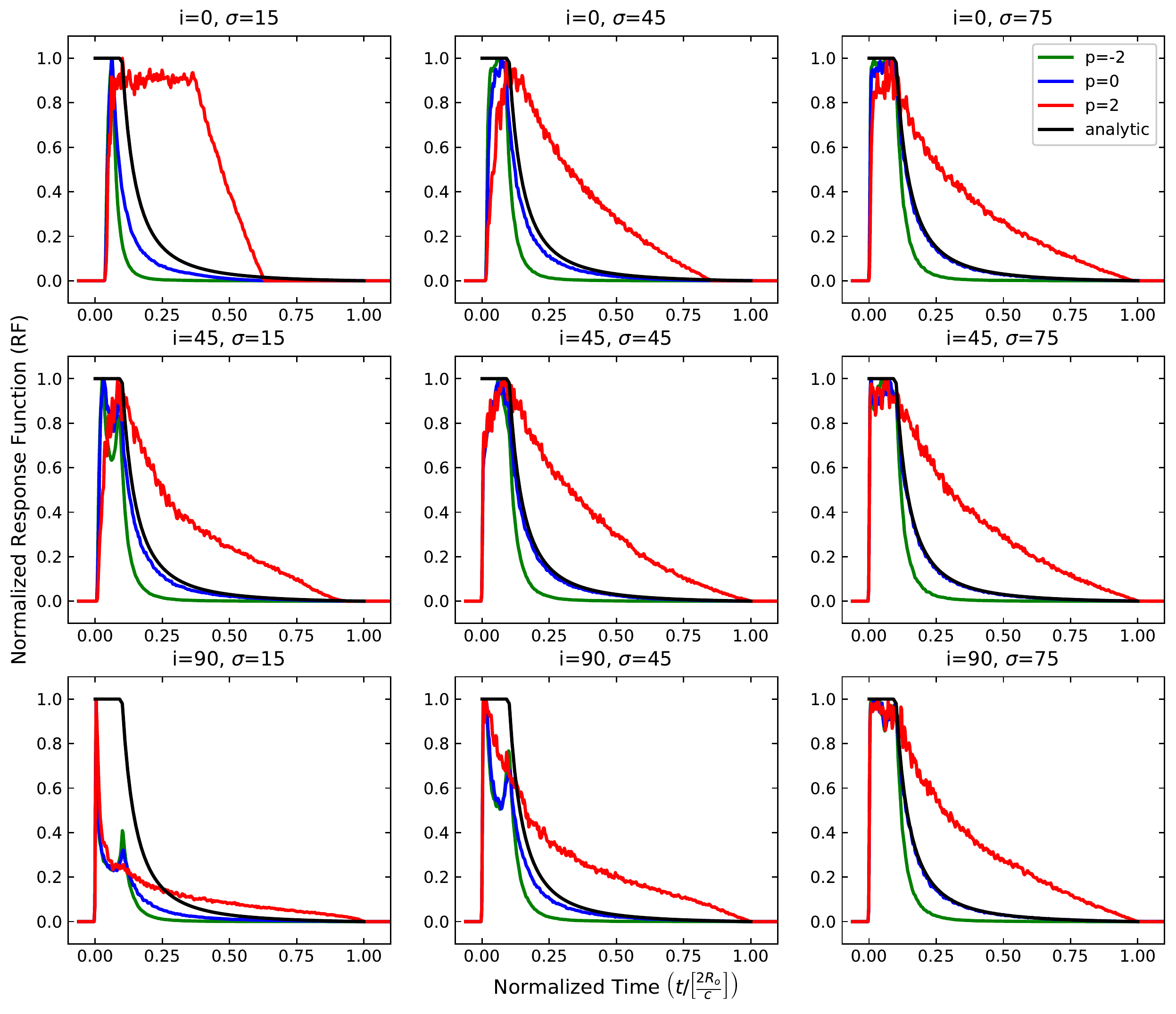}
\end{center}
\caption[RFs for BB simulations of different torus thickness cases]{\label{bbdisk3} \textbf{Torus of blackbody clouds versus $\sigma$:}  TORMAC RFs for a torus with $Y=10$ filled with isotropically emitting blackbody clouds for $p=-2$ (green), 0 (blue), and 2 (red). Each row represents a different inclination from top ($i$=0$^{\circ}$) to bottom ($i$=90$^{\circ}$). Each column represents a different $\sigma$ value increasing from left ($\sigma=15^{\circ}$) to right ($\sigma=75^{\circ}$). The black line represents the analytical transfer function for a spherical shell with $p=0$ for comparison. (The $p=2$ case appears below 1 due to numerical noise.)}
\end{figure*}

\subsection{Disk Response Functions for Blackbody Clouds}\label{sec:BBRF}

TORMAC was used to compute RFs for a disk containing isotropically emitting blackbody clouds.
Although the fuzzy torus case (Section~\ref{sec:model}) is probably a better representation of reality, only sharp-edged disk models are presented here as 
we are currently interested in how torus geometry affects the RFs.

Figures~\ref{bbdisk2} and~\ref{bbdisk3} illustrate the effect on the RFs of a sharp-edged disk when radial depth ($Y$), inclination ($i$), the angular width ($\sigma$), and the radial cloud distribution ($p$) are varied. Representative analytical solutions for the spherical shell transfer function are plotted (in black) for comparison. In Figure~\ref{bbdisk2}, the values of the descriptive quantities RWD, LWR, and CTAR are also indicated for each RF.

Just as for the spherical shell transfer function, the disk RFs reach their maxima for $\tau < 1/Y$ and have declining ``tails'' for $\tau \gtrsim 1/Y$, with the value of $p$ determining the rate at which the RF tail decays. However, whereas the spherical shell transfer function reaches its maximum at $\tau=0$ and is flat-topped for $\tau \leq 1/Y$, the structure of the disk RF, particularly within the core for delays $\tau \leq 1/Y$, also depends on  $i$ and $\sigma$.  
As discussed in A17, there is a delay before the onset of the response for  $i < 90^{\circ} -\sigma$ (Figures~\ref{bbdisk2} and~\ref{bbdisk3}; top row). This
results from the fact that the observer's line of sight falls within the hollow cone aligned with the polar axis (i.e.,  $\theta \le 90^{\circ} -\sigma$) so that there are no clouds 
that can respond with the shortest delays, $\tau\sim0$ (see Appendix~\ref{app:ltdelay}).

In addition, the RF becomes double-peaked for larger inclinations, with these features being most prominent when the disk is edge-on ($i=90^\circ$; Figures~\ref{bbdisk2} and~\ref{bbdisk3}; bottom row). Double peaks arise because the isodelay surface corresponding to the continuum pulse  intersects fewer clouds as it passes through the central cavity of the torus. 
The first peak is due to the response of clouds on the  near side of the torus relative to the observer; the second, lower amplitude, peak results from the more delayed 
response from its far side.
This double-peaked structure is suppressed, even for edge-on inclinations, at larger values of $p$ or for larger values of $\sigma$ (Figure~\ref{bbdisk3}). 
Increasing $p$ gives more weight to the outer regions in the response, resulting, as for the spherical shell, in a slower decay for $\tau > 1/Y$. 
As $\sigma$ is increased, the RFs more closely resemble those for a spherical shell, as expected (Figure~\ref{bbdisk3}). 

An interesting departure from the general trends seen in Figure~\ref{bbdisk3} is the case of the thin ($\sigma = 15^{\circ}$), face-on ($i=0^{\circ}$) disk with $p=+2$ (top left panel; red line). For this case, the RF is flat-topped with a steep decay, reminiscent of the spherical shell response for small $Y$ and $p<0$. This occurs because, for a thin disk with $p=+2$, elemental radial shells within the torus resemble annuli with constant luminosity -- even though the volume emissivity decreases as $1/r^2,$ the number of clouds per radial shell increases as $r^2$. The intersection of the isodelay surface with a thin, face-on disk is a circle, so the response to the continuum pulse propagates through the torus as an expanding annulus, which has a constant luminosity. 

In summary, the disk RFs show trends in the core width and the slope of the decay tail that are similar to those seen in the spherical shell models. However, they also exhibit features characteristic of a disk response, including a delay before the response onset at face-on inclinations and a double-peaked structure within the core for more edge-on inclinations. The presence and prominence of these features in general depend on $p$, $i,$ and $\sigma$.

\subsection{Descriptive Quantities}

\begin{figure}[!t]
\begin{center}
	\includegraphics[scale=0.49]{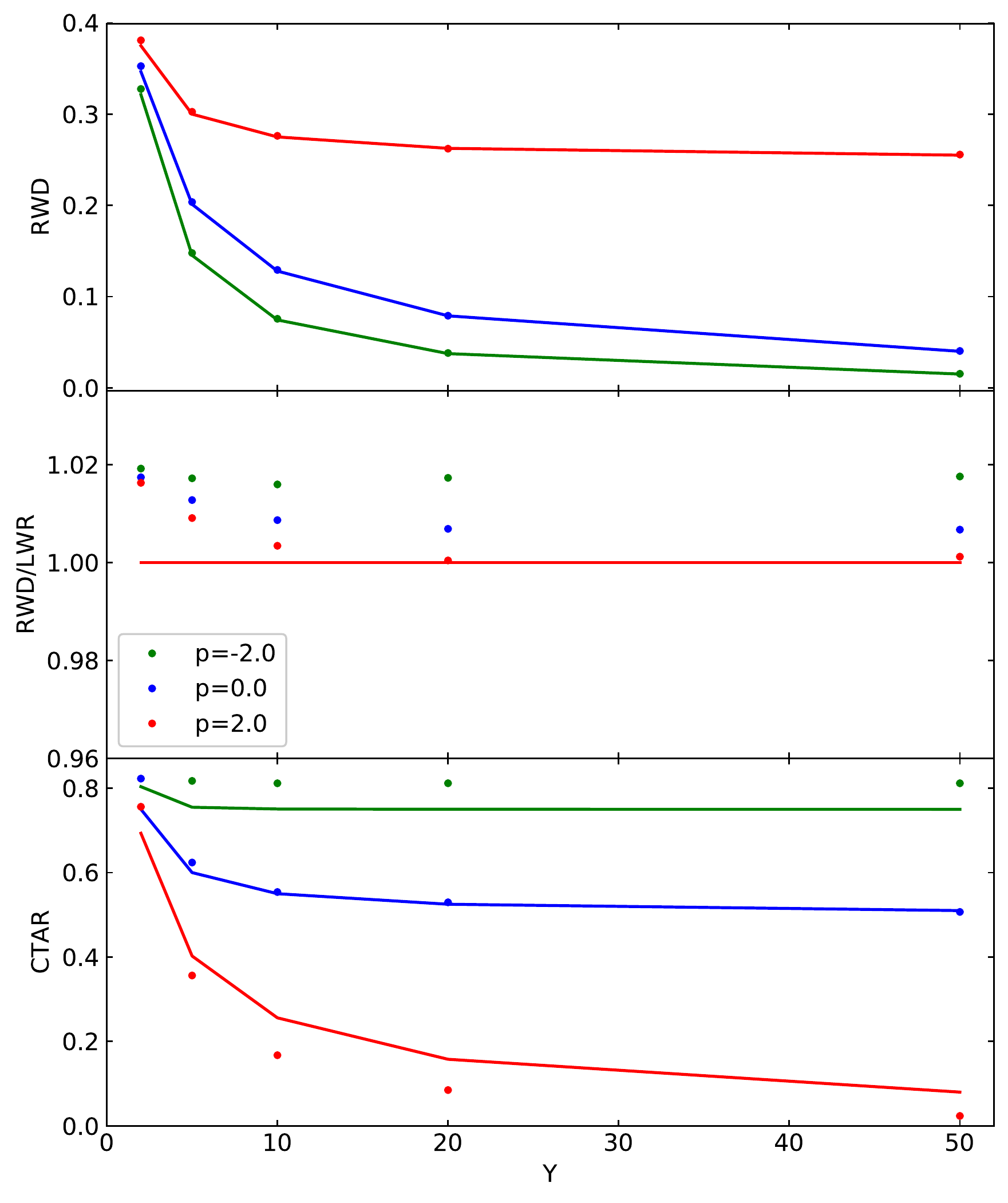}
\end{center}
\caption[DPs for BB simulations of a disk ($\sigma=45^{\circ}$) versus $Y$]{\label{rfdpBBdisk} \textbf{Torus of blackbody clouds versus $Y$:} Response function descriptive quantities for a torus filled with blackbody clouds with $\sigma=45^{\circ}$ and $i=0^{\circ}$, with respect to different values of $Y$ comparing different $p$ values. The solid lines represent the analytic solution for a spherical shell, and the circles represent the results from simulated RFs from TORMAC.  The colors of the symbols and the solid lines indicate the $p$ value.}
\end{figure}

\begin{figure}[!h]
\begin{center}
	\includegraphics[scale=0.49]{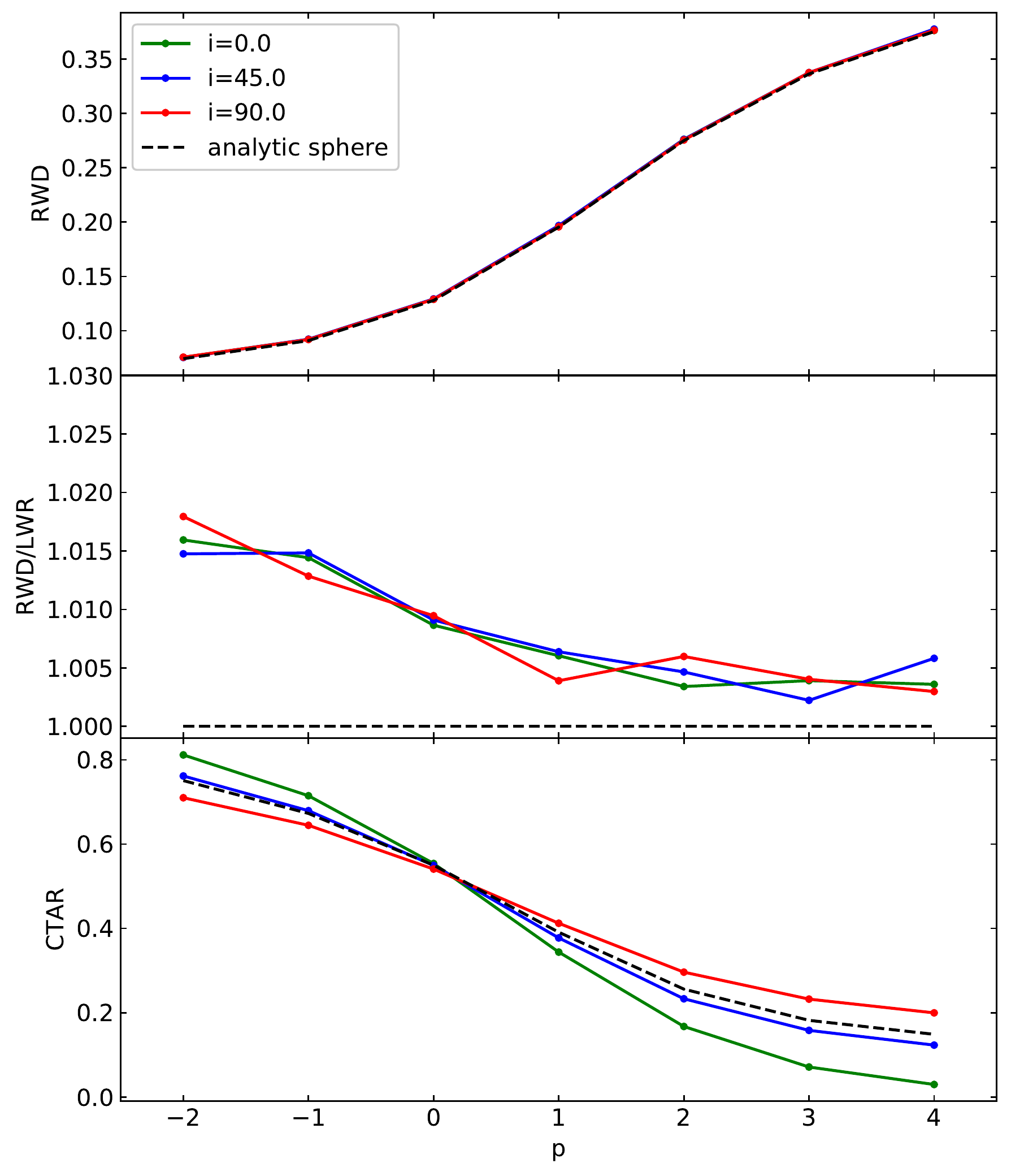}
\end{center}
\caption[DPs for BB simulations of a $Y=10$ disk ($\sigma=45^{\circ}$) versus vs all $p$]{\label{rfdpBBdisk_Y10} \textbf{Torus of blackbody clouds versus $p$ for different $i$:} Response function descriptive quantities for a torus filled with blackbody clouds with $\sigma=45^{\circ}$  and $Y=10$ with respect to different values of $p$ comparing different $i$ values. The black dashed lines represent the analytic solution for a spherical shell. The colored solid lines and circles represent the results from simulated RFs from TORMAC.  The colors correspond to the $i$ values.}
\end{figure}

\begin{figure}[!h]
\begin{center}
	\includegraphics[scale=0.49]{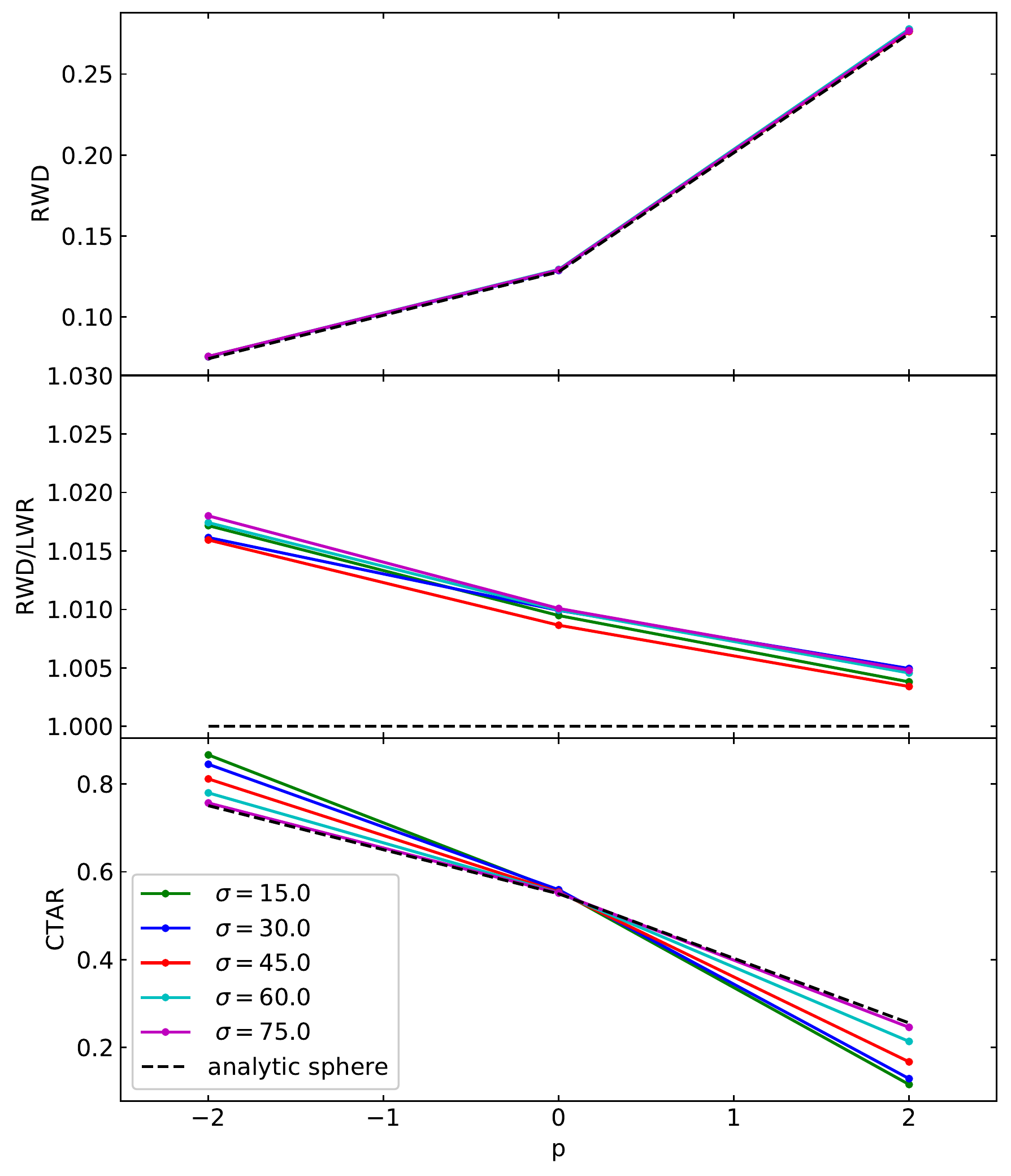}
\end{center}
\caption[DPs for BB simulations of a $Y=10$ face-on disk versus vs $p$]{\label{rfdpBBdisk_Y10i0}  \textbf{Torus of blackbody clouds versus $p$ for different $\sigma$ values:} Response function descriptive quantities for a face-on torus filled with blackbody clouds with $Y=10$ with respect to different values of $p$ comparing different $\sigma$ values. The solid lines and symbols represent the results from simulated RFs from TORMAC. The colors correspond to the $\sigma$ value that was varied. The black dashed lines represent the analytic solution for a spherical shell. }
\end{figure}

Figures~\ref{rfdpBBdisk} --~\ref{rfdpBBdisk_Y10i0} show the variations with $Y$, $p$, $i$, and $\sigma$ of the descriptive quantities for the computed RFs of a disk containing blackbody clouds. The corresponding values for a spherical shell calculated using the analytical formulae (Equations~\ref{eqn:LWR_ss} 
and~\ref{eqn:CTAR_ss}) are also plotted for comparison. Recall that the values of RWD, LWR, and CTAR for the disk models are also indicated in Figure~\ref{bbdisk2} so that they can easily be compared with the corresponding RFs. 

It can be seen in the top panels of Figures~\ref{rfdpBBdisk} --~\ref{rfdpBBdisk_Y10i0} that the RWD for the {\em disk} shows the same behavior  as the spherical shell RWD when  $Y$ and $p$ are varied, i.e., decreasing as $Y$ increases but increasing as $p$ increases. Indeed, the RWD values obtained from the computed disk RFs are essentially identical to those of the analytic spherical shell solutions.  Furthermore, the RWD for the disk is independent of $i$ and $\sigma$ (Figures~\ref{rfdpBBdisk_Y10} and~\ref{rfdpBBdisk_Y10i0}) -- it depends {\em only} on $Y$ and $p$ just as the analytical spherical shell case does.

The ratio RWD/LWR is plotted in the middle panel of Figures~\ref{rfdpBBdisk} --~\ref{rfdpBBdisk_Y10i0}. As noted in Section~\ref{sec:anaDP}, RWD/LWR = 1 for the spherical shell transfer function. The computed values for the disk RFs are also $\approx 1$, to within 2\%. The disk values are systematically slightly higher ($>1$) and also show systematic decreasing trends with $Y$ and $p$. These small differences result from the finite width of the input pulse used in TORMAC to compute the RFs, which has a larger effect for smaller $Y$ and $p$. 

The quantity that describes the RF shape, CTAR, is plotted in the bottom panels of Figures~\ref{rfdpBBdisk} --~\ref{rfdpBBdisk_Y10i0}. The CTAR values for the disk RF follow the same general trends when $Y$ and $p$ are varied as for the spherical shell model, decreasing as both $Y$ and $p$ increase. In detail, however, the behavior of the disk CTAR is not the same as for the spherical shell and depends on $p$, $i$, and $\sigma$. For example, for a face-on disk ($i=0^{\circ}$), CTAR is higher than the spherical shell value for $p<0$, but lower for $p>0$, whereas the edge-on disk ($i=90^\circ$) exhibits the opposite behavior (Figure~\ref{rfdpBBdisk_Y10}; green and red lines, respectively). These differences arise because, for $p\leq 0$, the core of the RF is more prominent when the disk is face-on, but less prominent when it is edge-on, than is the case for $p>0$ (Figure~\ref{bbdisk3}). The RF has a higher CTAR value for $p\leq 0$ when the disk is face-on than when it is edge-on because the RF is double-peaked in the latter case, causing it to have a smaller core area. 

CTAR also varies slightly with $\sigma$ {(Figure~\ref{rfdpBBdisk_Y10i0})}. Not surprisingly, the difference between the disk and spherical shell CTAR values at a given $p$ value increases as $\sigma$ is decreased, since the disk is becoming thinner, i.e., less ``spherical''. Conversely, as $\sigma$ is increased, the disk CTAR values converge to the spherical shell case. 

In summary, the disk RF descriptive quantities behave in a very similar way to the analytical solutions for a spherical shell.
The disk RWD exhibits essentially identical variations to those of the spherical shell model. It depends only on $Y$ (decreasing with increasing $Y$) and $p$ (increasing with increasing $p$), and it is independent of $i$ and $\sigma.$ In addition, within the limitations of the numerical approximation, RWD/LWR $= 1$ for the disk in all cases, as predicted by the analytical solution for a spherical shell, and indicating that, for this idealized torus, the reverberation lag is a precise indicator of its effective radius. Finally, CTAR is mainly dependent on $Y$ and $p$, decreasing as both parameters increase, as for the spherical shell model, but it also varies slightly with both $\sigma$ and $i$.


\section{Dust Emission Response Functions: Individual Effects}\label{sec:DustRF} 

 The blackbody cloud models considered in Section~\ref{sec:rBBRF} serve to establish the general features of the response of a disk with respect to the basic geometrical parameters ($Y$, $p$, $i$, $\sigma$) and the radial cloud distribution ($p$). In this section and Section~\ref{sec:combo} we analyze the multiwavelength IR dust emission response from a clumpy torus, now composed of dust clouds whose emission is determined using a grid of radiative transfer models as described in Section~\ref{sec:model}. The IR response will be explored at selected NIR and MIR wavelengths ($\lambda=$2.2, 3.6, 4.5, 10, 30 $\mu$m). These wavelengths were chosen to represent a range of dust temperatures characteristic of the torus, from the hottest dust near the sublimation radius to the cooler dust emitting near the peak of the AGN IR SED.

In Section~\ref{sec:combo}, we will investigate the responses of ``globally optically thick'' torus models, including the effects of cloud orientation, cloud shadowing, and cloud occultation.
However, in order to illustrate how each of these effects individually influences the dust emission response, they will be introduced and summarized one at a time in this section, starting with a ``globally optically thin'' torus, in which 
the individual clouds are optically thick but the global opacity effects (cloud shadowing and occultation) are omitted.

As the AGN central source is generally assumed to be a geometrically thin, optically thick accretion disk, we will also consider the effects of anisotropic illumination of the torus. 

The results discussed in this, and the following section are based on a large set of models computed for selected combinations of the parameter values listed in Table~\ref{tormac_param}. 
As noted in Section~\ref{sec:model}, we have explored the behavior of
the RFs along each axis of the multi-dimensional parameter space rather than uniformly sampling all possible combinations. 
In these models, the radial depth of the torus is varied over the range $Y=2 - 50$, representing compact to radially extended configurations. 
The inclination of the torus axis to the observer's line of sight ranges from $i=0$ (face-on) to $i=90^{\circ}$ (edge-on), and its angular width ranges from $\sigma =15$ (thin disk) to $\sigma =60^{\circ}$ (thick disk). 

The power-law index of the radial cloud distribution has values in the range $p=-2$ to 4, corresponding to cloud number density distributions $n(r)\propto r^{-4}$ to $r^{2}$. 
The individual dust clouds have optical depths ranging from  $\tau_{V}=5$ to $100,$ and the torus volume filling factor is varied between $\Phi=0.001$ and 0.1. 
Finally, the degree of anisotropy of the illuminating radiation field is controlled by the parameter $s$, which has values ranging from $s=0.01$ (highly anisotropic) to 0.99 (essentially isotropic). With the exception of $s$, the adopted ranges in the model parameters are loosely based on those found by \cite{Nenkova:2008ab} to generally reproduce the IR SEDs of AGN.

Also, as mentioned in Section~\ref{sec:model}, we define a reference model defined by a standard set of parameter values: $Y=10$, $i=0^{\circ}$, $p=0$, 
$\sigma=45^{\circ}$, and $\tau_{V}=40$.  
For models that include cloud shadowing or occultation or both, we adopt $\Phi = 0.01$ as the standard value of the torus volume filling factor. For this value of $\Phi$, the average number of clouds intercepting an equatorial ray is $\approx$ 5.
The radiation field asymmetry parameter is $s=1$ for models where the torus is illuminated isotropically, and it has a standard value $s=0.1$ for models where the torus is illuminated anisotropically. We have adopted the sharp-edged disk geometry as the standard case, for simplicity, but the RFs of the ``fuzzy-edged'' torus case are also discussed in Section~\ref{sec:combofuz}.

Recall that all RFs are the average of 5 simulation runs, each including 50,000 clouds, and are normalized as described in Section~\ref{sec:rBBRF}.


\subsection{Globally Optically Thin Torus}\label{sec:aniscloud}

Here we consider the effects on the torus response of radiative transfer in individual dust clouds. As these effects have already been discussed for a limited set of torus parameters in A17, here we will briefly recap the main points for completeness and investigate how the response changes with respect to the radial cloud distribution and optical depth.
In these models, the torus is considered 
to be globally optically thin in the sense that all clouds are directly illuminated by the AGN and their emission reaches the observer unattenuated by intervening clouds. This situation corresponds to the case of a  torus that has a negligible volume filling factor, but in practice, cloud shadowing and cloud occultation effects are simply turned off. This is equivalent to setting $\Phi \leq 0.001$  in the globally optically thick models of Section~\ref{sec:combo}.

\subsubsection{Isotropic Illumination}

\begin{figure*}[!h]
\begin{center}
	\includegraphics[scale=0.6]{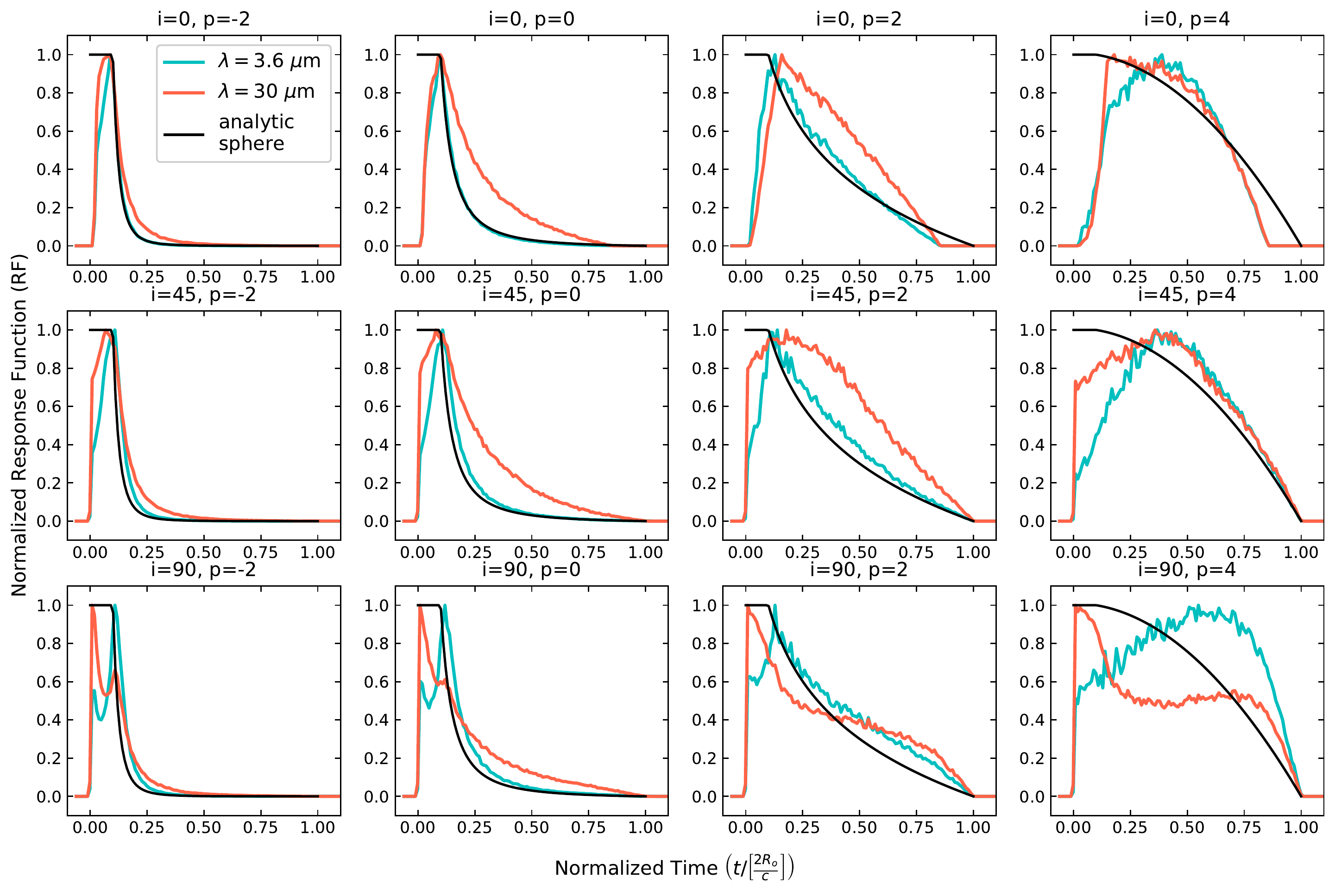}
\end{center}
\caption{\label{clorient_p36} \textbf{Globally optically thin torus versus $p$:} Response functions for an isotropically illuminated torus with $\sigma$=45$^{\circ}$ and $Y$=10 at 3.6 (blue) and 30 $\mu$m (red/orange), where each row represents a different inclination from top ($i$=0$^{\circ}$) to bottom ($i$=90$^{\circ}$). Each column represents a different $p$ value increasing from left ($p=-2$) to right ($p=4$). The torus is filled with dust clouds that emit anisotropically (i.e., including cloud orientation), and the black line is the analytic transfer function for a spherical shell. }
\end{figure*}

We first consider the case in which the torus is illuminated isotropically by the AGN continuum. As discussed in A17, the emitted flux decreases more rapidly with cloud surface temperature (and hence with radius) at shorter wavelengths, with the result that the  torus IR emission responds to the AGN continuum variations more quickly and with a larger amplitude at shorter wavelengths than at longer wavelengths. The effect is similar to that of varying $p$; at shorter wavelengths, the RF decays more steeply for $\tau > 1/Y$ (as it does when $p \le 0$) than for longer wavelengths (much as for $p > 0$; see Figure~\ref{bbdisk3}). 

In addition, as a result of the temperature gradient within individual clouds, the dust cloud emission is also strongly anisotropic at shorter wavelengths,  with the difference between the  fluxes emerging from the illuminated and non-illuminated sides of the cloud becoming smaller as wavelength increases. Therefore, the contribution of a given cloud to the torus response, especially at wavelengths $\lesssim 10\,\mu$m, depends on the fraction of its illuminated surface that is visible to the observer, which varies with its position relative to the central source and the line of sight to the observer. The main effect is to reduce the response amplitude at short delays $\tau \sim 0$, which tends to increase the RWD. 

This ``cloud orientation'' effect is much weaker at longer wavelengths ($\gtrsim 10\,\mu$m), since the clouds are effectively optically thin at these wavelengths and thus emit essentially isotropically.  

These wavelength-dependent effects can be seen in Figure~\ref{clorient_p36}, which shows RFs at 3.6 and 30 $\mu$m for an isotropically illuminated torus for several values of $p$ and $i$.
The tail of the RF decays much more gradually at $30 \,\mu$m  than at 3.6 $\mu$m  for $p\leq 0$, due to the slower variation of the dust cloud emissivity with torus radius at longer wavelengths. 
In the core,  the response at $3.6\,\mu$m is reduced at short delays ($\tau \lesssim 1/(2Y)$) but enhanced at larger delays, with the largest differences occurring at larger inclinations and larger $p$ values. The strength of this effect increases with inclination because for the clouds on the side of the torus closer to the observer a larger fraction of their cooler, non-illuminated side is displayed to the observer, decreasing the response at shorter delays, whereas the far-side clouds display a larger fraction of their illuminated side, increasing the response at longer delays. 
At 30 $\mu$m, the clouds emit effectively isotropically, so cloud orientation effects are insignificant.

The variations of the descriptive quantities with $p$ are compared with those for the globally optically thick case in Section~\ref{sec:gotp} (see  Figure~\ref{rfdpdiffpcombo0}).  As a result of the partial suppression of the response at short delays for shorter wavelengths, due to the cloud orientation effect, the RFs for the globally optically thin models typically reach their maxima sooner for wavelengths $>10\,\mu$m than for shorter wavelengths. Nevertheless, the RWD generally increases with wavelength and also increases systematically with $p$ as in the blackbody case (Figure~\ref{rfdpdiffpcombo0}, top panel, stars and dashed lines). 
The RWD/LWR ratio is not very sensitive to $p$ but is generally  $> 1$, with larger values at shorter wavelengths, reaching $\approx 1.5$ for $\lambda =2.2\,\mu$m (Figure~\ref{rfdpdiffpcombo0}, middle panel).

\begin{figure*}[!ht]
\begin{center}
	\includegraphics[scale=0.6]{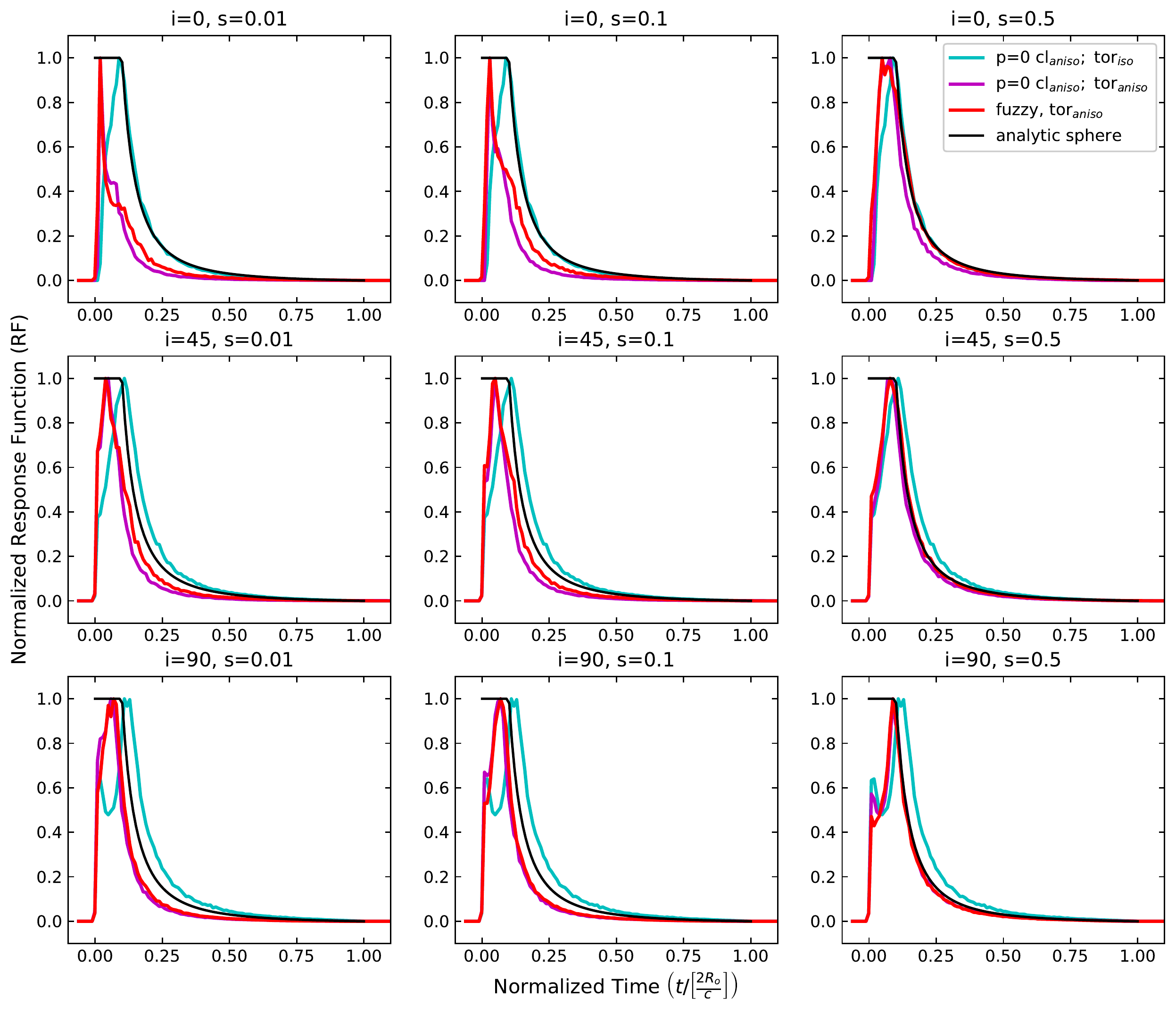}
\end{center}
\caption{\label{clorientvsani_sp36}  \textbf{Globally optically thin torus versus $s$:} Response functions for an isotropically or anisotropically illuminated torus with $p$=0, $\sigma=45^{\circ}$, and $Y$=10 at 3.6 $\mu$m, where each row represents a different inclination from top ($i$=0$^{\circ}$) to bottom ($i$=90$^{\circ}$). Each column represents a different $s$ value increasing from left ($s=0.01$) to right ($s=0.5$). The blue lines represent a sharp-edged isotropically illuminated torus. The purple lines represent a sharp-edged torus, and the red lines represent a fuzzy-edged torus, both of which are anisotropically illuminated. The black line is the analytic spherical transfer function.}
\end{figure*}

As would be expected, the strength of the cloud orientation effect also depends on the optical depth, $\tau_V$, of the individual clouds. This will be discussed, in comparison with the corresponding globally optically thick torus models, in Section~\ref{sec:combo}.


	\subsubsection{Anisotropic Illumination}

As discussed in Section~\ref{sec:model}, the AGN UV/optical continuum illuminating the torus is expected to be anisotropic due to edge darkening in a thin accretion disk. To model this
in TORMAC, the inner radius of the torus (i.e., the dust sublimation radius) has a polar angle dependence given by Equation~\ref{Rdtheta}, which is controlled by the parameter $s$. Anisotropic illumination ($s < 1$) results in a dust sublimation surface that resembles a figure eight in a vertical plane through the center of the torus (A17, Figure~10). This allows dust clouds to reside closer to the AGN continuum source near the equatorial plane than is the case for isotropic illumination.

This has an important effect on the RF at all wavelengths. Response functions for an anisotropically illuminated, globally optically thin torus (purple line) are shown in Figures~\ref{clorientvsani_sp36} and~\ref{clorientvsani_sp30} (Appendix~\ref{app:figs}) for 3.6 and 30 $\mu$m, respectively. The shorter light-travel times to the innermost clouds result in RFs that peak at shorter lags and  exhibit sharper and narrower features than their counterparts for isotropic illumination (blue line). The effects are most dramatic at shorter wavelengths, where the emission is dominated by the hotter, innermost clouds.

The effect of anisotropic illumination on the RFs is controlled by $s$. The isotropic case is recovered as $s\rightarrow 1$, but there is little difference between the RFs for $s=0.01$ and $s=0.1$,
showing that the main effects of anisotropic illumination are already present at $s =0.1$. 

The variations of the descriptive quantities with $s$ are shown in Figure~\ref{rfdpdiffspvsfuzi45} (Appendix~\ref{app:figs}). In general, the RWD decreases as $s$ is decreased, reaching a value at 3.6 $\mu$m (dark blue line) that is $\sim 30$\% smaller than that of the isotropically illuminated case ($s=1$) at the lowest value of $s=0.01$; at longer wavelengths, the  change is somewhat smaller.
The ratio RWD/LWR (middle panel Figure~\ref{rfdpdiffspvsfuzi45}) is relatively insensitive to $s$, increasing only slightly as $s$ decreases.


\begin{figure*}[!ht]
\begin{center}
	\includegraphics[scale=0.6]{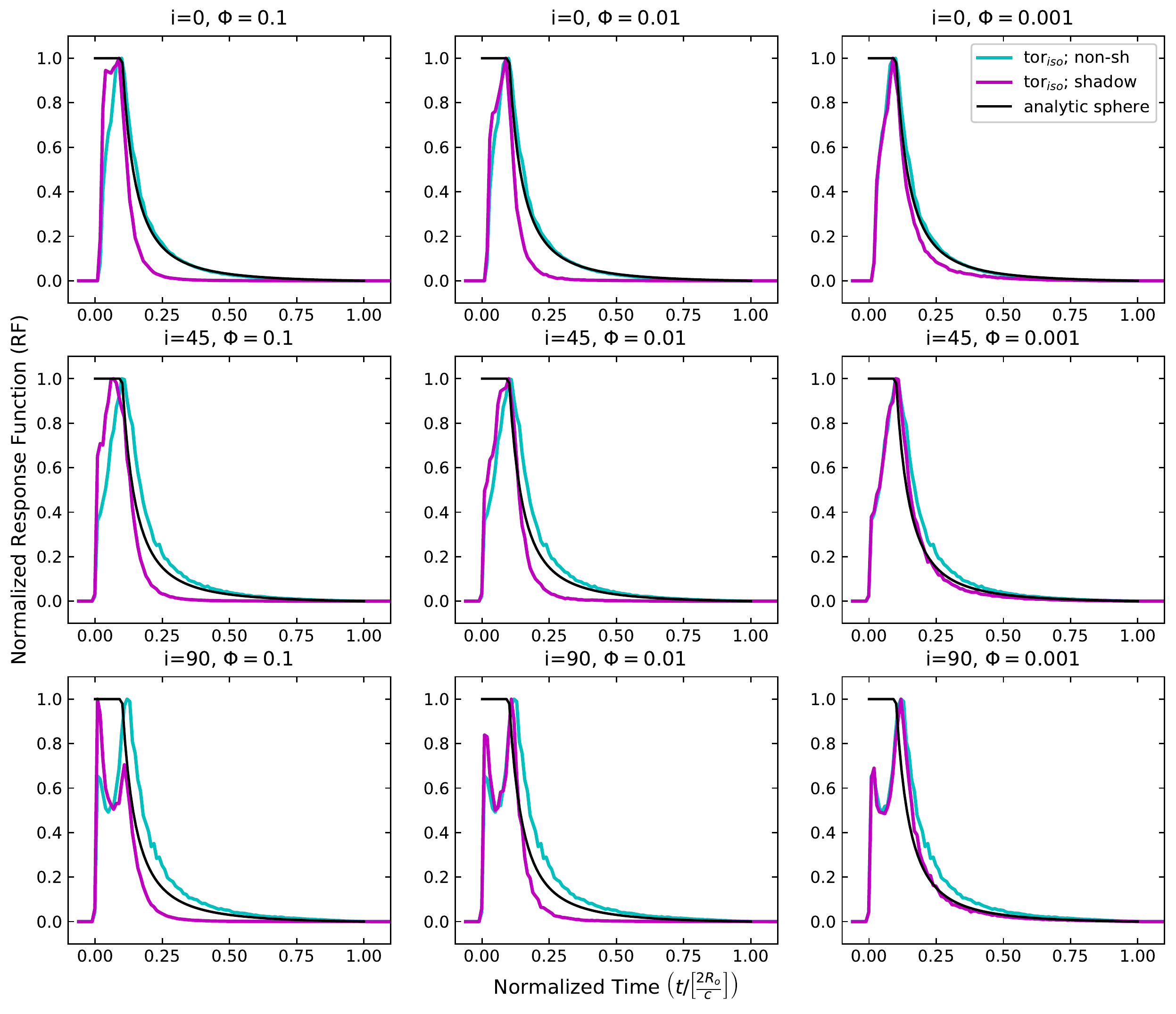}
\end{center}
\caption{\label{clshadow_vff36}  \textbf{Torus with cloud shadowing versus $\Phi$:} Response functions for an isotropically illuminated torus with $\sigma=45^{\circ}$; $Y=10$; $p=0$; for $i=0,$ 45, 90$^{\circ}$ at 3.6 $\mu$m. The blue lines represent simulations without cloud shadowing, the purple lines represent the simulations with cloud shadowing for values of the average volume filling factor ranging from $\Phi$=0.001 (right) to 0.1 (left), and the black line is the analytic spherical shell transfer function. } 
\end{figure*}

\subsection{Cloud Shadowing}\label{sec:shadowing}
	
There is a chance that any given cloud will have the light from the central source blocked by intervening clouds at smaller radii and thus will not be directly illuminated by the AGN continuum source.
These shadowed clouds are heated instead by the diffuse radiation field of nearby directly heated clouds and therefore emit isotropically. 
The effects of cloud shadowing on the RF were briefly discussed in A17, where it was noted that it has a dramatic effect on the RFs at shorter wavelengths (3.6 $\mu$m  and $\lambda<10\,\mu$m, in general), but a relatively small effect at longer wavelengths ($\lambda  = 10$, 30 $\mu$m).

The average number of clouds intercepting a given radial ray increases with the volume filling factor, $\Phi$. For example, in the cases shown in Figures~\ref{clshadow_vff36} and~\ref{clshadow_vff30} (Appendix~\ref{app:figs}), an equatorial ray is intercepted by $\approx$ 22, 5, and 1 cloud(s) for $\Phi$=0.1, 0.01, and 0.001, respectively\footnote{Note that for $\Phi \gtrsim 0.1$ the approximation used to compute the spectra of the indirectly heated clouds breaks down, as noted by \cite{Nenkova:2008aa}. This case is included here as an extreme limiting case but should not be considered ``realistic''.}. The probability that a given cloud will be shadowed also increases with its radial distance, $r$, and also depends on the radial cloud distribution.  Therefore, for given values of $Y$ and $p$, more clouds are shadowed as $\Phi$ increases, resulting in a more centrally concentrated distribution of directly heated clouds. This has an effect similar to that of decreasing $p$ and results in a higher response amplitude in the core at short delays ($\tau\sim 0$) as compared with longer delays ($\tau\sim 1/Y$), thus partly offsetting the effects of cloud orientation (purple lines in Figure~\ref{clshadow_vff36}). The RF tail also decays more steeply as $\Phi$ increases. These effects are strongest at shorter wavelengths, where the emission is dominated by directly heated clouds (Figure~\ref{clshadow_vff36}). 
 At longer wavelengths (e.g., $30\,\mu$m; Figure~\ref{clshadow_vff30}; Appendix~\ref{app:figs}), the introduction of cloud shadowing has relatively little effect on the RFs. At these wavelengths, even the directly heated clouds emit essentially isotropically, and as more clouds are shadowed, heating by the diffuse radiation field is sufficient to maintain the dust emissivity.

Reflecting the changes in the RFs, the RWD decreases and the CTAR increases with increasing $\Phi$, for wavelengths $\leq 10\,\mu$m (Figure~\ref{rfdpdiffvffvsi045p0}; Appendix~\ref{app:figs}).  
For $\Phi \gtrsim 0.01$, RWD $\sim 1/Y$, the light-crossing time of the torus inner cavity, for $\lambda =2.2$, 3.6, and 4.5 $\mu$m. 
The increase in CTAR reflects the steeper decay of the RF tail as shadowing effects become more important (as seen in Figure~\ref{clshadow_vff36}). 
The RWD/LWR ratio shows little variation with $\Phi$ at $i=0^{\circ}$ (circles and solid lines) but for the shorter wavelengths tends to decrease slowly with $\Phi$ at $i=45^{\circ}$ (stars and dashed lines).

In general, the introduction of cloud shadowing has the effect of decreasing both the RWD and the RWD/LWR ratio compared to the values of the globally optically thin torus models.  

\subsection{Cloud Occultation}\label{sec:cloccrf}
\begin{figure}[!t]
\begin{center}
	\includegraphics[scale=0.5]{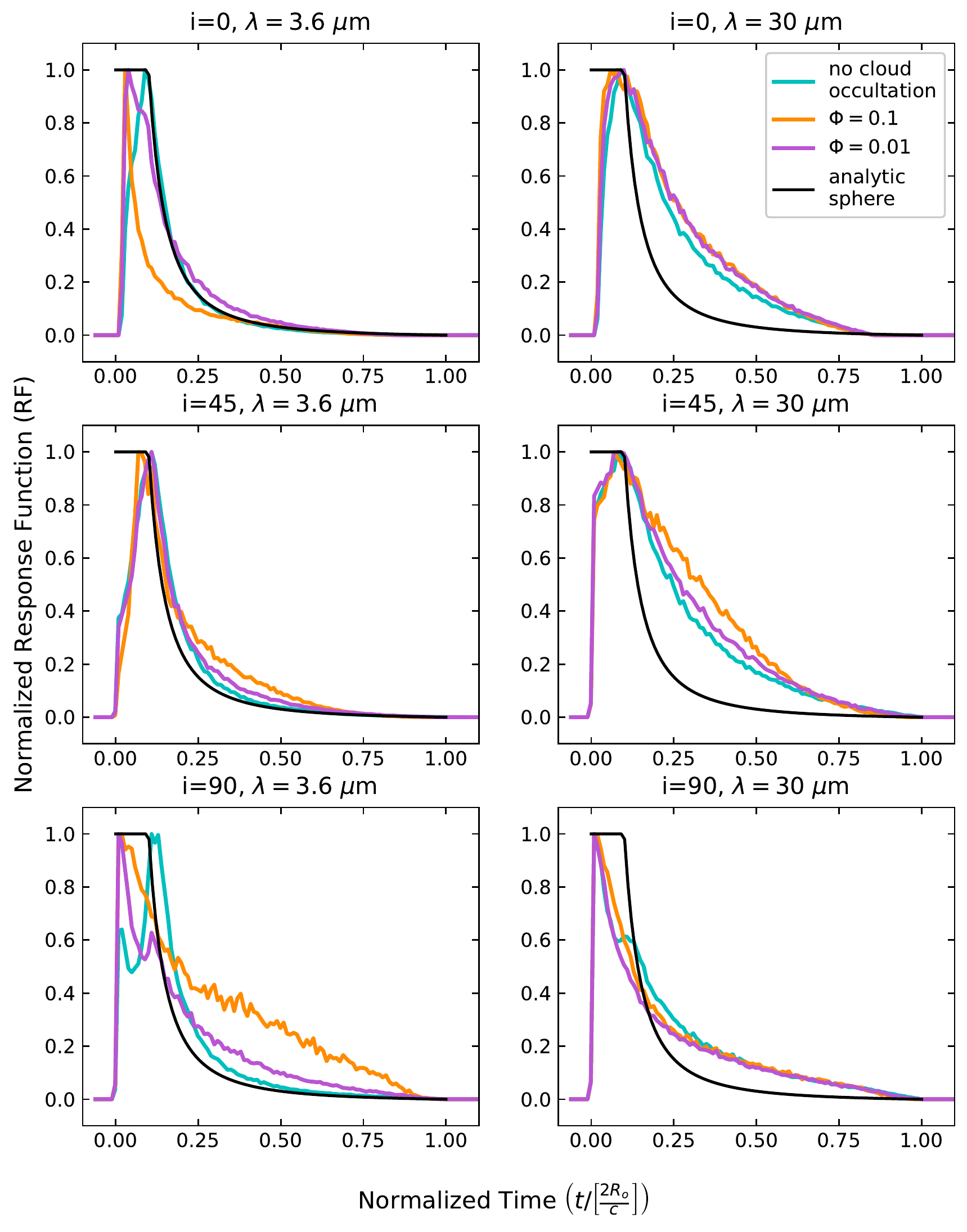} 
\end{center}
\caption{\label{clocc_vff3630}  \textbf{Torus with cloud occultation versus $\Phi$:} Response functions for an isotropically illuminated torus with $\sigma$=45$^{\circ}$; $Y=10$; $p=0$; for $i=0,$ 45, 90$^{\circ}$ at 3.6 (left) and 30 $\mu$m (right).  The blue lines represent simulations without cloud occultation (or $\Phi\sim0.0001$); the orange and purple lines represent the simulations with cloud occultation for $\Phi$=0.1 and 0.01, respectively; and the black line is the analytic spherical model. } 
\end{figure}

The dust emission from any one cloud may be intercepted and attenuated by other intervening clouds that happen to lie along the same line of sight to the observer. 
 The strength of the attenuation depends on the cloud optical depth and the number of occulting clouds, which itself depends on the path length through the torus along the observer's line of sight, the radial cloud distribution ($p$), and the volume filling factor, $\Phi$ (Section~\ref{sec:model_clocc}). 
 
Cloud occultation has the strongest effect at short wavelengths, where the optical depth is relatively large (e.g., $\tau_{3.6\,\mu\rm{m}} \approx 1.4$ for the standard value of $\tau_V = 40$). 
However, in contrast to the effects of cloud orientation and cloud shadowing, cloud occultation can also have significant effects even at $30\,\mu$m (where $\tau_{30\,\mu\rm{m}} \approx 0.4$ for $\tau_V = 40$) as a result of cumulative extinction by multiple clouds (Figure~\ref{clocc_vff3630}, right column).

For given values of $\Phi$, $p$, and $Y$, the effect on the RF is strongly dependent on inclination (and, to a lesser extent, the angular width, $\sigma$), since this determines the path length through the torus that each cloud's emission must traverse to reach the observer, and hence the regions of the torus that suffer least occultation (Figure~\ref{clocc_vff3630}).

At $i=0^{\circ}$, the early response of the RF at the shorter wavelengths is dominated by clouds around the inner ``wall'' of the torus, closest to the observer (i.e., $r\approx R_{\rm d, iso}$, $\beta\approx \sigma$; points A and A$^\prime$ in Figure~\ref{torus_isodelay}). This results, if $\Phi$ is large, in a sharp peak at a delay $\tau\approx (1-\sin(\sigma))/(2Y)\approx 0.015$ (Equation~\ref{eq:resbegin}). For smaller $\Phi$, clouds around $R_{\rm d, iso}$ on the far side of the inner ``wall'' (i.e., around points C and C$^\prime$ in Figure~\ref{torus_isodelay}) are less likely to suffer occultation, resulting in a broader peak, with a shoulder at a delay  $\tau\approx (1+\sin(\sigma))/(2Y)\approx 0.085$ (see the orange and purple lines in Figure~\ref{clocc_vff3630}, top left panel).

At $i=45^{\circ}$, cloud occultation has relatively little effect on the response, since (for $\sigma\leq 45^{\circ}$) the clouds along the inner ``wall'' of the torus farthest from the observer ($r\approx R_{\rm d, iso}$, $-\sigma \leq \beta\leq \sigma$; between points A$^\prime$ and C$^\prime$ in Figure~\ref{torus_isodelay}) are not occulted. These clouds, which have their illuminated faces toward the observer, dominate the response at delays $\tau\lesssim 1/Y$, as for the globally optically thin models.

Lastly, at $i=90^{\circ}$, the path lengths are maximized for most clouds. The early response at short wavelengths is, nevertheless, dominated by the hot clouds around $R_{\rm d, iso}$. Although a large fraction are occulted, and therefore their short-wavelength emission is attenuated, they are still brighter at these wavelengths than the cooler clouds at larger radii in the near side of the torus, even though the latter are less likely to be occulted. However, at larger values of $\Phi$ attenuation of emission from the inner clouds becomes more severe, with the result that the amplitude of the RF core is reduced relative to that of the tail (see Figure~\ref{clocc_vff3630}, bottom left panel).

The strength of both the cloud orientation and cloud occultation effects depends on the cloud optical depth, but in fact the RFs (and the RWD) are relatively insensitive to $\tau_V$, except at $i=90^{\circ}$ (Figures~\ref{clocc_tau3630} and~\ref{rfdpvstaui0clocc}, in Appendix~\ref{app:figs}). At a given wavelength, cloud occultation tends to become more important as $\tau_V$ increases, until the clouds become individually optically thick at that wavelength, 
when cloud orientation effects take over as the main factor in modifying the RF. 
For instance, at 3.6 $\mu$m the effects of cloud orientation become more important than cloud occultation for $\tau_{V}=100$ (for which $\tau_{3.6\,\mu\rm{m}} \approx 3.4$ and $\tau_{30\,\mu\rm{m}} \approx 1.1$), causing the peak response to shift to longer delays (Figure~\ref{clocc_tau3630}, purple line).
 
 The RWD varies slowly with $\Phi$ at  $i=0^{\circ}$ (Figure~\ref{rfdpdiffvffvsi0p0clocc}, Appendix~\ref{app:figs}; top panel, circles and solid lines), despite the large changes in the RF. The sharper and earlier peak that develops as $\Phi$ increases is evidently balanced by the slower decay of the tail. At higher inclinations, the RWD increases monotonically but relatively slowly as $\Phi$ increases (stars and dashed lines for $i=45^{\circ}$ in Figure~\ref{rfdpdiffvffvsi0p0clocc}).
 
The RWD/LWR ratio generally decreases as $\Phi$ increases for all $i$ (middle panel, Figure~\ref{rfdpdiffvffvsi0p0clocc}), although the change is more gradual for $i=45^{\circ}$ (stars and dashed lines) than at $i=0^{\circ}$ (circles and solid lines). 
The decrease in RWD/LWR is mainly due to the LWR, which increases as a result of cloud occultation effects. It is also worth noting that RWD/LWR can reach values $<1$ as $\Phi$ increases at most wavelengths; thus, the RWD can underestimate the LWR when cloud occultation is included. 

\begin{figure}[!t]
\begin{center}
	\includegraphics[scale=0.5]{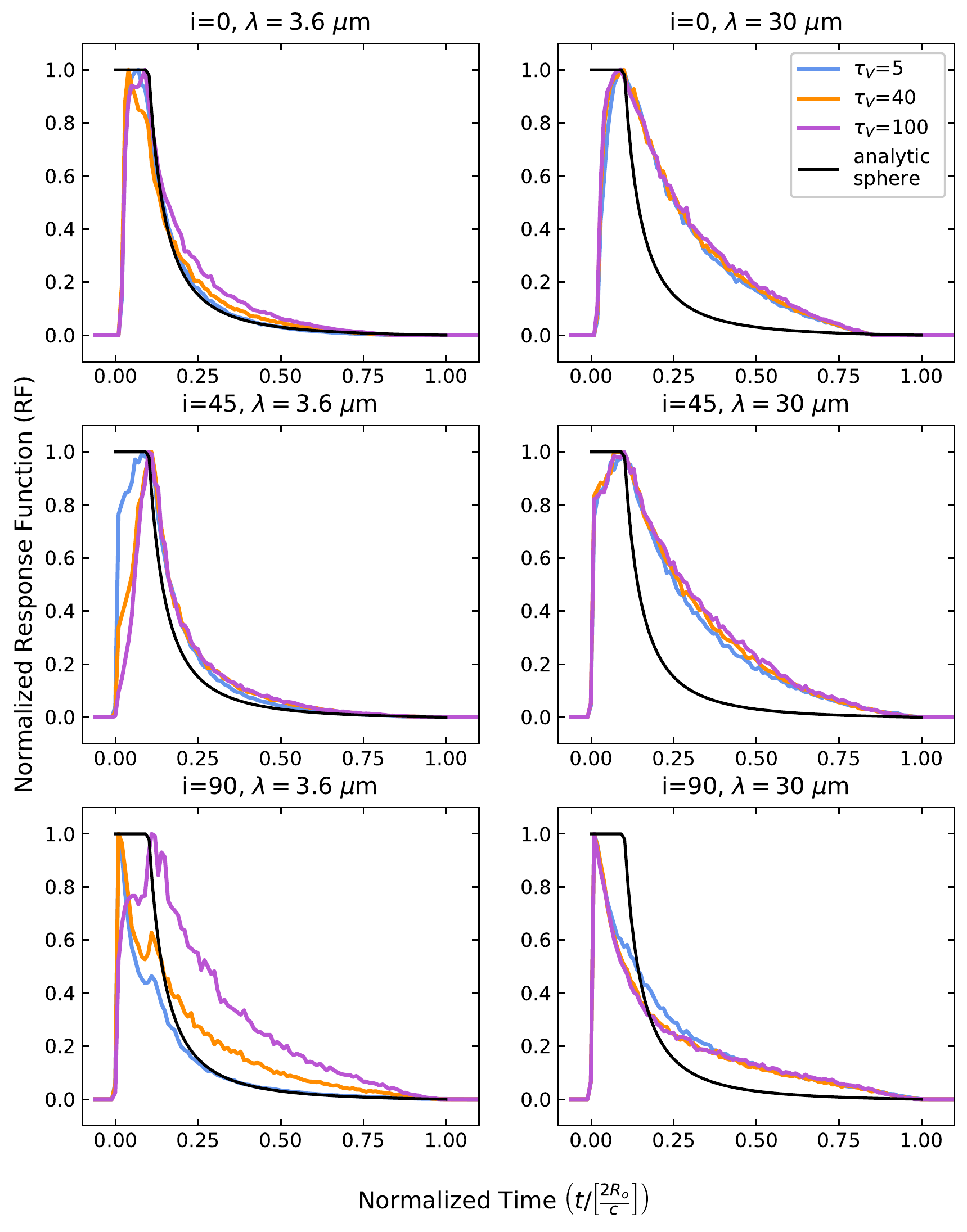}
	
\end{center}
\caption{\label{clocc_tau3630}  \textbf{Torus with cloud occultation versus $\tau_{V}$:} Response functions for an isotropically illuminated torus with $\sigma$=45$^{\circ}$; $Y=10$; $p=0$; $i$=0, 45, 90$^{\circ}$; and $\Phi=0.01$ at 3.6 (left) and 30 $\mu$m (right).
Each panel compares the tori models for clouds with $\tau_{V}=5$ (blue), 40 (orange), and 100 (purple).  The black line represents the analytic transfer function for a spherical shell.  All models include cloud orientation and cloud occultation.}
\end{figure}

\begin{figure*}[!t]
\begin{center}
	\includegraphics[scale=0.6]{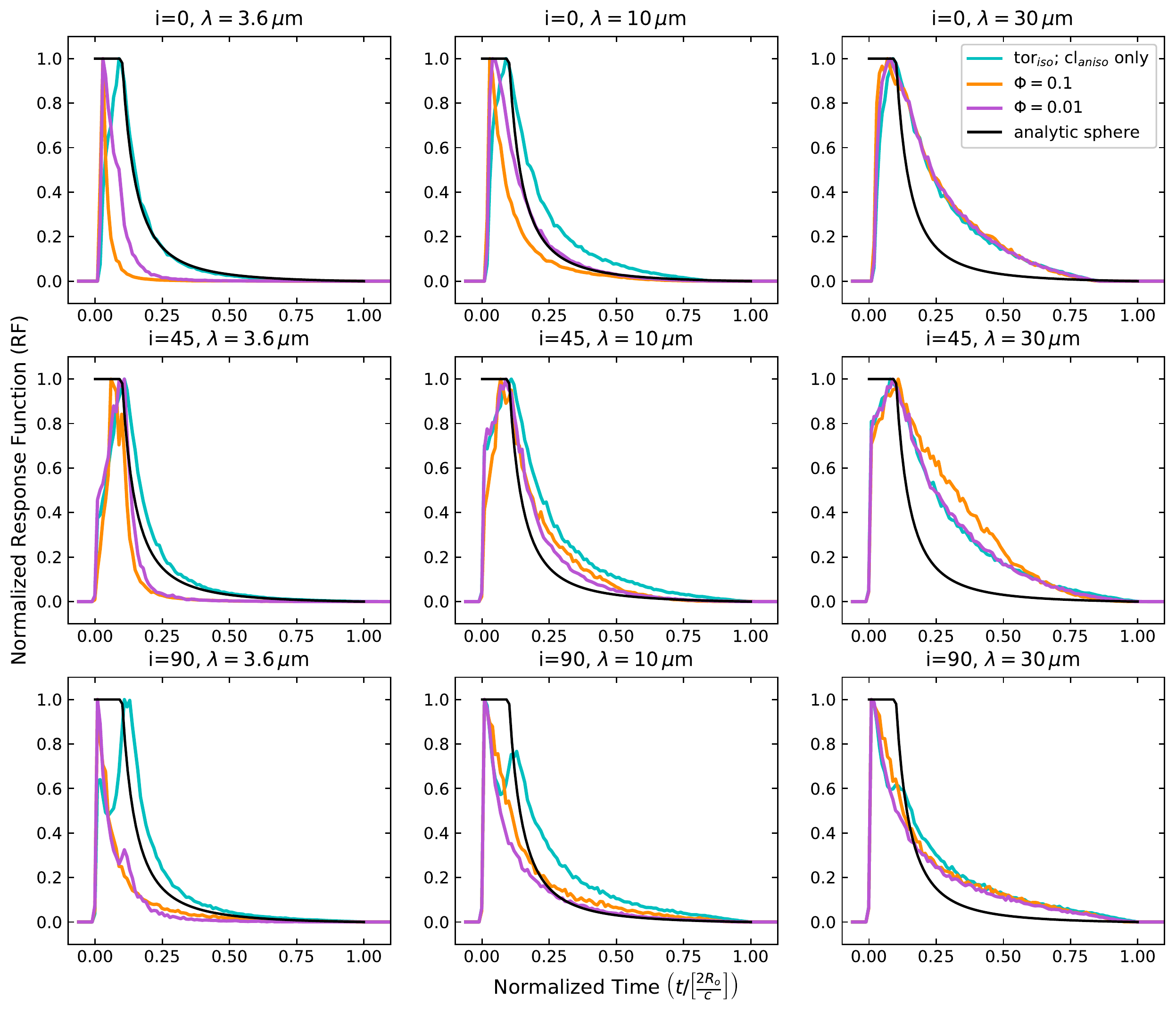}
\end{center}
\caption{\label{combo_ivslambda}  \textbf{GOT torus versus $i$:} Response functions for an isotropically illuminated torus with $\sigma=45^{\circ}$; $Y=10$; $\Phi=0.01$; $p=0$; when varying $i$ from $i$=0$^{\circ}$ (top) to $i$=90$^{\circ}$ (bottom) at 3.6 (left), 10 (middle), and 30 (right) $\mu$m. The blue lines represent simulations with only cloud orientation included (i.e., the globally optically thin case), the orange and purple lines represent the simulations with all of the radiative transfer treatments included (i.e. the globally optically thick (GOT) case), and the black line is the analytic spherical transfer function.}
\end{figure*}

The RWD and CTAR quantities show similar trends as $\tau_{V}$ increases for models with and without cloud occultation: the RWD generally increases, and CTAR decreases for $\tau_V > 10$; the behavior at smaller $\tau_V \,(\lesssim 10$) depends on wavelength (top and bottom panels of Figure~\ref{rfdpvstaui0clocc}; Appendix~\ref{app:figs}). The models that include cloud occultation (circles and solid lines) generally have higher RWD values but lower values of the RWD/LWR ratio than their globally optically thin counterparts (stars and dashed lines).

In summary, the effects of cloud occultation can either increase or decrease the response lag of the torus dust emission relative to the globally optically thin case, depending on the parameters $\Phi$ and $i$, with the size of the effect generally being comparable at all wavelengths.


\section{Globally Optically Thick Torus}\label{sec:combo}

In the previous section, we have individually discussed the effects of the different radiative transfer treatments that are included in TORMAC in order to identify their role in modifying the torus RFs. In this section, all of these treatments are included in the models in order to explore the response of a globally optically thick (GOT) torus, i.e., when the torus has a volume filling factor that is large enough that cloud shadowing and cloud occultation effects are important. In most cases, the RFs of the globally optically thick torus models are compared with equivalent models that include only cloud orientation effects (that is, a globally optically thin torus filled with optically thick dust clouds). 
Unless otherwise noted, all models assume an isotropically illuminated, sharp-edged torus. 

	\subsection{Volume Filling Factor, $\Phi$}\label{sec:gotvff}

The features identified in Section~\ref{sec:DustRF} allow us to approximately determine what radiative transfer processes have most influence on the RFs as $\Phi$ increases. 
For instance, the steepening of the decay tail for $\tau > 1/Y$ is predominantly due to cloud shadowing, whereas the sharper and earlier peak within the core $\tau < 1/Y$ can generally be attributed to cloud occultation. Thus, at wavelengths $\lesssim 10\,\mu$m (purple and orange lines in the left column of Figure~\ref{combo_ivslambda}), increasing $\Phi$ generally results in sharper, narrower RFs, which peak at short delays ($\tau\sim 0$ for $i=0^{\circ}$ and $i=90^{\circ}$), 
while at $i=90^{\circ}$, the second peak (at $\tau\sim 1/Y$) is heavily reduced as a result of cloud occultation. 

As $\Phi$ decreases, fewer clouds are occulted and/or shadowed, so cloud orientation effects become more important, especially at shorter wavelengths. The globally optically thin RFs are recovered at small values of $\Phi$ ($\lesssim 0.001$, depending on wavelength). 
Cloud orientation effects also play a larger role at intermediate inclinations, since the response is dominated by clouds along the inner radius on the side farthest from the observer, which are neither shadowed nor occulted. This is why the RFs are least sensitive to $\Phi$ at $i=45^{\circ}$ (Figure~\ref{combo_ivslambda}, middle row).

At 30 $\mu$m, the RFs for the globally thin and thick cases are very similar, except at $i=90^{\circ}$, where the second peak is reduced in the latter case (bottom row, right column of Figure~\ref{combo_ivslambda}). As previously discussed, although the clouds are individually optically thin at this wavelength, with longer path lengths through the edge-on torus, the cumulative effect of attenuation by several occulting clouds is sufficient to partially suppress the response at longer delays.

\begin{figure}[t]
\begin{center}
	\includegraphics[scale=0.49]{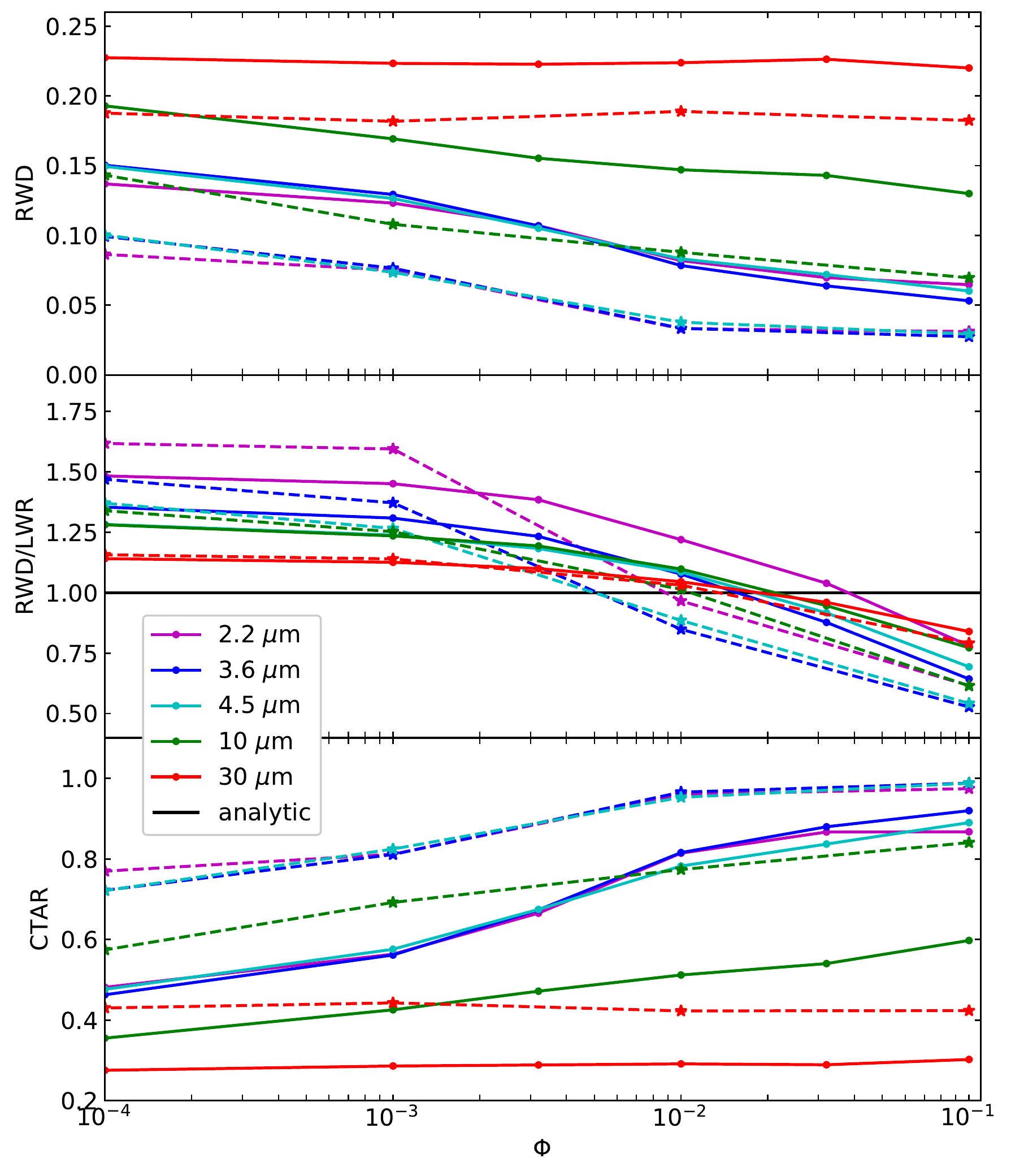}
\end{center}
\caption{\label{rfdpdiffvffvsi0p0_combo}  \textbf{GOT torus versus $\Phi$:} Response function descriptive quantities for a globally optically thick torus with $p=0$, $i=0^{\circ},$ $\sigma=45^{\circ},$ and $Y=10$ with respect to different values of $\Phi$ at select wavelengths. The circles and solid lines represent the models where the torus is illuminated isotropically, and the stars and dashed lines represent the models where the torus is illuminated anisotropically. Both tori have a sharp surface boundary. The limiting globally optically thin case is represented at a nominal value of $\Phi = 0.0001$. The colors represent the different wavelengths (2.2, 3.6, 4.5, 10, and 30 $\mu$m). The black line represents the analytic solution for RWD/LWR for a spherical shell with blackbody clouds.}
\end{figure}

The RF descriptive quantities follow generally similar behavior as $\Phi$ is varied for both isotropic and anisotropic illumination of the torus (Figures~\ref{rfdpdiffvffvsi0p0_combo}, and ~\ref{rfdpdiffvffvsi45p0_combo} and~\ref{rfdpdiffvffvsi90p0_combo}, Appendix~\ref{app:figs}), although the RWD values are usually smaller and the CTAR values larger for anisotropic illumination.  The RWD generally decreases and CTAR increases as $\Phi$ increases, but in detail the behavior depends on inclination, with the largest changes occurring for $i=0^{\circ}$ (Figure~\ref{rfdpdiffvffvsi0p0_combo}). 
For instance, at 3.6 $\mu$m the RWD decreases by a factor of $\sim 2$ at $i=0^{\circ}$ as $\Phi$ is increased from $0.0001$ to 0.1.

 It is also notable that in both illumination cases the RWD at wavelengths $\leq 4.5\,\mu$m reaches minimum values corresponding to $\sim R_{\rm d}/c$ for higher values of $\Phi$ ($>0.01$) and lower inclinations ($i\le 45^{\circ}$) (top panels of Figures~\ref{rfdpdiffvffvsi0p0_combo} and~\ref{rfdpdiffvffvsi45p0_combo}; Appendix~\ref{app:figs}). For example, for $i=0^{\circ}$ (Figure~\ref{rfdpdiffvffvsi0p0_combo}) and $\Phi=0.1$, at 3.6 $\mu$m, RWD $\approx 0.03$ for anisotropic illumination and $\approx 0.05$ for isotropic illumination, compared to the light-crossing times of the inner dust-free cavity of 0.032 and 0.1, respectively.

The  RWD/LWR ratio also exhibits rather different behavior as $\Phi$ is varied at different inclinations. At both $i=0^{\circ}$ and $90^{\circ}$ (Figures~\ref{rfdpdiffvffvsi0p0_combo} and~\ref{rfdpdiffvffvsi90p0_combo}, respectively), a sharp decrease occurs for $\Phi > 0.001$, by factors of $2-3$ at shorter wavelength (e.g., 3.6 $\mu$m). As a result, at face-on and edge-on inclinations, RWD/LWR generally has values $>1$ for $\Phi\lesssim 0.001$ but is $< 1$ for $\Phi\gtrsim 0.01$, with the largest changes occurring at shorter wavelengths. On the other hand, RWD/LWR shows only slow variations with $\Phi$ at $i=45^{\circ}$ (Figure~\ref{rfdpdiffvffvsi45p0_combo}), and is always $>1$, a consequence of the fact the RFs do not change much at this inclination, even between the globally optically thin and thick models (Figure~\ref{combo_ivslambda}).

In general, neither RWD nor CTAR shows a particularly strong dependence on inclination for either illumination case (Figure~\ref{rfdpvsip0vsani_combo}; Appendix~\ref{app:figs}). For $\Phi =0.01$, the RWD increases slowly and CTAR decreases slowly with  $i$, reflecting the broadening of the RF core as $i\rightarrow 45^{\circ}$ and the slower decline of the tail as $i\rightarrow 90^{\circ}$. The RWD/LWR ratio, however, exhibits a peak around $i\approx 45^\circ$ in Figure~\ref{rfdpvsip0vsani_combo}, which results from the sharp decreases with $\Phi$ at $i\approx 0^{\circ}$ and $i\approx 90^{\circ}$.

Overall, however, none of the descriptive quantities show a strong dependence on either $\Phi$ or $i$, typically varying by factors of $\lesssim 2$ over the ranges  $10^{-4} \leq \Phi \leq 10^{-1}$ and $0^{\circ} \leq i \leq 90^{\circ}$.

		\subsection{Radial Cloud Distribution, $p$}\label{sec:gotp}

\begin{figure*}[!h]
\begin{center}
	\includegraphics[scale=0.6]{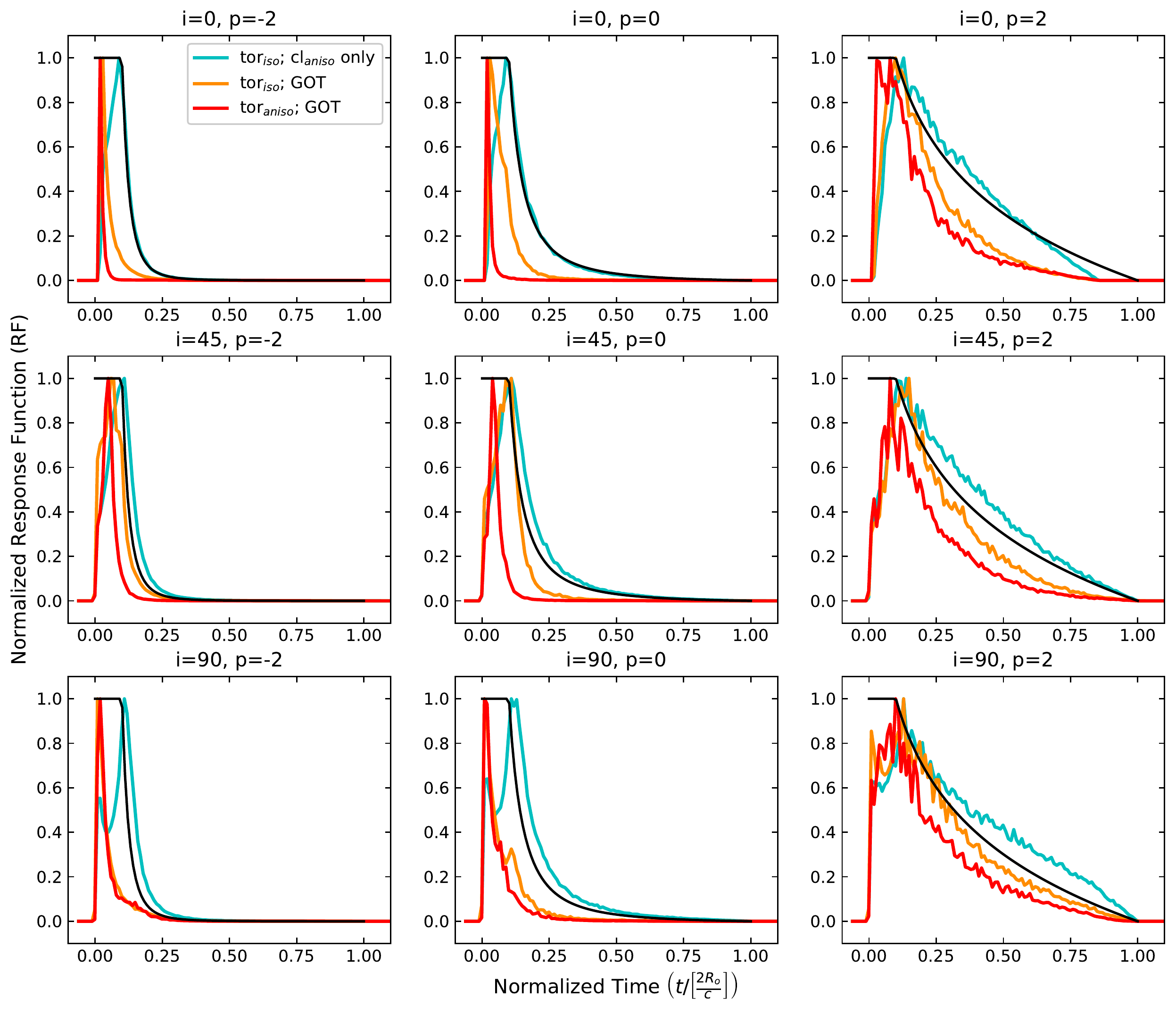}
\end{center}
\caption{\label{combo_p36}  \textbf{GOT torus versus $p$ at 3.6 $\mu$m:} Response functions for a torus with $\sigma=45^{\circ}$; $Y=10$; $\Phi=0.01$; for $i=0,$ 45, 90$^{\circ}$ at 3.6 $\mu$m.  Each column represents a different $p$ value increasing from left ($p=-2$) to right ($p=2$). The blue lines represent simulations with only cloud orientation included, and the orange lines represent the simulations with all of the radiative transfer treatments included with $\Phi=0.01$; in both cases the torus is illuminated isotropically by the central source. The red lines represent the simulations with all of the radiative transfer treatments (i.e., the globally optically thick (GOT) models) included with $\Phi=0.01$, where the torus is illuminated anisotropically by the central source. The black line is the analytic spherical transfer function. } 
\end{figure*}

As shown in Sections~\ref{sec:BBRF} and \ref{sec:aniscloud}, the value of $p$ determines the rate at which the RF decays for delays $\tau \gtrsim 1/Y$, with more positive values producing more gradual decays at a given wavelength.  
The 3.6 $\mu$m RFs of a globally optically thick torus are shown for several values of $p$ and both illumination cases in Figure~\ref{combo_p36}.
In the globally optically thin case (blue lines), the RF peak occurs at $\tau\approx 1/Y$ at all inclinations and $p$ values. In contrast, for the globally optically thick torus (orange lines), the effects on the RF of both cloud shadowing and cloud orientation depend on the radial distribution of the clouds, controlled by $p$.

For $p \leq 0$, the clouds are more centrally concentrated, strengthening cloud shadowing and occultation effects in the RF core.  On the other hand, for $p>0$, the decay tail is more affected, resulting in a steeper decay as compared to the globally optically thin case. As a result, the peak response of the globally optically thick torus occurs at delays ranging between $\tau \approx 0$ and $\approx 1/Y$, depending on $p$ and $i$.
Similar behavior is seen at other wavelengths $\lesssim 10\,\mu$m. At $30\,\mu$m, shadowing and occultation have relatively little effect on the RF, and hence the differences with respect to the globally optically thin torus are relatively small. 

The combination of anisotropic illumination with cloud shadowing and cloud occultation for $p \leq 0$ results in extremely narrow and sharply peaked RFs at $i=0$ and 90$^{\circ}$ (red lines in Figure~\ref{combo_p36}). Indeed, the response at $i=0^{\circ}$ becomes a spike at $\tau\approx (1-\sin(\sigma))/(2Y)\sim 0.015$ (see Equation~\ref{eq:resbegin}), corresponding to the intersection of the isodelay surface with the inner edge of the torus on the side nearest to the observer.

\begin{figure}[ht]
\begin{center}
	\includegraphics[scale=0.49]{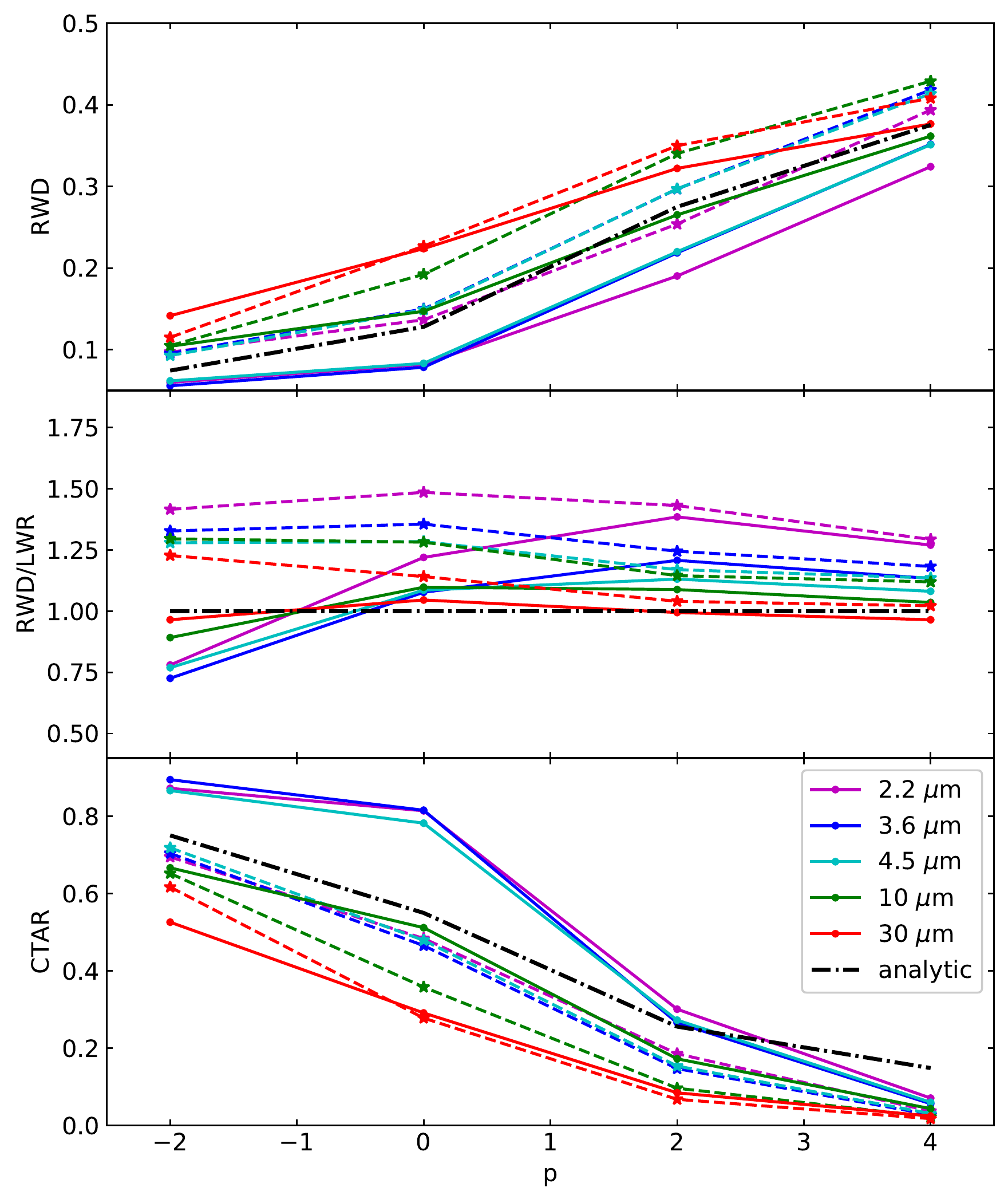}
\end{center}
\caption[DPs with the ``complete" radiative transfer treatment that is illuminated isotropically varying $p$ for $Y$=10 at $i$=0$^{\circ}$ and $\Phi=0.01$]{\label{rfdpdiffpcombo0} \textbf{GOT torus varying $p$:} Response function descriptive quantities for an isotropically illuminated torus with $\sigma = 45^{\circ}$, $\Phi=0.01$, $i$=0$^{\circ},$ and $Y$=10 with respect to different values of $p$ at select wavelengths. The circles and solid lines represent the globally optically thick torus models, and the stars and dashed lines represent the torus models with only cloud orientation. The colors represent the different wavelengths (2.2, 3.6, 4.5, 10, and 30 $\mu$m). The analytic spherical shell solution is plotted as a dashed--dotted black line for comparison.}
\end{figure}

The RF descriptive quantities for globally optically thick (with $\Phi$=0.01) and globally optically thin tori are compared as a function of $p$ in Figure~\ref{rfdpdiffpcombo0}.
The RWD increases strongly with $p$ in both cases, particularly at shorter wavelengths. For the globally optically thick torus models, the RWD generally has lower values than in the globally optically thin case but undergoes a larger increase, reflecting the larger changes in the RF as $p$ increases (Figure~\ref{combo_p36}). 

The CTAR also exhibits similar behavior for both the globally optically thin and thick models, decreasing strongly as $p$ is increased, with the globally optically thick torus models generally having larger values, reflecting the sharper and relatively strong RF peaks.

In comparison, the RWD/LWR ratio exhibits relatively small changes with $p$ in both cases, even at the shorter wavelengths. 
However, it is notable that the globally optically thick torus models have much smaller RWD/LWR values than the globally optically thin case for $p\leq0$, reflecting the sharper RFs, but the two cases converge to similar values for $p>0$.

		\subsection{Torus Geometry, Structure, and Cloud Optical Depth}\label{sec:gottor}

\subsubsection{Ratio of Outer to Inner Radius, $Y$}
\begin{figure*}[!ht]
\begin{center}
	\includegraphics[scale=0.49]{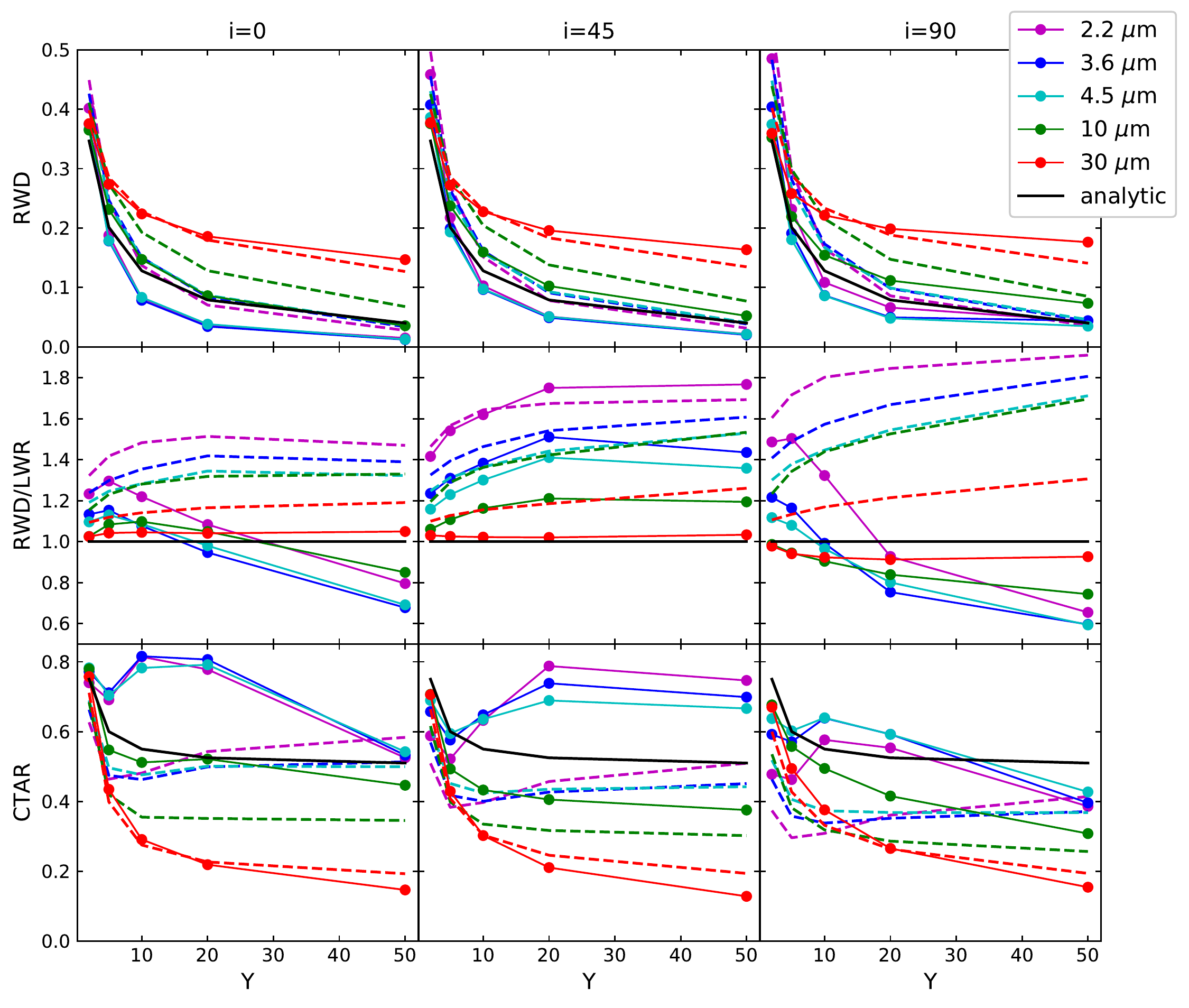} 
\end{center}
\caption[DPs for an isotropically illuminated disk torus with the ``complete" radiative transfer treatment versus $Y$]{\label{DPY_combo} \textbf{GOT torus varying $Y$:}  Response function descriptive quantities for an isotropically illuminated globally optically thick (circles and solid lines) or globally optically thin (dashed lines) torus with $\sigma = 45^{\circ},$ $p=0$, and $\Phi=0.01$ with respect to different values of $Y$ at select wavelengths. Each column represents the quantities when $i=0^{\circ}$ (left), $i=45^{\circ}$ (middle), and $i=90^{\circ}$ (right). The colors represent the different wavelengths (2.2, 3.6, 4.5, 10, and 30 $\mu$m). The black line represents the analytic solution for a spherical shell with blackbody clouds.}
\begin{center}
	\includegraphics[scale=0.48]{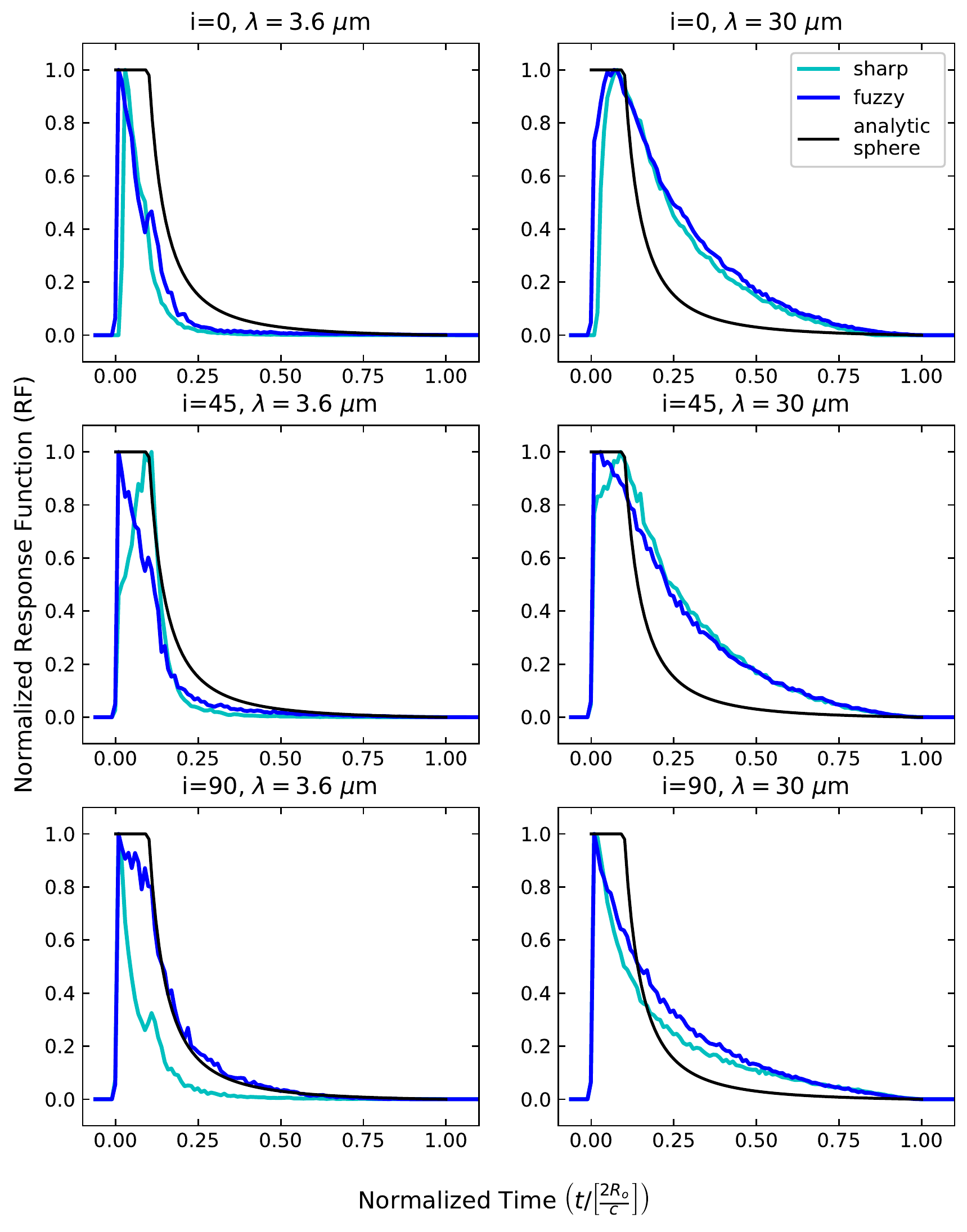}
\end{center}
\caption{\label{combo_ivslambda_fuzzy} \textbf{GOT torus with either a sharp or fuzzy-edge:}  Response functions for an isotropically illuminated, globally optically thick torus with $\sigma=45^{\circ}$; $Y=10$; $\Phi=0.01$; $p=0$; when varying $i$ from from $i$=0$^{\circ}$ (top) to $i$=90$^{\circ}$ (bottom) at 3.6 (left) and 30 (right) $\mu$m. The light blue lines represent the simulations for a torus with a sharp-edge, the dark blue lines represent the simulations for a torus with a fuzzy-edge, and the black lines are the analytic spherical transfer function.} 
\end{figure*}

 Figure~\ref{DPY_combo} shows the variations of the descriptive quantities with $Y$ for isotropically illuminated globally optically thick ($\Phi = 0.01$) and thin torus models for $i=0$, 45, and 90$^{\circ}$. 
 The major effect of increasing $Y$ is to narrow the core of the RF relative to the decay tail (see Figure~\ref{bbdisk2}) and as a result, the RWD varies strongly with $Y$ at all wavelengths for $p=0$.  At shorter wavelengths, the RWD loci for the torus models roughly follow the RWD$\sim 1/Y$ relationship of the analytical spherical shell solution. The dependence is weaker for longer wavelengths and for less centrally concentrated cloud distributions ($p>0$). Similar behavior is seen at all three inclinations.

Both RWD/LWR and CTAR also show quite strong variations with $Y$, although there are substantial differences in behavior between the globally optically thick and thin models, and in detail the behavior 
is also dependent on inclination. In RWD/LWR, the two cases exhibit similar behavior at $i  = 45^{\circ}$ (middle row of Figure~\ref{DPY_combo}), since the RFs are similar for the globally optically thick and thin models at this inclination for the reasons discussed in Section~\ref{sec:gotvff}. However, the trends diverge at $i  = 0^{\circ}$ and $i  = 90^{\circ}$, with RWD/LWR remaining $> 1$ in the globally optically thin models, but dropping to values $< 1$ in the globally optically thick models, due to the effects of cloud shadowing and cloud occultation on the LWR.

In CTAR (bottom row of Figure~\ref{DPY_combo}), both torus models typically exhibit larger and more complex variations at shorter wavelengths than the spherical shell solution, with the detailed behavior differing considerably between the globally thin and thick cases, and also with wavelength. The CTAR reaches higher values in the globally optically thick models than in the globally optically thin case for $Y = 10 - 20$, reflecting the much more sharply peaked RFs that result from cloud shadowing and occultation effects. At 10 and 30 $\mu$m, on the other hand, where radiative transfer effects have less influence on the RF shapes, CTAR  exhibits quite similar behavior in both cases, exhibiting a monotonic decrease with increasing $Y$, similar to the analytical solution.

\subsubsection{Angular Width, $\sigma$}
The main effect of increasing $\sigma$ is to broaden the RF core, but this causes only minor changes in the descriptive quantities (Figure~\ref{rfdpdiffsigcombo0}; Appendix~\ref{app:figs}). 
In the globally optically thick model, RWD is almost independent of $\sigma$, exhibiting variations of $< 10$\% as $\sigma$ varies from $\sigma = 15^{\circ}$ to $60^{\circ}$, whereas for the globally optically thin case, RWD increases gradually with $\sigma$, but only by $\approx 20$\%. 
The RWD/LWR ratio exhibits similarly small changes in both cases ($\lesssim 20$\%). 
As might be expected, CTAR is most sensitive to $\sigma$ in both models, decreasing ($\lambda \leq 4.5\,\mu$m) or increasing gradually ($\lambda = 10, 30\,\mu$m) as $\sigma$ increases. 
The variations in all three quantities with respect to $\sigma$ are also relatively small at the other inclinations.

\begin{figure*}[!ht]
\begin{center}
	\includegraphics[scale=0.6]{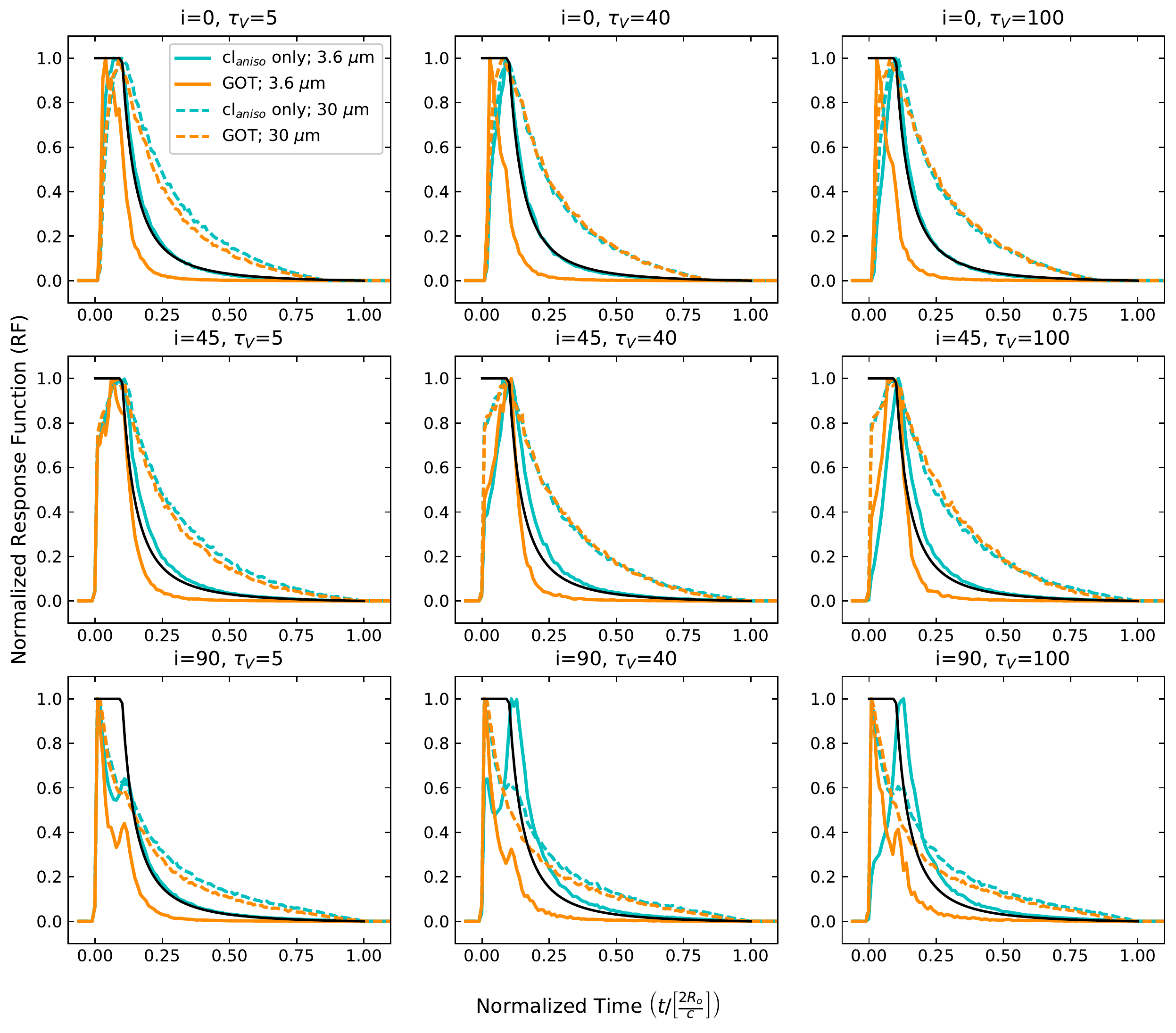}
\end{center}
\caption{\label{combo_tau} \textbf{GOT torus varying $\tau_V$:} Response functions for an isotropically illuminated torus with $\sigma=45^{\circ}$; $Y=10$; $p=0$; $\Phi=0.01$; for $i=0,$ 45, 90$^{\circ}$ at 3.6 (solid lines) and 30 $\mu$m (dashed lines).  Each column represents a different $\tau_{V}$ value increasing from left ($\tau_{V}=5$) to right ($\tau_{V}=100$). The blue lines represent simulations with only cloud orientation included, the orange lines represent the simulations with all of the radiative transfer treatments included (i.e., the globally optically thick (GOT) models). The black line is the analytic spherical transfer function. } 
\end{figure*}

\subsubsection{Torus Surface: Sharp versus Fuzzy}\label{sec:combofuz}

 In the globally optically thin models, the RFs differ only slightly between the sharp-edged and fuzzy-edged  torus models (Figure~\ref{clorientvsani_sp36}, purple and red lines, respectively). However, in the globally optically thick case, the effects of cloud occultation are reduced in the fuzzy-edged torus, since the clouds are distributed within an effectively larger volume. Thus, for example, clouds at higher altitudes above the disk midplane (i.e.,  $\beta \gtrsim \sigma$) are occulted by fewer intervening clouds, if at all, and this can quite substantially modify the RFs, as shown in  Figure~\ref{combo_ivslambda_fuzzy}. The largest differences occur at the
higher inclinations, for which the path length through the torus is typically longer for a given cloud. 
At $i=90^{\circ}$, the fuzzy torus RF is much broader than the sharp-edged counterpart, due to emission from high-altitude clouds, which are likely to be occulted (or shadowed) by fewer clouds.

As a  result of the  broadening of the RF due to the high-latitude clouds, with $\beta \gtrsim \sigma$, the fuzzy-edge torus
 typically has higher values of RWD and RWD/LWR, and lower CTAR, 
 with the differences being larger at shorter wavelengths (stars and dashed lines in Figure~\ref{rfdpfuzzycombo0}; Appendix~\ref{app:figs}).  Much larger differences occur at higher inclinations and higher volume filling factors, when the fuzzy torus RFs at short wavelengths are substantially broader than those for the sharp-edged case. For example, for 3.6 $\mu$m and $\Phi = 0.01$, the RWD for the fuzzy torus is $\sim 60$\% longer than for the sharp-edged torus at $i=90^{\circ}$, compared with $\sim 20$\% at $i=0^{\circ}$.

		\subsubsection{Cloud Optical Depth, $\tau_{V}$}

The effects of varying the cloud optical depth, $\tau_{V}$, are illustrated in Figure~\ref{combo_tau}, which compares the RFs for globally optically thin and thick torus models
at several values of $\tau_{V}$ ($=5$, 40, 100) for 3.6  and 30  $\mu$m. 

As already discussed in Section~\ref{sec:cloccrf}, although both the cloud orientation and cloud occultation effects depend on $\tau_{V}$, it has only a relatively modest effect on the shapes of the RFs.
As $\tau_{V}$ is increased, the cloud emission at shorter wavelengths becomes more anisotropic, and in the globally optically thin case (blue lines), therefore, the response amplitude is more strongly reduced at shorter delays within the core, which results in the RF peaking at $\tau\sim 1/Y$. 
In the globally optically thick case (orange lines), increasing $\tau_{V}$ also tends to reduce the response at longer delays within the core due to cloud occultation. As a result, the 3.6 $\mu$m RF (solid lines) peaks at  $\tau\approx 0$ for $i=0^{\circ}$ and $90^{\circ}$  for all values of $\tau_V$ (top and bottom rows). Cloud occultation has the least effect at $i=45^{\circ}$ (middle row), where the core response is predominantly shaped by cloud orientation effects (see Section~\ref{sec:cloccrf}). However, in comparison with the RFs for the globally optically thin models, the globally optically thick models typically have narrower RFs with steeper decay tails. This is the result of cloud shadowing, which is independent of $\tau_V$ (for the range considered here) since even at $\tau_V=5$ the clouds are still effectively opaque to the AGN's UV/optical radiation.

At 30 $\mu$m, the RFs are very similar for both the globally optically thick and thin torus models, and change only slightly as $\tau_V$ increases. The clouds emit almost isotropically, even for $\tau_V = 100$,
and it is only when the torus is edge-on that small differences between the globally optically thick and thin cases appear. 

None of the RF descriptive quantities are particularly sensitive to $\tau_V$ for either the globally optically thick or globally optically thin torus. For example, at $i=0^{\circ}$, the RWD changes by $\lesssim 10$\% overall and shows slight increases or decreases depending on wavelength (Figure~\ref{rfdpdifftaucombo0}; Appendix~\ref{app:figs}).
This is generally also the case even at $i=90^{\circ}$, where the effects of both cloud occultation and cloud orientation on the RFs are strongest. 
Evidently, even large variations in $\tau_V$ do not strongly affect any of the descriptive quantities.

\subsection{Response-weighted Delay and Luminosity-weighted Radius}\label{sec:rwdlwr}

The descriptive quantities introduced in this paper provide us with a way to characterize and quantitatively compare the RFs. As mentioned in Section~\ref{sec:anaDP}, the RWD is equivalent to the lag measured from  IR reverberation mapping campaigns, which in turn is interpreted as the light-travel time associated with the characteristic radius of the dust emitting at the observed IR wavelength (i.e, the LWR). This is the cornerstone behind observational reverberation mapping studies. In the simplest, idealized case, it is indeed true that RWD=LWR (Section~\ref{sec:BBRF}). The examples presented for more realistic dust emission models earlier in this section (Section~\ref{sec:combo}) have shown that this relationship approximately holds over wide ranges of the torus model parameters. Here we will explore the RWD--LWR relationship in more detail. Note that in this subsection we will consider only the RF models for the globally optically thick torus since it is the most realistic case. 

\begin{figure}[!t]
	\includegraphics[scale=0.49]{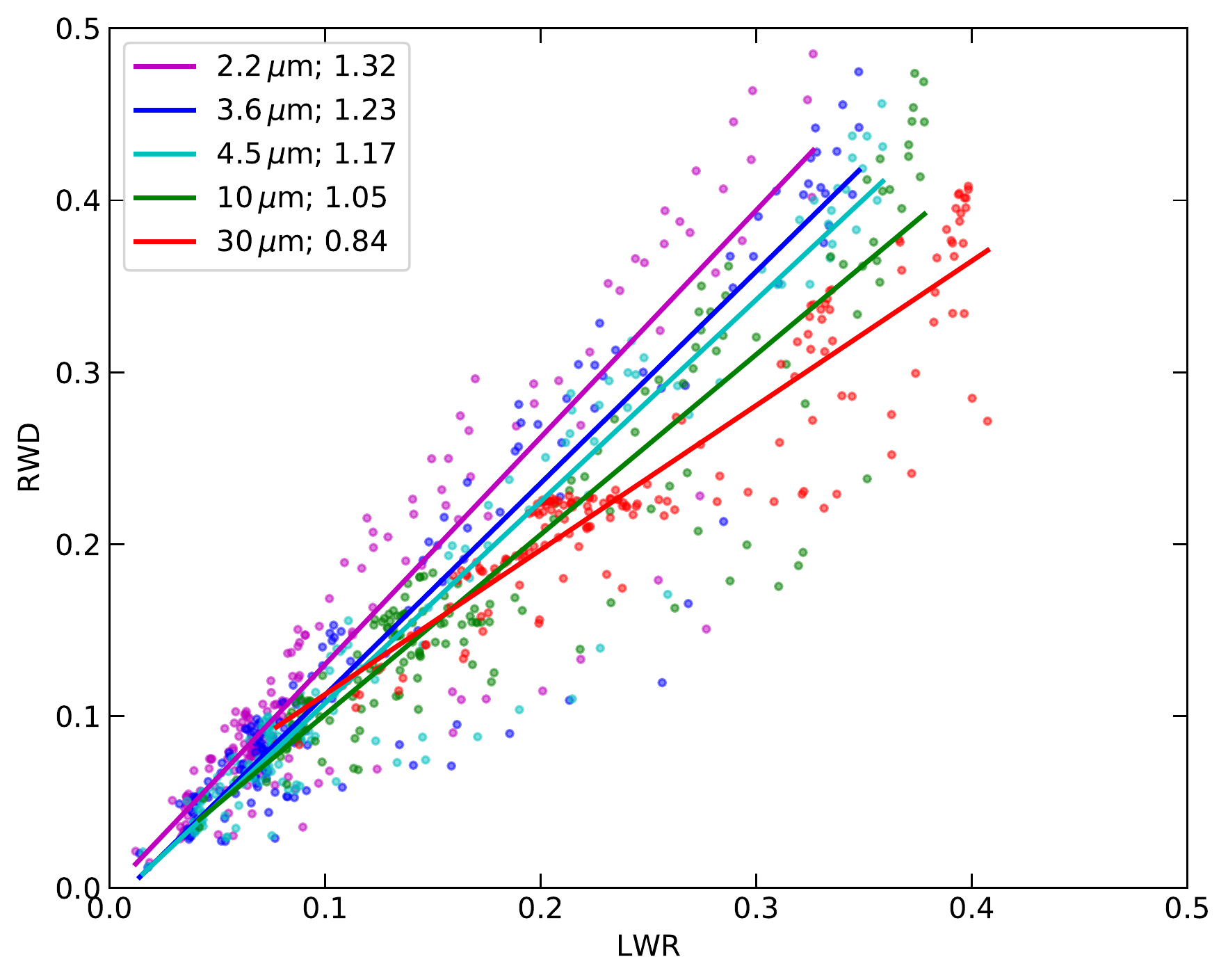}	
\caption{RWD and LWR values for all 5 selected wavelengths for all of the globally optically thick models included in this study. The best fits at each wavelength are plotted as solid colored lines, with the slope of the line shown in the legend.}
\label{gotrwdlwr}
\end{figure}

Figure~\ref{gotrwdlwr} shows the RWD--LWR relationship for all of the globally optically thick models included in this study, including all the selected wavelengths. The linear best fits are also shown for each wavelength. 
In general, the RWD increases approximately linearly  as LWR increases. However, there is quite a large scatter, which is partly due to wavelength, and the relationship is not exactly 1:1. 
At a specific wavelength, a given value of LWR is generally associated with a wide range in RWD. It is also clear that the slope of the best-fit lines of the RWD--LWR relationship decreases as wavelength increases, being $> 1$ for $\lambda \le 4.5\,\mu\rm m$,  $\approx 1$ when $\lambda = 10\,\mu\rm m$, and $< 1$ when $\lambda =30\,\mu\rm m$.

In order to characterize the scatter seen in Figure~\ref{gotrwdlwr} we investigated the relationship between RWD and LWR with respect to the different torus parameters. The radial depth, $Y$, radial cloud distribution, determined by $p$, and the volume filling factor, $\Phi$, have the largest influence on the RFs and therefore the descriptive quantities. 

The RWD--LWR relationship remains approximately linear as $Y$ varies, with only minor scatter. As seen in Section~\ref{sec:combo} Figure~\ref{DPY_combo} and Section~\ref{sec:rBBRF}, both the RWD and LWR decrease as $Y$ increases. However, the RWD--LWR relationship is not exactly 1:1; the RWD tends to decrease faster than the LWR, resulting in a slightly larger gradient. 

\begin{figure}[!t]
\begin{center}
	\includegraphics[scale=0.42]{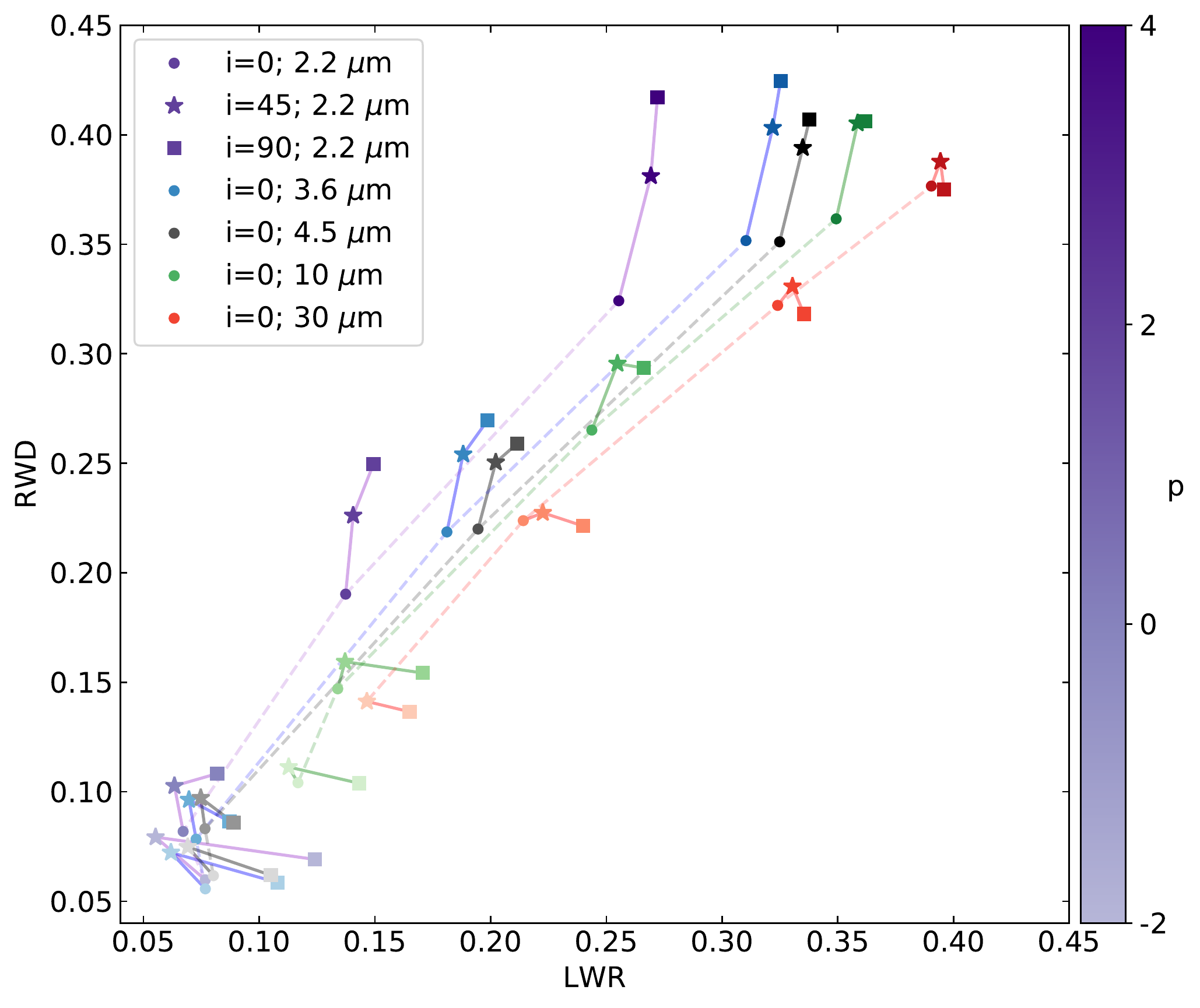}
\end{center}
\caption{\label{rwdlwrp} RWD and LWR values for a globally optically thick torus with $Y=10$, $\Phi=0.01$, and $\sigma=45^{\circ}$. The different symbols represent the values at different inclinations: $i=0^{\circ}$ (circle), $i=45^{\circ}$ (star), and $i=90^{\circ}$ (square). The colors represent the different wavelengths, with the shade of the color increasing as $p$ increases from $-2$ to $+4$ as shown in the color bar on the right. The solid lines connect sequences in inclination with the same wavelength and $p$ value. The light dashed lines connect the different $p$ values at one inclination ($i=0^{\circ}$). }
\end{figure}

Figure~\ref{rwdlwrp} shows the relationship between RWD and LWR at each wavelength when $p$ is varied from $-2$ (lighter shades) to $4$ (darker shades) for $Y=10$ and inclinations $i=0{\circ}$ (filled circles), $45^{\circ}$ (stars), and $90^{\circ}$ (squares). 
Both the RWD and LWR vary strongly with $p$, with both increasing as $p$ increases, yielding approximately diagonal trajectories in the RWD--LWR plane for a given inclination and wavelength. This, together with the variations with $Y$, is the main factor in producing the general correlation that is seen in Figure~\ref{gotrwdlwr}. 

The trend with wavelength is also apparent, with larger variations occurring for shorter wavelengths. However, we can see that the variations in RWD (and to a lesser extent in LWR) at a particular $p$ value also depend on $i$.
For instance, when $p>0$, the LWR barely changes with inclination at any wavelength, whereas the RWD changes dramatically at short wavelengths, increasing as $i$ increases, so as to produce the steeply rising inclination loci seen in the top right of the RWD--LWR plane in Figure~\ref{rwdlwrp}. The opposite is true when $p\le 0$; the changes in RWD are comparable to or smaller than the changes in LWR, and the behavior with $i$ is not, in general, monotonic, leading to clusters of points in the bottom left of the figure.

\begin{figure}[!t]
\begin{center}
	\includegraphics[scale=0.42]{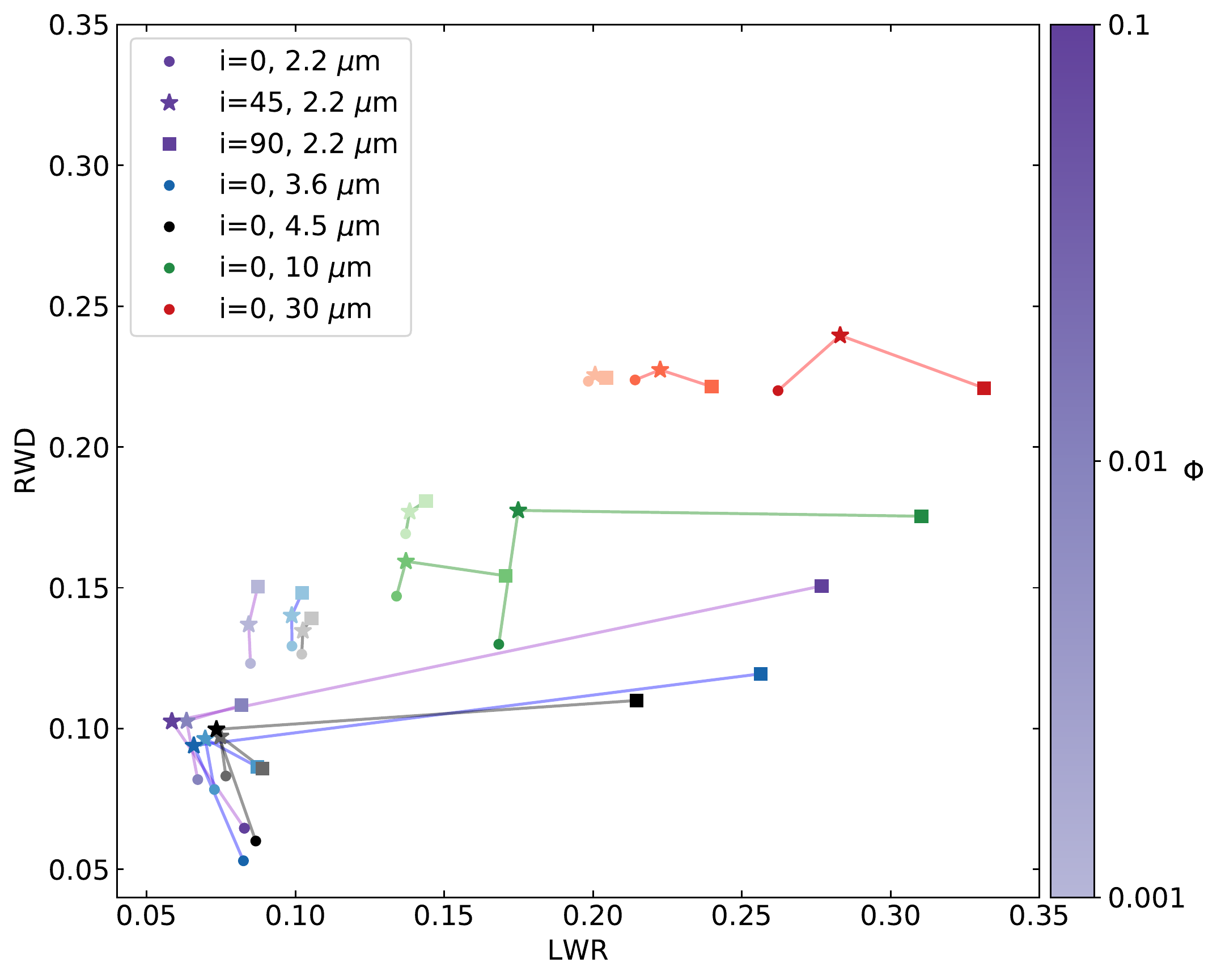}
\end{center}
\caption{\label{rwdlwrvff} RWD and LWR values for a globally optically thick torus with $Y=10$, $p=0$, and $\sigma=45^{\circ}$. The different symbols represent the values at different inclinations: $i=0^{\circ}$ (circle), $i=45^{\circ}$ (star), and $i=90^{\circ}$ (square). The colors represent the different wavelengths, with the shade of the color increasing as $\Phi$ increases from 0.001 to 0.1 as shown in the color bar on the right. The solid lines connect the sequences in inclination with the same wavelength and $\Phi$ value.}
\end{figure}

Figure~\ref{rwdlwrvff} shows the RWD--LWR relationship when  $\Phi$ is varied from 0.001 (lighter shades) to 0.1 (darker shades) for $p=0$,  $Y=10$, and the same inclinations. Here, there is almost no correlation between RWD and LWR as $\Phi$ varies, and the detailed behavior depends strongly on wavelength. For instance, at 30\,$\mu$m, the variations in RWD with both $\Phi$ and $i$ are relatively small, whereas LWR increases with both parameters, resulting in approximately horizontal trajectories (top right in the figure). 

On the other hand, at the shorter wavelengths ($\lambda \le 4.5\,\mu\rm m$), RWD tends to vary more than LWR. For $i=0^{\circ}$ (filled circles) the RWD decreases sharply as $\Phi$ increases, whereas the change in LWR is relatively small. The behavior is somewhat similar for $i=45^{\circ}$ (filled stars), except that there is little change between the intermediate and largest values of $\Phi$ (0.01 and 0.1, respectively). The largest change in LWR occurs for $\Phi=0.1$, as $i$ increases from $45^{\circ}$ (stars) to $90^{\circ}$ (filled squares), whereas the change in RWD is relatively small, leading to large almost horizontal excursions across the RWD--LWR plane. 

Recall that Figures~\ref{rfdpdiffvffvsi0p0_combo}, and ~\ref{rfdpdiffvffvsi45p0_combo} and~\ref{rfdpdiffvffvsi90p0_combo} (Appendix~\ref{app:figs}) show that models with large $\Phi$ can have  values of RWD/LWR$<1$. We can now see from Figure~\ref{rwdlwrvff} that at $i=0^{\circ}$ this is due mainly to the decrease in RWD as $\Phi$ increases and at $i=90^{\circ}$ to the increase in LWR.

Figures~\ref{rwdlwrp} and~\ref{rwdlwrvff} show that although a fairly tight correlation is maintained between RWD and LWR as $p$ is varied, albeit with slopes deviating slightly from unity depending on $\lambda$ and $i$,  this is not the case when $\Phi$ is varied. Referring back to Figure~\ref{gotrwdlwr}, the general $\sim$ linear relation between RWD and LWR is mostly due to $p$ and $Y$, with the scatter due mainly to $\Phi$ and $i$. 

\begin{table*}
\caption{{RWD/LWR}\\ The minimum, median, and maximum RWD/LWR values for each wavelength for both isotropic and anisotropic illumination and for the various cloud-level and global optical depth effects.}
\begin{center}
\label{rwdlwrtab}
{\renewcommand{\arraystretch}{1.6}

\scriptsize{
\begin{tabular}{lccc | ccc | ccc | ccc | ccc}
\hline
\hline
Model & \multicolumn{15}{c}{$\lambda$ $\left({\mu{\rm m}}\right)$}\\ 
 \hline
 &  \multicolumn{3}{c}{2.2 $\mu$m} & \multicolumn{3}{c}{3.6 $\mu$m} & \multicolumn{3}{c}{4.5 $\mu$m} & \multicolumn{3}{c}{10 $\mu$m} & \multicolumn{3}{c}{30 $\mu$m} \\
\hline
 &  min & med & max & min & med & max & min & med & max & min & med & max & min & med & max  \\
\hline

cl$_{\rm aniso}$; tor$_{\rm iso}$  & 1.12 & \bf{1.60} & 1.91 & 1.10 & \bf{1.40} & 1.81 & 1.07 & \bf{1.31} & 1.71 & 1.05 & \bf{1.29} & 1.70 & 1.02 & \bf{1.14} & 1.31 \\
cl$_{\rm aniso}$; tor$_{\rm aniso}$ & 1.18 & \bf{1.63} & 1.95 & 1.10 & \bf{1.44} & 1.77 & 1.09 & \bf{1.34} & 1.69 & 1.08 & \bf{1.30} & 1.77 & 1.02 & \bf{1.16} & 1.41  \\

\hline
cl$_{\rm aniso}$;  cl$_{\rm shadow}$; tor$_{\rm iso}$ & 1.23 & \bf{1.53} & 1.87 & 1.20 & \bf{1.36} & 1.56 & 1.18 & \bf{1.30} & 1.48 & 1.14 & \bf{1.29} & 1.48 & 1.04 & \bf{1.16} & 1.33  \\
cl$_{\rm aniso}$; cl$_{\rm shadow}$; tor$_{\rm aniso}$ & 1.18 & \bf{1.59} & 2.26 & 1.20 & \bf{1.39} & 1.71 & 1.21 & \bf{1.30} & 1.53 & 1.19 & \bf{1.31} & 1.58 & 1.04 & \bf{1.18} & 1.46  \\
\hline
cl$_{\rm aniso}$; cl$_{\rm occult}$; tor$_{\rm iso}$ & 0.89 & \bf{1.39} & 1.73 & 0.78 & \bf{1.19} & 1.50 & 0.73 & \bf{1.11} & 1.36 & 0.69 & \bf{1.07}  & 1.31 & 0.65 & \bf{1.00} & 1.11 \\ 
cl$_{\rm aniso}$; cl$_{\rm occult}$; tor$_{\rm aniso}$ & 0.96 & \bf{1.47} & 1.73 & 0.86 & \bf{1.27}  & 1.40 & 0.80 & \bf{1.17} & 1.27 & 0.89 & \bf{1.09}  & 1.20 & 0.87 & \bf{0.96} & 1.02 \\
 
\hline
GOT; tor$_{\rm iso}$ & 0.54 & \bf{1.38} & 1.77 & 0.45 & \bf{1.16} & 1.51 & 0.51 & \bf{1.12} & 1.41 & 0.57 & \bf{1.06} & 1.28 & 0.65 & \bf{0.99} & 1.13  \\
GOT; tor$_{\rm aniso}$ & 0.40 & \bf{1.34} & 1.80 & 0.38 & \bf{1.16} & 1.48 & 0.40 & \bf{1.11} & 1.34 & 0.60 & \bf{1.06}  & 1.28 & 0.69 & \bf{1.00} & 1.14  \\

\hline
\end{tabular}}}

\end{center}
\end{table*}

\begin{table*}
\begin{center}
\caption{{RWD$\times$Y}\\ The minimum, median, and maximum values of RWD in units of the normalized inner cavity light-crossing time, $\tau_{\rm d}$.} 

\label{RWDtab}

{\renewcommand{\arraystretch}{1.6}

\scriptsize{
\begin{tabular}{lccc | ccc | ccc | ccc | ccc}
\hline
\hline
Model & \multicolumn{15}{c}{$\lambda$ $\left({\mu{\rm m}}\right)$}\\ 
 \hline
 &  \multicolumn{3}{c}{2.2 $\mu$m} & \multicolumn{3}{c}{3.6 $\mu$m} & \multicolumn{3}{c}{4.5 $\mu$m} & \multicolumn{3}{c}{10 $\mu$m} & \multicolumn{3}{c}{30 $\mu$m} \\
\hline
 &  min & med & max & min & med & max & min & med & max & min & med & max & min & med & max  \\
\hline
cl$_{\rm aniso}$; tor$_{\rm iso}$  & 0.82 & \bf{1.44} & 4.72 & 0.80 & \bf{1.56} & 4.87 & 0.78 &  \bf{1.55} & 4.68 & 0.78 & \bf{2.00} & 4.86 & 0.75 & \bf{2.29} & 7.04  \\
cl$_{\rm aniso}$; tor$_{\rm aniso}$ & 0.50 & \bf{0.92} & 4.18 & 0.51 & \bf{0.95} & 4.58 & 0.49 &  \bf{0.92} & 4.53 & 0.51 & \bf{1.22} & 4.91 & 0.50 &  \bf{1.76} & 5.71  \\

\hline
cl$_{\rm aniso}$; cl$_{\rm shadow}$; tor$_{\rm iso}$ & 0.73 & \bf{1.03} & 2.97 & 0.74 & \bf{0.95} & 3.33 & 0.76 & \bf{0.96} & 3.23 & 0.80 & \bf{1.48} & 3.71 & 0.79 & \bf{2.18} & 5.62 \\
cl$_{\rm aniso}$; cl$_{\rm shadow}$; tor$_{\rm aniso}$ & 0.40 & \bf{0.63} & 2.18 & 0.40 & \bf{0.58} & 2.84 & 0.41 & \bf{0.59} & 2.90 & 0.50 & \bf{0.98} & 3.54 & 0.63 & \bf{1.80} & 4.20  \\

\hline
cl$_{\rm aniso}$; cl$_{\rm occult}$; tor$_{\rm iso}$ & 0.84 & \bf{1.67} & 4.69 & 0.79 & \bf{1.84} & 5.98 & 0.76 & \bf{1.77} & 6.28 & 0.75 & \bf{2.22} & 11.42 & 0.73 & \bf{2.35} & 10.12 \\ 
cl$_{\rm aniso}$; cl$_{\rm occult}$; tor$_{\rm aniso}$ & 0.78 & \bf{1.80} & 3.95 & 0.75 & \bf{2.15} & 4.31 & 0.69 & \bf{2.13}  & 4.24 & 1.00 & \bf{2.64}  & 4.51 & 1.14 & \bf{2.70}  & 3.90 \\

\hline
GOT; tor$_{\rm iso}$ & 0.58 & \bf{1.08} & 4.64 & 0.53 & \bf{0.95} & 4.75 & 0.57 & \bf{0.96} & 4.56 & 0.70 & \bf{1.61} & 4.69 & 0.72 & \bf{2.25} & 8.81 \\ 
GOT; tor$_{\rm aniso}$ & 0.29 & \bf{0.75} & 3.94 & 0.27 & \bf{0.67} & 4.42 & 0.29 & \bf{0.67} & 4.37 & 0.56 & \bf{1.26} & 4.74 & 0.83 & \bf{2.05} & 4.04 \\ 
 
\hline
\end{tabular}}}
\end{center}
\end{table*}

\begin{figure}[!t]
\begin{center}
	\includegraphics[scale=0.52]{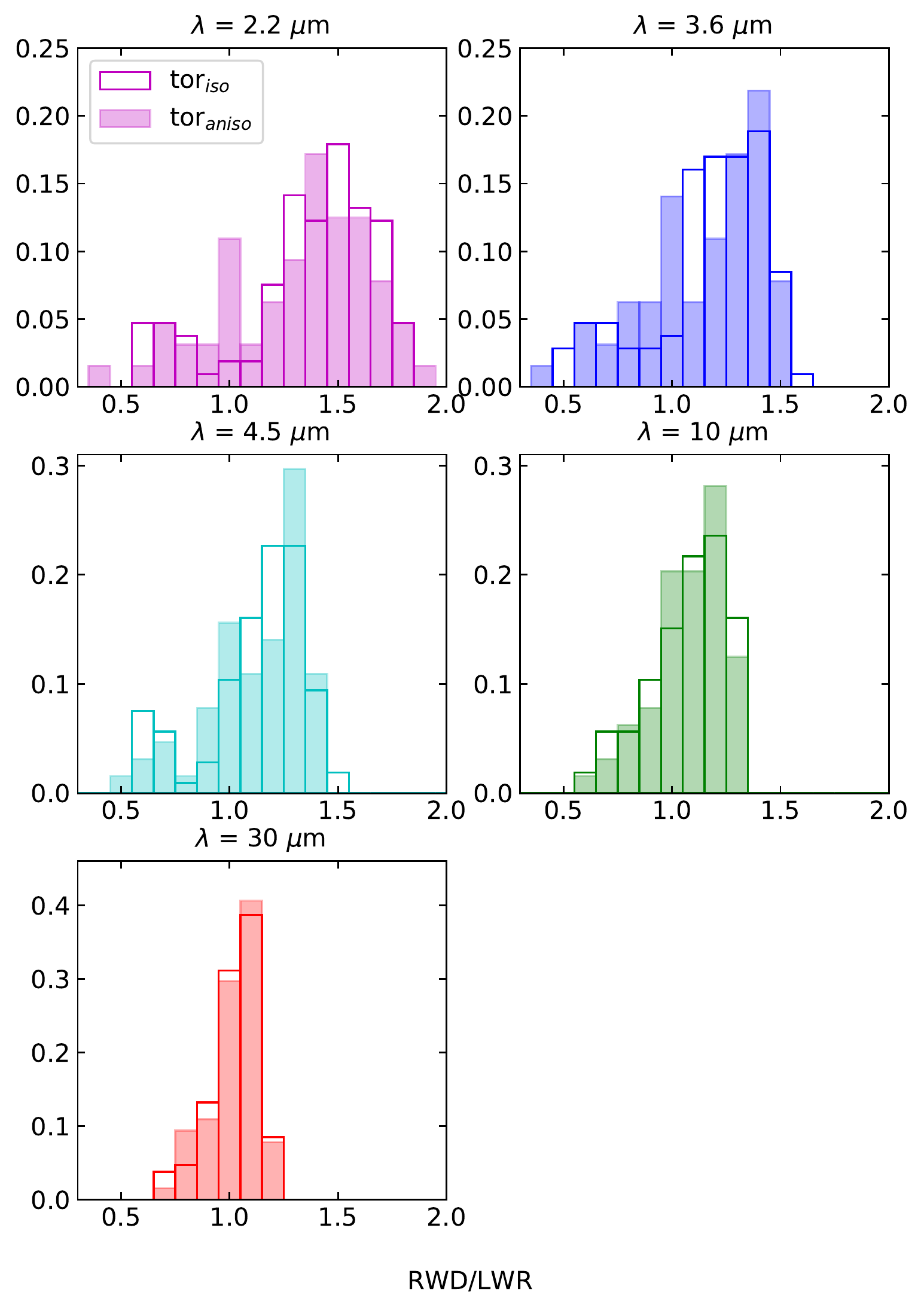}
	
\end{center}
\caption{\label{gothistogram}The normalized distributions of RWD/LWR with respect to wavelength for the complete set of globally optically thick torus models. Each panel shows the distribution with respect to isotropic or anisotropic illumination of the torus as empty and shaded bars, respectively. For better comparison, the distributions have been normalized by the total number of models run for the two illumination cases (106 and 64 for isotropic and anisotropic illumination, respectively).} 
\end{figure}

Figure~\ref{gothistogram} shows the RWD/LWR distributions for the globally optically thick models shown in Figure~\ref{gotrwdlwr}, broken down by wavelength. The models where the torus is illuminated either isotropically or anisotropically are represented by open and shaded bars, respectively.   At each wavelength, the distributions peak at RWD/LWR $ > 1$ but are generally asymmetric, with a tail extending to values $<1$. The widths of the RWD/LWR distributions increase as wavelength decreases, with the peak also shifting to higher values. The median values, in fact, increase systematically as wavelength decreases, as can be seen in Table~\ref{rwdlwrtab}. 

Interestingly, the RWD/LWR distributions are very similar for isotropically and anisotropically illuminated tori, even though, as discussed earlier, the latter typically have much smaller RWDs. 
In fact, the two-sample Kolmogorov--Smirnov test, whether applied to the complete sets of models for both illumination cases (106 models for isotropic and 64 for anisotropic illumination) or to matched sets of models (54), indicates that the distributions are indistinguishable at the 95\% confidence level.
This is because anisotropic illumination causes comparable decreases in both the RWD and the LWR, so there is no systematic shift in the RWD/LWR distribution relative to isotropic illumination.

However, the most important general result is that, even at $2.2\,\mu\rm m$, at which the RWD/LWR distribution is broadest, almost all RWD/LWR values are within a factor of $\sim 2$ of the ideal case RWD/LWR=1. 

The minimum, median, and maximum RWD/LWR values for each wavelength are listed in Table~\ref{rwdlwrtab} for both isotropic and anisotropic illumination and for the various cloud-level and global optical depth effects considered in Sections~\ref{sec:DustRF} --~\ref{sec:gottor}. In general, the median RWD/LWR at a given wavelength is largest for the globally optically thin torus and smallest for the globally optically thick torus. We also see that cloud occultation has a greater effect in decreasing this ratio than cloud shadowing. As the globally optically thick models typically have the smallest median values but also exhibit the widest range in RWD/LWR, the general result that the RWD/LWR values are within a factor of $\sim 2$ of unity holds in all cases. 
More precisely, if we consider the minimum and maximum values for the most extreme case (the anisotropically illuminated globally optically thick torus), then at the short wavelengths (2.2, 3.6, and $4.5\,\mu\rm m$) the RWD underestimates the LWR by at most a factor of $\approx 2.5$, or overestimates it by at most a factor of 1.4--1.8. At 10 and  $30\,\mu\rm m$, the RWD is within 10--40\% of the LWR. 
However, the median values indicate that more typically the RWD only overestimates the LWR by 30--40\% at $2.2\,\mu\rm m$ and by $< 20$\% at longer wavelengths. 

It is also of interest to compare the RWD to the light-crossing time of the inner torus cavity, 
which is the same for all models. Table~\ref{RWDtab} lists the minimum, median, and maximum values of RWD in units of the normalized inner cavity light-crossing time, $\tau_{\rm d}$, which is given by RWD$\times Y$. The models that include only cloud shadowing have the lowest median values of RWD$\times Y$ but those of the globally optically thick models are only slightly larger median values, while the models including only cloud occultation have the largest values. This indicates that cloud shadowing has the largest effect in modifying RWD$\times Y$ relative to the globally optically thin case. The range in RWD$\times Y$ is also much larger than the range in RWD/LWR, particularly for the globally optically thick models. For example, the maximum values of RWD$\times Y$ are typically $\sim 4-5$, implying that the RWD$\sim 4-5\times \tau_{\rm d}$, even at $2.2\,\mu\rm m$. 
However, the median values for $\lambda \leq 4.5\,\mu\rm m$ are $\sim 1$. 

At a given wavelength, the dust cloud distribution parameters such as $p$ and $\Phi$, global opacity effects, and anisotropic illumination all play important roles in determining the RWD and LWR of a torus. However, despite the range of values in RWD, LWR, and therefore RWD/LWR, Figure~\ref{gotrwdlwr} and Table~\ref{rwdlwrtab} show that the RWD is a reliable proxy of the LWR for the globally optically thick torus. The correlation between RWD and LWR becomes tighter as wavelength increases. Regardless of wavelength and illumination, the median RWD/LWR  for all the globally optically thick models is 1.09, with minimum and maximum values of 0.38 and 1.80, respectively. Thus, the RWD is always within a factor of 3 (at most) of the LWR.


\section{Discussion}\label{sec:dis}
In A17, we described the basic capabilities of our dust reverberation code, TORMAC, and discussed the dust emission RFs for a limited range of torus parameters, in particular $Y$, $p$, and $i$. In that paper, we focused on the effects of dust radiative transfer within individual clouds (i.e., the cloud orientation effect) for cases where the torus is illuminated either isotropically or anisotropically by the AGN continuum. We also briefly discussed the effects of cloud shadowing on the RFs.

 In this paper, we have introduced cloud occultation and expanded the parameter space to include several additional parameters, specifically $\Phi$, $s$, $\sigma$, and $\tau_V$, as well as wider ranges in the parameters discussed in A17. We  separately investigated the effects on the RFs of the different radiative transfer treatments included in TORMAC (cloud orientation, shadowing, and occultation) and then presented globally optically thick models in which all these effects are included. 
In addition, we considered the effects of anisotropic illumination and compared RFs computed for a sharp-edged torus (the standard model) with the RFs for more realistic ``fuzzy'' torus models.
We have also introduced quantities that characterize the lag (RWD), effective torus radius (LWR), and RF shape (CTAR) and investigated their dependence on the model parameters. 

	\subsection{Which Torus Properties Most Strongly Influence the Response?}\label{sec:dis_rt}
The blackbody cloud models discussed in Section~\ref{sec:rBBRF} illustrate how the shape of the RF depends on the basic torus parameters $Y$, $i$, $\sigma$, and $p$. The response reaches its maximum amplitude within the inner cavity light-crossing time ($\tau_{\rm d} = 1/Y$), and therefore the radial depth parameter, $Y$, sets the width in $\tau$ of the RF ``core'' relative to the overall duration of the response but does not otherwise affect its shape. The shape of the decay ``tail'' segment of the RF is controlled by the power-law index ($p$) of the radial cloud distribution, with more centrally concentrated distributions resulting in steeper decays. The inclination ($i$) and angular width ($\sigma$) affect the shape (and width) of the core response (for $\tau < 1/Y$), which, for example, is flat-topped with a delayed onset for $i=0^\circ$ but double-peaked for $i=90^\circ$. 

For this idealized case of a disk with constant angular width, the RWD depends only on $Y$ and $p$;  it is {\em independent} of $i$ and $\sigma$, just as for a spherical shell (Section~\ref{sec:rBBRF}). In addition, and again similar to the spherical shell case, RWD = LWR independently of the model parameters. That is, the dust reverberation lag (RWD) is, in this ideal case, an exact measure of the effective radius of the torus (LWR).

However, the RFs can be substantially modified, 
in a wavelength-dependent way, both by dust radiative transfer within individual clouds and also by extinction effects within the body of the torus. 

The emission from an individual cloud at a particular wavelength depends not only on $T_{\rm cl}$ and $\tau_V$ but also, because of the temperature gradient between the illuminated and non-illuminated sides, on $\alpha$,  its position angle within the torus with respect to the observer.  The emission is therefore orientation dependent and the effect is to reduce the response at short delays ($\tau\sim0$), particularly for $i>0$ and for shorter wavelengths. The change in the response increases the RWD,  which results in the RWD overestimating the LWR for all wavelengths in  ``globally optically thin" models.

Also, recall that the change in the emitted flux of a cloud is highly wavelength dependent, with the emission at shorter wavelengths decreasing much faster with respect to distance from the central source than the emission at larger wavelengths. Thus, the RFs for the shorter wavelengths are generally much more sensitive to the model parameters. 

Cloud shadowing, that is, shielding of clouds from the AGN continuum source by intervening clouds, also causes wavelength-dependent modifications to the RF. The fraction of clouds that is shadowed is a function of $Y$, $p$, and $\Phi$. As $\Phi$ increases, more clouds are shadowed, leading to changes in the RF shape similar to changing $p$. In other words, the RF is less severely reduced at short delays (partly offsetting the effects due to cloud orientation), and the tail decays more steeply as more clouds are shadowed. This effect tends to decrease the RWD, although it still, slightly, overestimates the LWR.  

As noted in A17, the contribution of the indirectly heated clouds is likely overestimated at higher values of $\Phi$, since we assume that the diffuse radiation field comes from 
neighboring directly heated clouds. In reality, at higher values of $\Phi$, most directly heated clouds are concentrated near the inner edge of the torus, and therefore the diffuse radiation produced by these clouds 
would be attenuated by geometrical dilution and extinction, and also subject to significant light-travel delays.

Cloud occultation refers to the extinction of IR emission from a given cloud due to intervening clouds along the line of sight to the observer. The amount of attenuation depends on $\Phi$, $p$, $i$, $\sigma$, and $\tau_V$. In this case, the RFs are significantly affected at all wavelengths, not just the shorter ones. The RFs are typically more sharply peaked at earlier delays, offsetting the effects of cloud orientation, but have shallower tail decays. Overall, this leads to larger RWDs. However, the ratio between RWD and LWR is smaller than when cloud shadowing is included or when the torus is globally optically thin, and can even become $<1$. 

The cloud orientation, cloud shadowing, and cloud occultation effects discussed above combine to produce the RFs of the globally optically thick torus. As would be expected, the importance of the global opacity effects (cloud shadowing and occultation) in shaping the RF increases with the volume filling factor, $\Phi$. In general, as $\Phi$ increases, the RFs  become narrower and more sharply peaked at $\tau < 1/Y$ (due mainly to cloud occultation), with more rapidly decaying tails (due to cloud shadowing). However, the way in which the RFs change as $\Phi$ varies is wavelength dependent, and it also depends on other torus parameters, particularly $p$, $i$, and, to a lesser extent, $\tau_V$.  

Reflecting the changes in the RF, the RWD generally decreases as $\Phi$ increases, with the shorter wavelengths being the most sensitive. 
However, in detail the behavior also depends on inclination. Also, 
certain parameter combinations result in values of RWD $<$ LWR. For example, for $\lambda \leq 4.5\,\mu$m RWD $\sim 0.4$LWR$-0.5$LWR when $\Phi \gtrsim 0.01$, $p=-2$, and $i=0$ or $90^\circ$.

As discussed in A17, when the torus is illuminated anisotropically by the AGN UV/optical continuum, the dust sublimation radius becomes a function of polar angle, reaching a minimum in the equatorial plane ($\theta = 90^\circ$).  The degree of anisotropy of the central source illumination is governed by the parameter $s$, which sets the luminosity at polar angle $\theta = 0^\circ$, relative to the isotropic luminosity. 

As a result of the smaller light-travel times to the innermost clouds, the RFs  exhibit narrower core widths and sharper peaks than for isotropic illumination. 
The degree to which anisotropic illumination affects the RFs also depends on $p$, $\Phi$, and $i$. In general, anisotropic illumination results in smaller values of RWD,
with the decrease relative to isotropic illumination being dependent on $i$ and on $p$, in particular, as well as wavelength. Anisotropic illumination also results in a smaller LWR, which, in fact, exhibits variations with $s$ similar to those of the RWD. As a result, even though the RWD decreases by up to a factor of $\sim 2$ at shorter wavelengths, the RWD/LWR ratio is only slightly smaller, if at all (typically $\lesssim 20$\%), for anisotropic illumination than for isotropic illumination. 

The distribution of clouds in $\beta$ (elevation angle measured from the equatorial plane; Section~\ref{sec:model})
can have important effects on the RFs of globally optically thick models, since for a Gaussian distribution clouds at higher latitudes ($\beta \gtrsim \sigma$) are less affected by shadowing and occultation. The general effect is to broaden the RF in comparison to the sharp-edged case, especially at $i=90^\circ$ when the contribution of high-latitude clouds to the response is maximized. This produces increases in the RWD, with inclination having a stronger influence on the changes in RWD. 

Overall, in the general globally optically thick case, the RWD is affected by all the parameters that determine the torus geometry ($Y$, $i$, $\sigma$), its illumination ($s$), the cloud distribution (both radially, $p$, and in $\beta$), and local or global opacity ($\tau_V$, $\Phi$). The strength of the dependence on any one of these parameters varies with wavelength and, usually, with other parameters. Nevertheless, it is possible to make some general statements. A general summary of how the descriptive quantities change with respect to each parameter can also be found in Table~\ref{RQT}.

First, the RWD is much more sensitive to all of the model parameters at shorter wavelengths (i.e., $\lambda \leq 4.5\,\mu$m) than at longer wavelengths. Considering specific parameters, the RWD is most strongly dependent on $Y$, depends strongly on $p$, depends more moderately on $\Phi$ and $s$, but is relatively insensitive to $i$, $\sigma$, and $\tau_V$. 

At $3.6\,\mu$m, the RWD decreases by a factor of $\sim 30$ as $Y$ is increased from $Y=2$ to 50, for $p=0$. However, this mainly reflects the $1/Y$ scaling of the torus inner radius since the LWR at $\lambda \leq 4.5\,\mu$m is generally within a factor of $\sim 2$ of $R_{\rm d}$.  Also at $3.6\,\mu$m, the RWD increases by a factor of $\sim 6$ as $p$ varies from $-2$ to $4$ and decreases by a factor of $\sim 3$ as $s$ decreases from $s=1$ (isotropic illumination) to 0.01 (highly anisotropic). The behavior of RWD with respect to $\Phi$ differs somewhat depending on $p$ and $i$, but it varies by up to a factor of $\sim 3$ at $3.6\,\mu$m over the range in $\Phi$ sampled by our models.

In contrast, the RWD varies by only $\approx 10 -20$\% with $i$, $\sigma$, and $\tau_V$. However, its behavior with other parameters, notably $\Phi$ and $s$, depends on how the torus is inclined relative to the line of sight. 

\begin{table*}
\begin{center}
\caption{{Response Function Quantities:}\\ Summary of dependences on torus parameter for a globally optically thick torus.} 
\label{RQT}

\scriptsize{

{\RaggedRight
\begin{tabular}{ C{1.5cm} | L{3.5cm} | L{3.5cm} | L{3.5cm} | L{3.5cm} }

        \hline
        \hline
        Torus Parameter & RWD & LWR & RWD$/$LWR & CTAR \\ 
        
        \hline
        $Y$ &   V. Strong decrease ($\rm R^{\rm a}_{3.6\,\mu \rm m} =$30; R$_{30\,\mu \rm m} =$1.6) &  V. Strong decrease (R$_{3.6\,\mu \rm m} = 18$; R$_{30\,\mu \rm m} =1.6$) & Increase then decrease at shorter $\lambda$; slow increase otherwise; $<1$ at larger $Y$ (R$_{3.6\,\mu \rm m} = 0.7$; R$_{30\,\mu \rm m} =0.02$) & Moderate-strong, overall decrease (R$_{3.6\,\mu \rm m} =0.5$ ; R$_{30\,\mu \rm m} =4.2$)   \\
        
        \hline
        $p$ &   V. Strong increase (R$_{3.6\,\mu \rm m} =$5.3; R$_{30\,\mu \rm m} =$1.7) &  Strong increase (R$_{3.6\,\mu \rm m} = 3$; R$_{30\,\mu \rm m} =1.7$)& Increase then decrease; slow increase at $i=90^{\circ}$; $<1$ for $p \le 0$ (R$_{3.6\,\mu \rm m} =0.7 $; R$_{30\,\mu \rm m} =0.08$)& V.Strong decrease (R$_{3.6\,\mu \rm m} = 15$; R$_{30\,\mu \rm m} =21$) \\
        \hline
        $i$ &   Slow inc./dec. (R$_{3.6\,\mu \rm m} =0.23$; R$_{30\,\mu \rm m} =$0.03) &  Slow inc/dec (R$_{3.6\,\mu \rm m} = 0.26$; R$_{30\,\mu \rm m} = 0.12$) & Slow increase then decrease (R$_{3.6\,\mu \rm m} =0.4$ ; R$_{30\,\mu \rm m} =0.1$) & Weak inc./dec. (R$_{3.6\,\mu \rm m} =0.3$ ; R$_{30\,\mu \rm m} =0.1$) \\
        \hline
        $\sigma$ &  Slow decrease  (R$_{3.6\,\mu \rm m} =0.1$; R$_{30\,\mu \rm m} =$0.01) &  Slow increase (R$_{3.6\,\mu \rm m} = 0.1$; R$_{30\,\mu \rm m} =0.1$) & Depends on $i$; decrease for $i \le 45^{\circ}$; V. slow increase at $i =90^{\circ}$ (R$_{3.6\,\mu \rm m} = 0.2$; R$_{30\,\mu \rm m} =0.1$) & Weak/moderate inc./dec. (R$_{3.6\,\mu \rm m} = 0.08$; R$_{30\,\mu \rm m} =0.54$) \\ 
        \hline
        $s$ &  Strong increase (R$_{3.6\,\mu \rm m} =1.7$; R$_{30\,\mu \rm m} =$0.24) & Moderate increase (R$_{3.6\,\mu \rm m} =0.96$ ; R$_{30\,\mu \rm m} =0.21$) & Slow increase; $<1$ at shorter $\lambda$ for $i=0^{\circ}$  (R$_{3.6\,\mu \rm m} =0.4$ ; R$_{30\,\mu \rm m} =0.02$) & Weak/moderate increase (R$_{3.6\,\mu \rm m} = 0.18$; R$_{30\,\mu \rm m} =0.56$) \\
        \hline
        $\tau_{V}$ &  Slow dec./inc.  (R$_{3.6\,\mu \rm m} =0.09$; R$_{30\,\mu \rm m} =$0.04) & Slow increase (R$_{3.6\,\mu \rm m} = 0.04$; R$_{30\,\mu \rm m} =0.15$) & Slow decrease for most cases; $<1$ for $\tau_{V}>40$ (R$_{3.6\,\mu \rm m} =0.1$ ; R$_{30\,\mu \rm m} =0.1$) & Weak increase (R$_{3.6\,\mu \rm m} =0.05$ ; R$_{30\,\mu \rm m} =0.03$) \\
        \hline
        $\Phi$ &  Moderate dec./inc.  (R$_{3.6\,\mu \rm m} =1.8$); R$_{30\,\mu \rm m} =0.06$) & Slow dec. then increase (R$_{3.6\,\mu \rm m} = 0.53$; R$_{30\,\mu \rm m} =0.32$) & Moderate-strong decrease, sometimes flat (R$_{3.6\,\mu \rm m} =1$; R$_{30\,\mu \rm m} =0.3$) & Moderate/weak increase (R$_{3.6\,\mu \rm m} = 0.64$; R$_{30\,\mu \rm m} =0.06$) \\
        \hline

\end{tabular}}}

\end{center}
 \footnotesize{$^a$ $R$ is the variability descriptor, where $\rm R=(Q_{max}/Q_{min})-1$, and $\rm Q$ represents the descriptive quantity (i.e., RWD, LWR, CTAR).  

 \begin{itemize}
 \begin{minipage}{0.4\linewidth}
 	\item R $=0$ \ \ \ \ \ \ \ \ \ Constant
	\item $0 < \rm R \le 0.5$ \ Weak variability
	\item $0.5 < \rm R \le 1$ \	Moderate variability
\end{minipage}
\begin{minipage}{0.4\linewidth}
	\item $1 < \rm R \le 4$ \	Strong variability
	\item R $>4$ \ \ \ \ \ \  Very strong variability
\end{minipage}	
 \end{itemize}

 }
 
\end{table*}

\subsection{How Well Does the RWD Measure the LWR?}\label{sec:dis_pars}

We have shown that the RWD can vary widely from model to model depending on the values of $Y$, $p$, $s$, and $\Phi$. Nevertheless, as discussed in Section~\ref{sec:rwdlwr}, the RWD still tracks the LWR fairly well. This is because the LWR itself depends on these parameters and exhibits similar, although not identical, behavior. That LWR varies with $Y$ and $p$ is to be expected from the analytical solution for the spherical models. Similarly, LWR depends on $\Phi$ because cloud shadowing causes directly illuminated clouds to be more centrally concentrated and with $s$ because of the polar angle dependence of $R_{\rm d}$. 
The weaker dependence of LWR 
on $i$ and $\sigma$ is due to the global opacity effects, particularly cloud occultation, which depends on the path length through the torus along the observer's line of sight.

Thus, the relationship between RWD and LWR is, in general, model dependent. The RWD/LWR ratio varies to a greater or lesser extent with all the model parameters and also depends on whether the torus is sharp-edged, or fuzzy.

There is a clear trend with wavelength in the distribution of RWD/LWR, with shorter wavelengths having higher median values and broader distributions. 
In general, at a given wavelength, a specific value of RWD corresponds to a range in LWR, with the range being wider at shorter wavelengths. 
The median RWD/LWR values imply that the RWD typically overestimates the LWR, but only by $\lesssim 50$\%, even at $2.2\,\mu$m. However, for certain parameter values (e.g., $p=-2$, $\Phi \geq 0.01$, $i=0^\circ$ for $\lambda \leq 4.5\,\mu$m) it is also possible for the RWD to underestimate the LWR by as much as a factor of $\sim 2$. 
Also, interestingly, the distributions in RWD/LWR are very similar for isotropic and anisotropic illumination.

Nevertheless, considering the entire parameter space covered by our model grid,\footnote{Although the model parameter space was not covered uniformly in this study, we chose values so as to cover the widest plausible range in each parameter. Therefore, we consider that the distribution of models in the RWD--LWR plane reliably represents the scatter due to the ranges in all of the torus parameters explored here, even though these were not fully sampled for each parameter.} which covers all (if not more) plausible possibilities within the clumpy torus paradigm, the RWD is always within a factor of $\sim 3$ of the LWR. Thus, the RWD, which is measurable via reverberation mapping, provides an estimate of the effective radius of the torus (i.e., LWR) at the observed wavelength to within a factor of $\sim 3$, irrespective of the geometrical configuration of the torus, its orientation, cloud distribution, or illumination by the AGN continuum.

Compared to the ligh-crossing time corresponding to $2R_{\rm d,iso}$, that is, the inner cavity of the isotropically illuminated torus, the RWD covers a much wider range (typically a factor of $\sim 10$), even at wavelengths $\lambda \leq 4.5\,\mu$m, where the emission is dominated by the hottest clouds in the inner regions of the torus. This is not surprising because the RWD tracks the LWR, not { $R_{\rm d, iso}$}. Depending on wavelength and the torus parameters, {$r_{\rm LW}$ (= LWR$\times R_o$)} can be a factor of $\sim 10$ larger than $R_{\rm d, iso}$, which is fixed by the AGN luminosity and so does not depend on these parameters. Thus, even at $2.2\,\mu$m, the RWD can exceed $\tau_{\rm d}$ by factors of $2-4$ for $p\geq 2$.  On the other hand, under anisotropic illumination with $s=0.1$ and for $p\leq 0$, the RWD at $2.2\,\mu$m can reach values as low as $\sim 0.3\tau_{\rm d}$, since in this case $R_{\rm d}(\theta = 90^\circ)\approx 0.3R_{\rm d,iso}$.
It is also worth noting that the median values and ranges of RWD relative to $\tau_{\rm d}$ are very similar at 2.2, 3.6, and $4.5\,\mu$m, so the latter two wavelengths are just as much representative of the ``inner radius'' of the torus as the K-band, at least for the ISM dust composition adopted in the models presented here.

	\subsection{RWD--LWR Relationship and Implications for Reverberation Mapping}\label{sec:dis_rl}
	
\begin{figure*}[ht]
\begin{center}
	\includegraphics[scale=0.5]{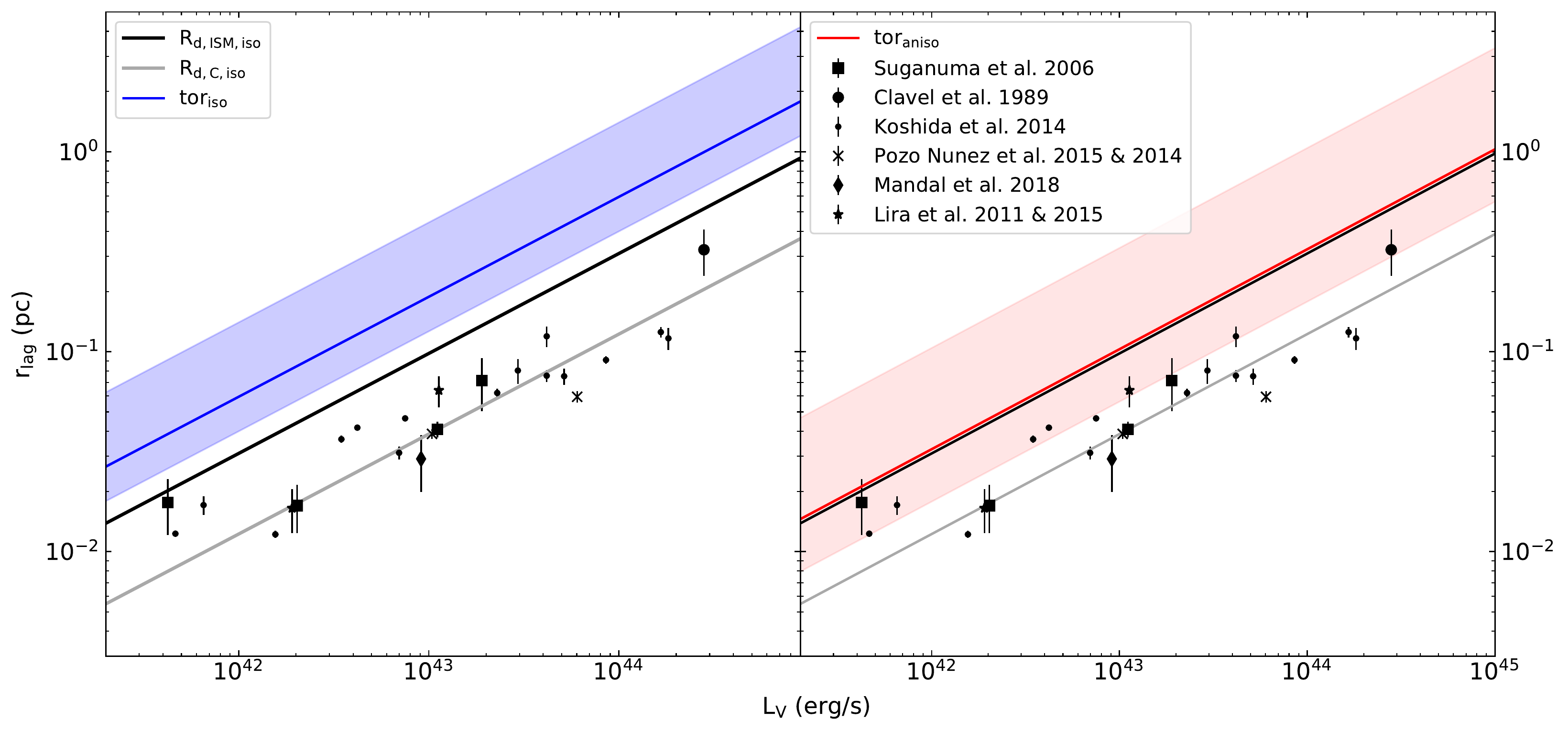}
\end{center}
\caption{\label{rad_lum} The K-band radius--luminosity relationships for the observationally determined values of $r_{\rm K}$ available in the literature (\citealp{Clavel:1989aa,Suganuma:2006aa,Lira:2011aa,Koshida:2014aa,Pozo-Nunez:2014aa,Lira:2015aa,Pozo-Nunez:2015aa,Mandal:2018aa}) and for $r_{\rm RW,K}$ values derived from a subset of our globally optically thick torus models. The values of $r_{\rm RW}$ corresponding to the median value of the RWD and the range between minimum and maximum RWD values at $\lambda = 2.2\,\mu$m are represented by a solid line and a shaded band, respectively, for both the isotropically (blue; left panel) and anisotropically (red; right panel) illuminated torus. The isotropically illuminated sublimation radii for the ISM dust mixture \citep[][and Equation~\ref{Rdtheta} where $s=1$]{Nenkova:2008ab} and graphite grains \citep{Mor:2012aa} are also plotted as the black and gray solid lines, respectively.}
\end{figure*}

As outlined in Section~\ref{intro}, the reverberation radius has been measured in the K-band for about 20 AGN.
An important result that has emerged from these studies is that the K-band reverberation radius ($r_{\rm K}$) correlates tightly with AGN luminosity, scaling as $r_{\rm K}\propto {L_{\rm AGN}}^{0.5}$ (\citealp{Suganuma:2006aa, Koshida:2014aa}), which is the  expected relationship given that the dust emitting in the K-band is heated by the AGN (Equation~\ref{Rdtheta}). 

However, it has also been pointed out that the measured reverberation radii are systematically smaller by a factor of $\sim 2$ than the theoretical dust sublimation radius for an ISM dust composition with an average grain size of 0.05 $\mu$m \citep{Kishimoto:2007aa}. Several explanations have been proposed to explain this discrepancy, including an inner component of hot graphite dust \citep{Mor:2009aa,Mor:2012aa} or large grains \citep{Kishimoto:2007aa}, or anisotropic illumination of the torus (\citealp{Kawaguchi:2010aa,Kawaguchi:2011aa}; A17). 

It is of interest to compare the radii corresponding to the range in the RWD values produced by our models with the observed K-band reverberation radii.
In principle, it is possible to recover the RWD directly from observations using the common reverberation mapping technique of cross-correlating  the optical and IR light curves to determine the time lag, $t_{\rm lag}$ (e.g., \citealp{Peterson:2004aa, Koshida:2014aa}). For well-sampled light curves covering a sufficiently long time baseline (i.e., long enough to sample the transfer function at all delays), the cross-correlation centroid (i.e., $t_{\rm lag}$) yields the centroid of the transfer function \citep{Koratkar:1991aa, Perez:1992aa}. In terms of our model RFs, this is equivalent to the time delay corresponding to the RWD: $t_{\rm RW} = 2R_{\rm o} \rm{RWD}/c = 2YR_d\rm{RWD}/c$. The corresponding radius is $r_{\rm RW} = ct_{\rm RW} = 2YR_{\rm d}\rm{RWD}$, where $R_d \simeq 0.4(L_{\rm AGN}/(10^{45} \,{\rm erg\, s^{-1}}))^{1/2}$\, pc (assuming $T_{\rm sub} = 1500$\,K and isotropic illumination).

Figure~\ref{rad_lum} shows the K-band radius--luminosity relationship for the observationally determined values of $r_{\rm K}$ available in the literature (see figure caption for references), compared with the $r_{\rm RW,K}$ values derived from the $2.2\,\mu$m RFs for a subset of our globally optically thick torus models. The values of $r_{\rm RW}$ corresponding to the median value of the RWD at $\lambda = 2.2\,\mu$m and the range between the minimum and maximum RWD values are represented by a solid line and a shaded band, respectively, for both the isotropically and anisotropically illuminated torus (left and right panels, respectively). In each case, the model subset used for this purpose covers restricted ranges in $p$ ($p = 0, 2$) and $\Phi$ (0.01--0.1), which approximately correspond to the models found by \cite{Nenkova:2008ab} to broadly reproduce observed AGN SEDs. In addition, we restrict the inclination to $i \leq 60^{\circ}$ since the reverberation radii can only be determined for broad-line (type 1) AGN (however, all values of $Y$, $\tau_V$, $s$, and $\sigma$ are included). The isotropically illuminated sublimation radii for the ISM dust mixture with $T_{\rm sub}=1500$ K \citep[][and Equation~\ref{Rdtheta} where $s=1$]{Nenkova:2008ab} and graphite grains $T_{\rm sub}=1800$ K \citep{Mor:2012aa} are also plotted for comparison. 

The observed reverberation radii ($r_K$) scale quite tightly with luminosity as $r_K\propto (\lambda L_{\rm V})^{1/2}$, as already noted, with the isotropically illuminated ISM dust sublimation radius lying above the observed trend by about a factor of 2. The band representing the isotropically illuminated torus models lies well above the observed radius--luminosity relationship (it also lies above the ISM dust sublimation line, but this is not surprising, as the torus inner radius is defined as $R_{\rm d, iso}$ and the median LWR is $r_{\rm LW}\approx 1.3R_{\rm d, iso}$ at $2.2\,\mu$m). The median radius for the models is a factor of $\sim 4$ higher than the observed reverberation radii. The range in $r_{\rm RW}$ for the anisotropically illuminated torus extends below the ISM dust sublimation line and in fact overlaps the larger values of $r_{\rm K}$. However, even in this case, the median $r_{\rm RW}$ line is still a factor of $\sim 2$ above the general observed trend (coincidentally close to the ISM dust sublimation line). The smallest lags obtained for the anisotropic illumination models are $t_{\rm RW}\sim 0.6R_{\rm d, iso}/c$, with these values being obtained for  $i=0^\circ$, $p=0$, and $s \leq 0.1$. Therefore, even though anisotropic illumination can produce lags substantially {\bf shorter} than that corresponding to the ISM dust sublimation radius (as previously noted by \citealp{Kawaguchi:2010aa,Kawaguchi:2011aa} and A17) this by itself is insufficient to completely explain the discrepancy between the sublimation radius for the ISM dust mixture and the measured K-band lags.

As previously noted by \citet{Vazquez:2015aa}, the majority of the $r_K$ values actually fall close to the sublimation line for graphite grains, suggesting that the reverberation response in the K-band could be dominated by such grains. It has also been suggested that a component of hot graphite dust residing within the ISM dust sublimation radius is required to account for a peak in the NIR SED of broad-line AGN (\citealt{Mor:2012aa}; see also \citealt{Koshida:2014aa}). Since the sublimation temperature is higher for graphite than silicate grains and increases with grain size, inner dust clouds that survive several continuum variability events are likely to evolve a grain composition that is dominated by large graphite grains \citep{Perna:2003aa}. Some of these graphite clouds may also be part of a polar wind, rather than the torus itself \citep{Honig:2017aa}. 
Infrared interferometry has revealed a strong polar component extended on scales of a few parsecs (i.e., a few $\times$ the torus inner radius)  in the MIR emission of several nearby AGN \citep[e.g.,][]{Honig:2013aa,Lopez-Gonzaga:2016aa}, and polar dust emission appears to be relatively common on scales of tens of parsecs \citep{Asmus:2016aa}. It has been proposed that these structures are associated with polar winds emerging from the dust sublimation region of the torus \citep{Honig:2013aa,Honig:2017aa}.
The effects of a population of clouds composed of large graphite grains and of a polar wind component on the reverberation response will be subjects of future papers.

Another notable feature of Figure~\ref{rad_lum} is that the scatter about the observed radius--luminosity relation ($\approx 0.3$ dex) is much smaller than the range in $r_{\rm RW}$ spanned by the model bands ($\sim 0.7$\,dex for both isotropic and anisotropic illumination), even for the restricted subset of models used here. This suggests that in reality there are relatively small variations between AGN in the parameters that most affect the torus response, namely,\footnote{Note that $r_{\rm RW}$ is relatively insensitive to $Y$, since the RWD varies approximately as $1/Y$} $p$, $\Phi$, and $s$. For example, if the $p=2$ models are excluded from the model subset, the range in $r_{\rm RW}$ for isotropic illumination is comparable to the observed scatter. Alternatively, it is possible that the relatively small observed scatter is a consequence of the fact that, as discussed above, the K-band response is dominated by hot graphite dust clouds located within a narrow range in radius, between the sublimation radii for graphite and silicate grains. 

The relatively small scatter in the observed $r_K\propto (\lambda L_{\rm V})^{1/2}$ relation promises that a more detailed comparison between torus response simulations and the observed light curves may place much tighter constraints on the torus properties as represented by the model parameters. In this paper, we have presented the torus responses to a short UV/optical continuum pulse to approximate the transfer function. However, TORMAC is also capable of computing the torus response to input light curves based on observed AGN optical light curves (A17). In future work, we will use TORMAC to model the IR (K-band; Spitzer 3.5 and 4.5 $\mu$m) light curves of previously monitored individual sources in detail, with the aim of generating the best-fit probability density distributions for the key torus parameters.

\subsection{The Effects of Saturation}\label{sec:dis_sat}

As the AGN luminosity increases, the surface temperature of clouds located near the inner edge of the torus is likely to exceed $T_{\rm sub}$. However, as outlined in Section~\ref{sec:model} (see also the discussion in A17), these clouds are optically thick and may be expected to survive much longer than the sublimation time for an individual grain. In the current version of TORMAC, this is handled simply by restricting the cloud surface temperature to a maximum value  $T_{\rm cl,max} =T_{\rm sub} = 1500$\,K, regardless of the computed value. The emission from such clouds does not change in response to the input continuum pulse that is used to compute the RFs. This inhibits the torus response at short delays, with the result that the RF tends to ``saturate''. In reality, however, even individual grains can survive at temperatures above $T_{\rm sub}$ for timescales comparable to characteristic AGN continuum variability timescales (e.g., $\sim 60$ days for a $1\,\mu$m grain at 1600\,K; \citealp{Waxman:2000aa}). Emission from these ``super-heated'' grains will boost the response amplitude, particularly at shorter wavelengths. 

In order to determine whether this saturation effect could significantly change our results, a comparison set of torus models was computed in which saturation does not occur. For this purpose, the temperature range covered by the grid of synthetic cloud spectra used to determine the cloud emission was extended to include values of  $T_{\rm cl}$ up to $1700$\,K. This is approximately the surface temperature that a cloud at radius $R_{\rm d}$ would reach for a factor of 2 increase in $L_{\rm AGN}$, the amplitude of the continuum pulse used to compute the RFs.  Response functions were then computed using the extended grid for a set of models in which $T_{\rm cl,max}=1700$\,K, sparsely sampling the ranges in the torus parameters that are considered in this paper.

The RFs produced by the $T_{\rm cl,max}=1700$\,K models have slightly shorter RWDs than the corresponding models for which the cloud surface temperature is capped at $T_{\rm cl,max} =1500$\,K (as used in this paper). This is because the innermost clouds, which respond at the shortest delays, reach higher temperatures and thus produce stronger emission in response to the continuum pulse than is the case when their temperatures are capped. This has the effect of enhancing the amplitude of the RF at shorter delays, resulting in a shorter RWD than when the RF is saturated. 
Therefore, comparison of the two sets of models shows that the main effect of saturation (that is, capping the cloud surface temperature) is to slightly increase the RWD.

The size of the effect is largest at the shorter wavelengths and also depends on the torus parameters, particularly $Y$, $p$, and, to a lesser extent, $\Phi$ and $i$.  However, even at 2.2 $\mu$m, the differences between the RFs for $T_{\rm cl,max}=1500$\,K (saturated) and $T_{\rm cl,max}=1700$\,K (non-saturated) are relatively small. The median increase in RWD at this wavelength is $\sim 10$\%, with the most extreme case being a 40\% increase for an anisotropically illuminated, face-on torus. 

As the LWR is computed for the initial state of the torus, prior to the continuum pulse, it is not affected by saturation, and therefore the RWD/LWR ratio increases by the same amount as the RWD. 
On the other hand, saturation tends to decrease CTAR, again because the amplitude at early delays (i.e., within the core of the RF) is reduced. The effect is again larger at shorter wavelengths but is relatively small even at these wavelengths; the median decrease in CTAR is $\sim 15$\% at 2.2 $\mu$m. 

Overall, saturation has a relatively small effect. Even at 2.2 $\mu$m, it only increases the RWD and hence the quantities derived from it (RWD/LWR ratio and $r_{\rm RW}$) by $\sim 10$\%. This does not significantly alter our main results.


\section{Conclusions}\label{sec:conc}
TORMAC simulates the multiwavelength IR response of the AGN torus to variations in the optical/UV continuum. It includes the effects of anisotropic emission from the clouds (cloud orientation) and accounts for cloud shadowing and heating by the diffuse IR radiation field. It also accounts for anisotropic illumination of the torus by the AGN continuum.  In this paper, we outlined the implementation of cloud occultation and presented a comprehensive exploration of the torus RFs at selected wavelengths. 
In the most realistic case of a globally optically thick torus, a given RF model is defined by parameters that determine the torus geometry ($Y$, $i$, $\sigma$), the degree of anisotropy of the AGN radiation field ($s$), the cloud distribution in radius ($p$) and in polar angle (uniform or Gaussian), the cloud optical depth ($\tau_{\rm V}$), and the volume filling factor ($\Phi$). 

We first explored the RFs for an idealized case in which the torus is populated by isotropically emitting blackbody clouds, in order to characterize the basic shape of the RF for a disk 
and to establish how the response-weighted delay (RWD), its ratio to the luminosity weighted radius of the torus (RWD/LWR), and the shape of the RF (CTAR) depend on $Y$, $p$, $\sigma$, and $i$. In general, we find that even though the RF {\em shape} depends on all 4 parameters, the RWD depends only on $Y$ (controlling the RF core width, $\sim1/Y$) and $p$ (controlling the steepness of tail decay) and is independent of $i$ and $\sigma$. In fact, the behavior of the computed RWD with $p$ and $Y$ is identical to that predicted by a simple analytical formula derived for a spherical shell (Equation~\ref{eqn:RWD}). In addition, as for the spherical solution, RWD = LWR (expressed in dimensionless units) for all values of the model parameters.

A large set of RFs\footnote{The RFs and tabulated RWD, LWR, and CTAR values for the models presented in this paper are available from the authors upon request.} was created for torus models in which the IR dust emission was determined using cloud spectra generated using the DUSTY radiative transfer code. 
This model set was used to investigate the effects on the response at selected wavelengths (2.2, 3.6, 4.5, 10, $30\,\mu$m) of the torus geometry, cloud orientation, cloud optical depth, global torus opacity effects (cloud shadowing and cloud occultation), and anisotropic illumination. 

In these models, a given cloud's emitted flux as a function of surface temperature (hence distance from the illuminating source) is highly wavelength dependent, with the emission at shorter wavelengths decreasing more rapidly as temperature decreases. Furthermore, due to an internal temperature gradient, the dust clouds emit anisotropically, with the emission being more strongly anisotropic at shorter wavelengths. As a result, the observed emission from a cloud depends on its orientation with respect to the central source and the observer. This tends to reduce the response amplitude at short delays, which results in larger RWDs for wavelengths $<10\,\mu$m than would be the case if the clouds emitted isotropically. While the RFs are still mainly dependent on $Y$ and $p$, with these radiative transfer effects, they also become more dependent on $\sigma$ and $i$, showing larger variations in shape with these parameters, especially at shorter wavelengths. 

The global opacity effects, namely, cloud shadowing and cloud occultation, become influential as the volume filling factor increases; more clouds are shadowed/occulted for higher volume filling factors.
Cloud shadowing tends to result in ``sharper''  RFs yielding shorter RWDs, but the effects are highly wavelength dependent, being much stronger at shorter wavelengths. At longer wavelengths, the effect is smaller due to the contribution of the cooler, shadowed clouds that are heated by the diffuse radiation field. 
However, as noted in Footnote 1 and discussed in A17, for high values of the volume filling factor the approximate treatment of diffuse radiation field heating overestimates the contribution of these clouds to the RF. The effects of cloud occultation are also wavelength dependent, with shorter wavelengths again being most strongly affected. As the number of occulting clouds along a given line of sight is partly determined by the path length through the torus, the way in which cloud occultation affects the RF shape and hence the RWD depends on the volume filling factor and also inclination.  

Overall, the globally optically thick torus models, i.e. those with $\Phi \gtrsim 0.001$,  typically produce narrower RF core widths and steeper decay tails, leading to much smaller RWDs compared to the globally optically thin case (in which cloud shadowing and cloud occultation are insignificant). As might be expected, the RWD generally increases with wavelength, with the largest range in RWD occurring for larger values of $Y$ and smaller values of $p$. 

Anisotropic illumination of the torus by the AGN radiation field also has the effect of narrowing the RF core width and steepening the decay tail.  Therefore, anisotropic illumination substantially decreases the RWD compared to the isotropic illumination case. The largest effect occurs at shorter wavelengths, where the emission is largely due to the hotter, innermost clouds, but it also depends on inclination, since these inner clouds are more likely to be occulted at larger inclinations. 

In general (see Table~\ref{RQT}), the RWD is most influenced by $Y$, $p$, $s$, and $\Phi$;  it decreases as $Y$ increases (approximately as $1/Y$) and as $\Phi$ increases (but with the exact behavior depending on $i$ and $p$), whereas it decreases as $p$ decreases (i.e., as the clouds become more centrally concentrated) and as $s$ decreases (as the radiation field becomes more anisotropic).
In each case, the changes in RWD are larger at shorter (2.2, 3.6, $4.5\,\mu$m) than at longer (10, $30\,\mu$m) wavelengths. On the other hand, the RWD is relatively insensitive to $\sigma$, $\tau_V$, and also $i$, even though the RF shape is quite strongly inclination dependent. 

Our models show that the RF shapes can be quite dramatically affected by the torus geometry and inclination, the cloud distribution, radiative transfer and global opacity effects, and anisotropic illumination.
Nevertheless, it can be generally stated that RWD$\sim$LWR within a factor of 3, even though the RWD by itself can vary by a factor of $\sim 10$ at shorter wavelengths. The ratio RWD/LWR is a function of wavelength: for the globally optically thick models, the median value ranges from $\approx 1.4$ at $2.2\,\mu$m to $\approx 1.0$ at $30\,\mu$m. The dispersion also decreases with wavelength, from $0.4 \lesssim$ RWD/LWR $\lesssim 1.8$ at $2.2\,\mu$m to $0.7 \lesssim$ RWD/LWR $\lesssim 1.1$ at $30\,\mu$m. The volume filling factor drives much of the dispersion in RWD/LWR, which becomes more sensitive to other parameters, notably $i$ for large values of $\Phi$. On the other hand, as compared to the isotropic illumination case, anisotropic illumination does not produce systematically different values of RWD/LWR.

The finding that RWD $\approx$ LWR over the wide range of parameter values that we have explored is important for the interpretation of results from IR reverberation mapping campaigns on multiple levels. 
It confirms that the reverberation lag, which is essentially the time delay corresponding to the RWD, provides a reasonably robust estimate, to within a factor of $\sim 3$, of the luminosity-weighted torus radius. This therefore confirms the basic assumption that the reverberation lag is a good measure of the torus ``size'' at that wavelength, even though we lack detailed knowledge of the torus structure and composition. This in turn is important for using AGN IR reverberation lags as cosmological standard candles, which is the main goal of the new VISTA Extragalactic Infrared Legacy Survey \citep[VEILS;][]{Honig:2017aa}.

It should also be noted that the strong wavelength dependence of the IR reverberation response may have implications for detailed studies of the IR SEDs of AGN, since the dust emission at short wavelengths responds quickly to short timescale UV/optical continuum variability, whereas at longer wavelengths the dust emission response will only follow longer-term variations. In a sample of IR SEDs for different AGN obtained from ``snapshot'' observations, this could result in a wider dispersion in the strength of the dust continuum in the K -- M bands relative to that at $\sim 30\,\mu$m (where the torus SED peaks) than that due only to object-to-object differences in the SED.
The effects of reverberation on the torus SED will be explored in a future paper.

A number of further refinements are necessary to fully capture the complexities associated with dust sublimation and distributions in cloud optical depth and dust composition. In particular, compared to the results from observational reverberation mapping in the K-band, our models predict radii that are larger by factors of $\sim 2$ and $4$ for anisotropic and isotropic illumination, respectively.   
The discrepancy between the observed and modeled reverberation radii is most likely due to the lack of a ``hot dust" component, i.e., large, carbonaceous grains, residing within the time-averaged sublimation radius ($r < \langle R_{\rm d} \rangle$), whose emission spectrum is brighter at shorter (NIR) wavelengths. This can be treated by implementing a more sophisticated treatment of time-dependent dust sublimation that incorporates a radial distribution in dust cloud composition. 
These features, a dynamic treatment of dust sublimation, a radial gradient in dust composition, and polar clouds, will be included in the next version of TORMAC. This will enable more accurate calculations of the reverberation response at short wavelengths (especially the J, H, and K bands), which are most accessible to ground-based observations.

We also plan to include polar dust clouds in TORMAC since interferometric observations have shown that polar dust emission is the dominant source of the MIR continuum in some AGN (e.g., \citealp{Honig:2013aa,Asmus:2016aa}). In Seyfert 1 galaxies, this dust will in general be closer (in polar angle) to the observer's line-of-sight than the torus itself and could significantly modify the MIR response at short delays.

The key advantage of NIR/MIR reverberation mapping over single-dish imaging or
interferometry is that it is not critically dependent on spatial resolution and so it can be used to
study much larger samples of AGN. Although only $\sim 20$ AGN have been ``reverberation
mapped'' in the IR to date, the VEILS campaign is predicted to recover K-band light
curves for $\sim 500$ type 1 AGN at redshifts ranging up to $z \approx 1$ \citep{Honig:2017aa}.
The current observed scatter about the  $r_{\rm K}\propto L_{\rm UV}^{0.5}$ relation is much smaller than the range spanned by the model results, indicating that AGN tori are relatively homogeneous in comparison to the parameter ranges sampled by our model grid. This suggests that the planned upgrades and detailed modeling of observed light curves with TORMAC will yield constraints not only on the size of the torus but also on other properties such as the radial depth, opening angle, and cloud distribution, all as a function of redshift.

\acknowledgments
This paper is based on work supported by NASA under award Nos. NNX12AC68G and NNX16AF42G. T.A .acknowledges support from the Horizon 2020 ERC Starting Grant DUST-IN-THE-WIND (ERC-2015-StG-677117). A.R. thanks S. H\"onig and the University of Southampton Astronomy Group for their hospitality. We also thank the anonymous referee for useful comments that improved the clarity of the paper.

\bibliography{paper_v2}

\begin{thebibliography}{60}
\expandafter\ifx\csname natexlab\endcsname\relax\def\natexlab#1{#1}\fi

\bibitem[{{Almeyda} {et~al.}(2017){Almeyda}, {Robinson}, {Richmond}, {Vazquez},
  \& {Nikutta}}]{Almeyda:2017aa}
{Almeyda}, T., {Robinson}, A., {Richmond}, M., {Vazquez}, B., \& {Nikutta}, R.
  2017, \apj, 843, 3, 3

\bibitem[{{Almeyda}(2017)}]{Almeyda:2017ab}
{Almeyda}, T.~R. 2017, PhD thesis, Rochester Institute of Technology 

\bibitem[{{Alonso-Herrero} {et~al.}(2018){Alonso-Herrero}, {Pereira-Santaella},
  {Garc{\'\i}a-Burillo}, {Davies}, {Combes}, {Asmus}, {Bunker},
  {D{\'\i}az-Santos}, {Gand hi}, {Gonz{\'a}lez-Mart{\'\i}n},
  {Hern{\'a}n-Caballero}, {Hicks}, {H{\"o}nig}, {Labiano}, {Levenson},
  {Packham}, {Ramos Almeida}, {Ricci}, {Rigopoulou}, {Rosario}, {Sani}, \&
  {Ward}}]{Alonso-Herrero:2018aa}
{Alonso-Herrero}, A., {Pereira-Santaella}, M., {Garc{\'\i}a-Burillo}, S.,
  {et~al.} 2018, \apj, 859, 144, 144

\bibitem[{{Antonucci}(1993)}]{Antonucci:1993aa}
{Antonucci}, R. 1993, \araa, 31, 473, 473

\bibitem[{{Asmus} {et~al.}(2016){Asmus}, {H{\"o}nig}, \&
  {Gandhi}}]{Asmus:2016aa}
{Asmus}, D., {H{\"o}nig}, S.~F., \& {Gandhi}, P. 2016, \apj, 822, 109, 109

\bibitem[{{Barvainis}(1987)}]{Barvainis:1987aa}
{Barvainis}, R. 1987, \apj, 320, 537, 537

\bibitem[{{Barvainis}(1992)}]{Barvainis:1992aa}
---. 1992, \apj, 400, 502, 502

\bibitem[{{Blandford} \& {McKee}(1982)}]{Blandford:1982aa}
{Blandford}, R.~D., \& {McKee}, C.~F. 1982, \apj, 255, 419, 419

\bibitem[{{Burtscher} {et~al.}(2013){Burtscher}, {Meisenheimer}, {Tristram},
  {Jaffe}, {H{\"o}nig}, {Davies}, {Kishimoto}, {Pott}, {R{\"o}ttgering},
  {Schartmann}, {Weigelt}, \& {Wolf}}]{Burtscher:2013aa}
{Burtscher}, L., {Meisenheimer}, K., {Tristram}, K.~R.~W., {et~al.} 2013, \aap,
  558, A149, A149

\bibitem[{{Clavel} {et~al.}(1989){Clavel}, {Wamsteker}, \&
  {Glass}}]{Clavel:1989aa}
{Clavel}, J., {Wamsteker}, W., \& {Glass}, I.~S. 1989, \apj, 337, 236, 236

\bibitem[{{Combes, F.} {et~al.}(2019){Combes, F.}, {Garc\'{\i}a-Burillo, S.},
  {Audibert, A.}, {Hunt, L.}, {Eckart, A.}, {Aalto, S.}, {Casasola, V.},
  {Boone, F.}, {Krips, M.}, {Viti, S.}, {Sakamoto, K.}, {Muller, S.}, {Dasyra,
  K.}, {van der Werf, P.}, \& {Martin, S.}}]{Combes:2019aa}
{Combes, F.}, {Garc\'{\i}a-Burillo, S.}, {Audibert, A.}, {et~al.} 2019, A\&A,
  623, A79, A79

\bibitem[{{Gallimore} {et~al.}(2016){Gallimore}, {Elitzur}, {Maiolino},
  {Marconi}, {O'Dea}, {Lutz}, {Baum}, {Nikutta}, {Impellizzeri}, {Davies},
  {Kimball}, \& {Sani}}]{Gallimore:2016aa}
{Gallimore}, J.~F., {Elitzur}, M., {Maiolino}, R., {et~al.} 2016, \apjl, 829,
  L7, L7

\bibitem[{{Garc{\'{\i}}a-Burillo} {et~al.}(2016){Garc{\'{\i}}a-Burillo},
  {Combes}, {Ramos Almeida}, {Usero}, {Krips}, {Alonso-Herrero}, {Aalto},
  {Casasola}, {Hunt}, {Mart{\'{\i}}n}, {Viti}, {Colina}, {Costagliola},
  {Eckart}, {Fuente}, {Henkel}, {M{\'a}rquez}, {Neri}, {Schinnerer}, {Tacconi},
  \& {van der Werf}}]{Garcia-Burillo:2016aa}
{Garc{\'{\i}}a-Burillo}, S., {Combes}, F., {Ramos Almeida}, C., {et~al.} 2016,
  \apjl, 823, L12, L12

\bibitem[{{H{\"o}nig} {et~al.}(2013){H{\"o}nig}, {Kishimoto}, {Tristram},
  {Prieto}, {Gandhi}, {Asmus}, {Antonucci}, {Burtscher}, {Duschl}, \&
  {Weigelt}}]{Honig:2013aa}
{H{\"o}nig}, S.~F., {Kishimoto}, M., {Tristram}, K.~R.~W., {et~al.} 2013, \apj,
  771, 87, 87

\bibitem[{{H{\"o}nig} {et~al.}(2017){H{\"o}nig}, {Watson}, {Kishimoto},
  {Gandhi}, {Goad}, {Horne}, {Shankar}, {Banerji}, {Boulderstone}, {Jarvis},
  {Smith}, \& {Sullivan}}]{Honig:2017aa}
{H{\"o}nig}, S.~F., {Watson}, D., {Kishimoto}, M., {et~al.} 2017, \mnras, 464,
  1693, 1693

\bibitem[{{Ichikawa} \& {Tazaki}(2017)}]{Ichikawa:2017aa}
{Ichikawa}, K., \& {Tazaki}, R. 2017, \apj, 844, 21, 21

\bibitem[{{Imanishi} {et~al.}(2016){Imanishi}, {Nakanishi}, \&
  {Izumi}}]{Imanishi:2016aa}
{Imanishi}, M., {Nakanishi}, K., \& {Izumi}, T. 2016, \apjl, 822, L10, L10

\bibitem[{{Imanishi} {et~al.}(2018){Imanishi}, {Nakanishi}, {Izumi}, \&
  {Wada}}]{Imanishi:2018aa}
{Imanishi}, M., {Nakanishi}, K., {Izumi}, T., \& {Wada}, K. 2018, \apjl, 853,
  L25, L25

\bibitem[{{Kawaguchi} \& {Mori}(2010)}]{Kawaguchi:2010aa}
{Kawaguchi}, T., \& {Mori}, M. 2010, \apjl, 724, L183, L183

\bibitem[{{Kawaguchi} \& {Mori}(2011)}]{Kawaguchi:2011aa}
---. 2011, \apj, 737, 105, 105

\bibitem[{{Kishimoto} {et~al.}(2011){Kishimoto}, {H{\"o}nig}, {Antonucci},
  {Barvainis}, {Kotani}, {Tristram}, {Weigelt}, \& {Levin}}]{Kishimoto:2011aa}
{Kishimoto}, M., {H{\"o}nig}, S.~F., {Antonucci}, R., {et~al.} 2011, \aap, 527,
  A121, A121

\bibitem[{{Kishimoto} {et~al.}(2007){Kishimoto}, {H{\"o}nig}, {Beckert}, \&
  {Weigelt}}]{Kishimoto:2007aa}
{Kishimoto}, M., {H{\"o}nig}, S.~F., {Beckert}, T., \& {Weigelt}, G. 2007,
  \aap, 476, 713, 713

\bibitem[{{Koratkar} \& {Gaskell}(1991)}]{Koratkar:1991aa}
{Koratkar}, A.~P., \& {Gaskell}, C.~M. 1991, \apjs, 75, 719, 719

\bibitem[{{Koshida} {et~al.}(2014){Koshida}, {Minezaki}, {Yoshii}, {Kobayashi},
  {Sakata}, {Sugawara}, {Enya}, {Suganuma}, {Tomita}, {Aoki}, \&
  {Peterson}}]{Koshida:2014aa}
{Koshida}, S., {Minezaki}, T., {Yoshii}, Y., {et~al.} 2014, \apj, 788, 159, 159

\bibitem[{{Krolik} \& {Begelman}(1988)}]{Krolik:1988aa}
{Krolik}, J.~H., \& {Begelman}, M.~C. 1988, \apj, 329, 702, 702

\bibitem[{{Li} \& {Draine}(2001)}]{Li:2001aa}
{Li}, A., \& {Draine}, B.~T. 2001, \apj, 554, 778, 778

\bibitem[{{Lira} {et~al.}(2011){Lira}, {Ar{\'e}valo}, {Uttley}, {McHardy}, \&
  {Breedt}}]{Lira:2011aa}
{Lira}, P., {Ar{\'e}valo}, P., {Uttley}, P., {McHardy}, I., \& {Breedt}, E.
  2011, \mnras, 415, 1290, 1290

\bibitem[{{Lira} {et~al.}(2015){Lira}, {Ar{\'e}valo}, {Uttley}, {McHardy}, \&
  {Videla}}]{Lira:2015aa}
{Lira}, P., {Ar{\'e}valo}, P., {Uttley}, P., {McHardy}, I.~M.~M., \& {Videla},
  L. 2015, \mnras, 454, 368, 368

\bibitem[{{L{\'o}pez-Gonzaga} {et~al.}(2016){L{\'o}pez-Gonzaga}, {Burtscher},
  {Tristram}, {Meisenheimer}, \& {Schartmann}}]{Lopez-Gonzaga:2016aa}
{L{\'o}pez-Gonzaga}, N., {Burtscher}, L., {Tristram}, K.~R.~W., {Meisenheimer},
  K., \& {Schartmann}, M. 2016, \aap, 591, A47, A47

\bibitem[{{Mandal} {et~al.}(2018){Mandal}, {Rakshit}, {Kurian}, {Stalin},
  {Mathew}, {Hoenig}, {Gandhi}, {Sagar}, \& {Pandge}}]{Mandal:2018aa}
{Mandal}, A.~K., {Rakshit}, S., {Kurian}, K.~S., {et~al.} 2018, \mnras, 475,
  5330, 5330

\bibitem[{{Mathis} {et~al.}(1977){Mathis}, {Rumpl}, \&
  {Nordsieck}}]{Mathis:1977aa}
{Mathis}, J.~S., {Rumpl}, W., \& {Nordsieck}, K.~H. 1977, \apj, 217, 425, 425

\bibitem[{{Merloni} {et~al.}(2014){Merloni}, {Bongiorno}, {Brusa}, {Iwasawa},
  {Mainieri}, {Magnelli}, {Salvato}, {Berta}, {Cappelluti}, {Comastri},
  {Fiore}, {Gilli}, {Koekemoer}, {Le Floc'h}, {Lusso}, {Lutz}, {Miyaji},
  {Pozzi}, {Riguccini}, {Rosario}, {Silverman}, {Symeonidis}, {Treister},
  {Vignali}, \& {Zamorani}}]{Merloni:2014aa}
{Merloni}, A., {Bongiorno}, A., {Brusa}, M., {et~al.} 2014, \mnras, 437, 3550,
  3550

\bibitem[{{Minezaki} {et~al.}(2004){Minezaki}, {Yoshii}, {Kobayashi}, {Enya},
  {Suganuma}, {Tomita}, {Aoki}, \& {Peterson}}]{Minezaki:2004aa}
{Minezaki}, T., {Yoshii}, Y., {Kobayashi}, Y., {et~al.} 2004, \apjl, 600, L35,
  L35

\bibitem[{{Mor} \& {Netzer}(2012)}]{Mor:2012aa}
{Mor}, R., \& {Netzer}, H. 2012, \mnras, 420, 526, 526

\bibitem[{{Mor} {et~al.}(2009){Mor}, {Netzer}, \& {Elitzur}}]{Mor:2009aa}
{Mor}, R., {Netzer}, H., \& {Elitzur}, M. 2009, \apj, 705, 298, 298

\bibitem[{{Nelson}(1996)}]{Nelson:1996aa}
{Nelson}, B.~O. 1996, \apjl, 465, L87, L87

\bibitem[{{Nenkova} {et~al.}(2008{\natexlab{a}}){Nenkova}, {Sirocky},
  {Ivezi{\'c}}, \& {Elitzur}}]{Nenkova:2008aa}
{Nenkova}, M., {Sirocky}, M.~M., {Ivezi{\'c}}, {\v Z}., \& {Elitzur}, M.
  2008{\natexlab{a}}, \apj, 685, 147, 147

\bibitem[{{Nenkova} {et~al.}(2008{\natexlab{b}}){Nenkova}, {Sirocky},
  {Nikutta}, {Ivezi{\'c}}, \& {Elitzur}}]{Nenkova:2008ab}
{Nenkova}, M., {Sirocky}, M.~M., {Nikutta}, R., {Ivezi{\'c}}, {\v Z}., \&
  {Elitzur}, M. 2008{\natexlab{b}}, \apj, 685, 160, 160

\bibitem[{{Netzer}(1987)}]{Netzer:1987aa}
{Netzer}, H. 1987, \mnras, 225, 55, 55

\bibitem[{{Netzer}(2015)}]{Netzer:2015aa}
---. 2015, \araa, 53, 365, 365

\bibitem[{{Oknyanskij} \& {Horne}(2001)}]{Oknyanskij:2001aa}
{Oknyanskij}, V.~L., \& {Horne}, K. 2001, in Astronomical Society of the
  Pacific Conference Series, Vol. 224, Probing the Physics of Active Galactic
  Nuclei, ed. B.~M. {Peterson}, R.~W. {Pogge}, \& R.~S. {Polidan}, 149

\bibitem[{{Perez} {et~al.}(1992){Perez}, {Robinson}, \& {de La
  Fuente}}]{Perez:1992aa}
{Perez}, E., {Robinson}, A., \& {de La Fuente}, L. 1992, \mnras, 255, 502, 502

\bibitem[{{Perna} {et~al.}(2003){Perna}, {Lazzati}, \& {Fiore}}]{Perna:2003aa}
{Perna}, R., {Lazzati}, D., \& {Fiore}, F. 2003, \apj, 585, 775, 775

\bibitem[{{Peterson}(1993)}]{Peterson:1993aa}
{Peterson}, B.~M. 1993, \pasp, 105, 247, 247

\bibitem[{{Peterson}(2014)}]{Peterson:2014aa}
---. 2014, \ssr, 183, 253, 253

\bibitem[{{Peterson} {et~al.}(2004){Peterson}, {Ferrarese}, {Gilbert}, {Kaspi},
  {Malkan}, {Maoz}, {Merritt}, {Netzer}, {Onken}, {Pogge}, {Vestergaard}, \&
  {Wandel}}]{Peterson:2004aa}
{Peterson}, B.~M., {Ferrarese}, L., {Gilbert}, K.~M., {et~al.} 2004, \apj, 613,
  682, 682

\bibitem[{{Pozo Nu{\~n}ez} {et~al.}(2014){Pozo Nu{\~n}ez}, {Haas}, {Chini},
  {Ramolla}, {Westhues}, {Steenbrugge}, {Kaderhandt}, {Drass}, {Lemke}, \&
  {Murphy}}]{Pozo-Nunez:2014aa}
{Pozo Nu{\~n}ez}, F., {Haas}, M., {Chini}, R., {et~al.} 2014, \aap, 561, L8, L8

\bibitem[{{Pozo Nu{\~n}ez} {et~al.}(2015){Pozo Nu{\~n}ez}, {Ramolla},
  {Westhues}, {Haas}, {Chini}, {Steenbrugge}, {Barr Dom{\'{\i}}nguez},
  {Kaderhandt}, {Hackstein}, {Kollatschny}, {Zetzl}, {Hodapp}, \&
  {Murphy}}]{Pozo-Nunez:2015aa}
{Pozo Nu{\~n}ez}, F., {Ramolla}, M., {Westhues}, C., {et~al.} 2015, \aap, 576,
  A73, A73

\bibitem[{{Prieto} {et~al.}(2014){Prieto}, {Mezcua}, {Fern{\'a}ndez-Ontiveros},
  \& {Schartmann}}]{Prieto:2014aa}
{Prieto}, M.~A., {Mezcua}, M., {Fern{\'a}ndez-Ontiveros}, J.~A., \&
  {Schartmann}, M. 2014, \mnras, 442, 2145, 2145

\bibitem[{{Robinson} \& {Perez}(1990)}]{Robinson:1990aa}
{Robinson}, A., \& {Perez}, E. 1990, \mnras, 244, 138, 138

\bibitem[{{Sanders} {et~al.}(1989){Sanders}, {Phinney}, {Neugebauer}, {Soifer},
  \& {Matthews}}]{Sanders:1989aa}
{Sanders}, D.~B., {Phinney}, E.~S., {Neugebauer}, G., {Soifer}, B.~T., \&
  {Matthews}, K. 1989, \apj, 347, 29, 29

\bibitem[{{Shen} {et~al.}(2016){Shen}, {Horne}, {Grier}, {Peterson}, {Denney},
  {Trump}, {Sun}, {Brandt}, {Kochanek}, {Dawson}, {Green}, {Greene}, {Hall},
  {Ho}, {Jiang}, {Kinemuchi}, {McGreer}, {Petitjean}, {Richards}, {Schneider},
  {Strauss}, {Tao}, {Wood-Vasey}, {Zu}, {Pan}, {Bizyaev}, {Ge}, {Oravetz}, \&
  {Simmons}}]{Shen:2016aa}
{Shen}, Y., {Horne}, K., {Grier}, C.~J., {et~al.} 2016, \apj, 818, 30, 30

\bibitem[{{Suganuma} {et~al.}(2006){Suganuma}, {Yoshii}, {Kobayashi},
  {Minezaki}, {Enya}, {Tomita}, {Aoki}, {Koshida}, \&
  {Peterson}}]{Suganuma:2006aa}
{Suganuma}, M., {Yoshii}, Y., {Kobayashi}, Y., {et~al.} 2006, \apj, 639, 46, 46

\bibitem[{{Telesco} {et~al.}(1984){Telesco}, {Becklin}, {Wynn-Williams}, \&
  {Harper}}]{Telesco:1984aa}
{Telesco}, C.~M., {Becklin}, E.~E., {Wynn-Williams}, C.~G., \& {Harper}, D.~A.
  1984, \apj, 282, 427, 427

\bibitem[{{Urry} \& {Padovani}(1995)}]{Urry:1995aa}
{Urry}, C.~M., \& {Padovani}, P. 1995, \pasp, 107, 803, 803

\bibitem[{{van Velzen} {et~al.}(2016){van Velzen}, {Mendez}, {Krolik}, \&
  {Gorjian}}]{vanVelzen:2016aa}
{van Velzen}, S., {Mendez}, A.~J., {Krolik}, J.~H., \& {Gorjian}, V. 2016,
  \apj, 829, 19, 19

\bibitem[{{Vazquez}(2015)}]{Vazquez:2015ab}
{Vazquez}, B. 2015, PhD thesis, Rochester Institute of Technology

\bibitem[{{Vazquez} {et~al.}(2015){Vazquez}, {Galianni}, {Richmond},
  {Robinson}, {Axon}, {Horne}, {Almeyda}, {Fausnaugh}, {Peterson}, {Bottorff},
  {Gallimore}, {Eltizur}, {Netzer}, {Storchi-Bergmann}, {Marconi}, {Capetti},
  {Batcheldor}, {Buchanan}, {Stirpe}, {Kishimoto}, {Packham}, {Perez},
  {Tadhunter}, {Upton}, \& {Estrada-Carpenter}}]{Vazquez:2015aa}
{Vazquez}, B., {Galianni}, P., {Richmond}, M., {et~al.} 2015, \apj, 801, 127,
  127

\bibitem[{{Waxman} \& {Draine}(2000)}]{Waxman:2000aa}
{Waxman}, E., \& {Draine}, B.~T. 2000, \apj, 537, 796, 796

\bibitem[{{Weingartner} \& {Draine}(2001)}]{Weingartner:2001aa}
{Weingartner}, J.~C., \& {Draine}, B.~T. 2001, \apj, 548, 296, 296

\end{thebibliography}

\appendix

\section{Isodelay Surfaces and Light-Travel Delays}\label{app:ltdelay}

As an aid for interpreting the response functions presented in this paper, we briefly discuss the isodelay surface corresponding to a short continuum pulse emitted by the AGN central source, as it propagates through the torus from the point of view of a distant observer. In general, a cloud located at radial distance $r$ from the central source will be observed to respond to the continuum pulse with a delay, $t$ (measured from the time at which the observer detects the continuum pulse), given by
\begin{equation}\label{eq:isodelay}
ct=r(1-\cos\alpha)
\end{equation}
where $\alpha$ is the angle between the cloud--central source and the observer--central source vectors. Equation~\ref{eq:isodelay} defines a paraboloid isodelay surface, whose axis is aligned with the observer's line of sight (LOS) and with its focus at the central continuum source. In the case of a cloud located within the torus at coordinates ($r$,$\beta$,$\phi$), 
\begin{equation}\label{eq:cosalpha}
\cos\alpha =\cos\beta\cos\phi\sin i + \sin\beta\cos i
\end{equation}
where $i$  is the inclination of the torus to the LOS. In terms of the normalized delay, $\tau = ct/2R_o$, the isodelay surface is given by 
\begin{equation}\label{eq:normisodelay}
\tau=\frac{\gamma}{2}(1-\cos\alpha)
\end{equation}
where $\gamma = r/R_o$.

\begin{figure*}[h]
\begin{center}
	\includegraphics[scale=0.6]{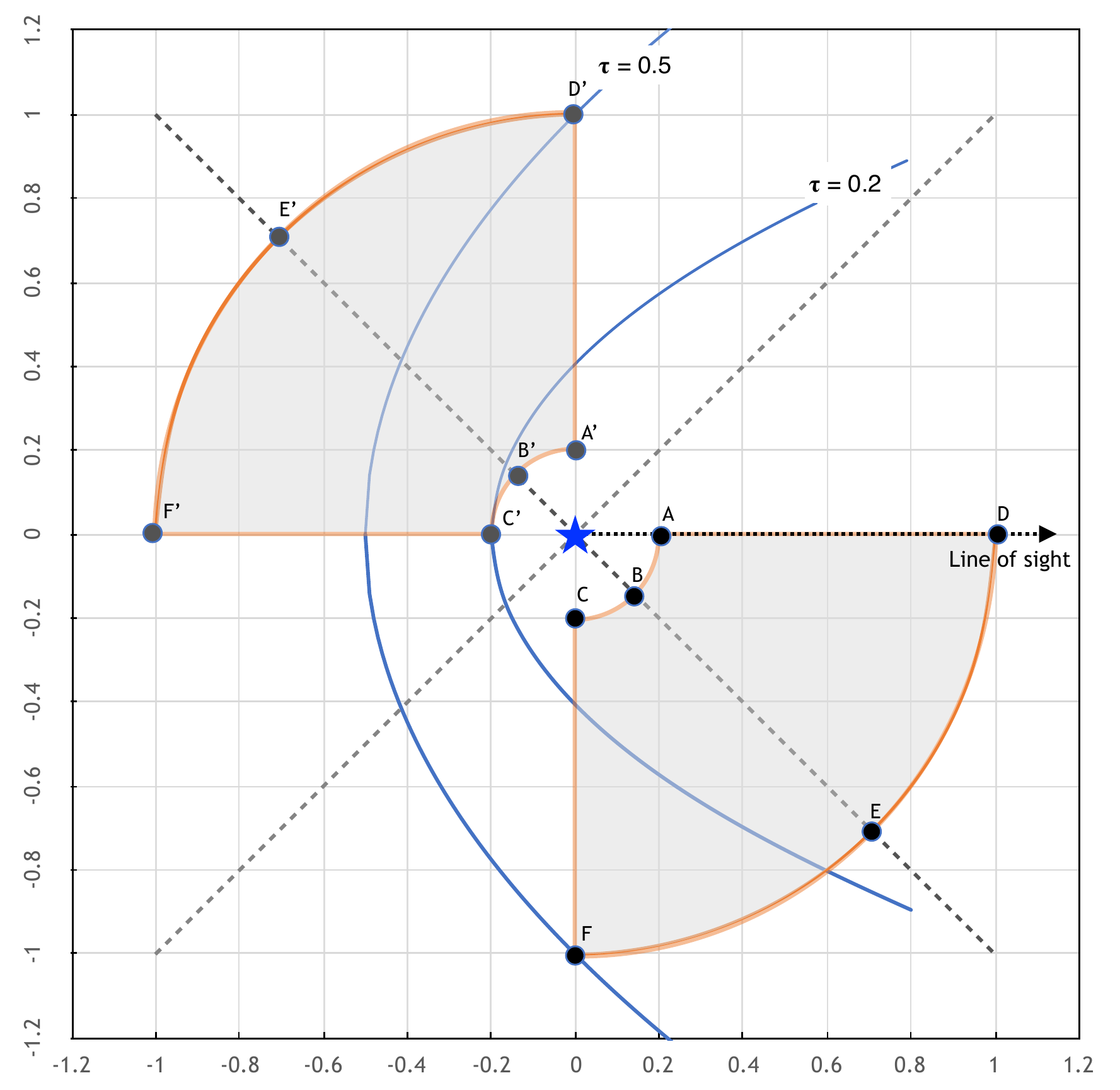}
\end{center}
\caption{\label{torus_isodelay} {Isodelay surfaces within a torus with $\sigma = 45^{\circ}$, inclined at $i = 45^{\circ}$ to the observer's line of sight (LOS). The shaded regions represent a slice through the torus in the plane containing the torus axis and the LOS. Isodelay surfaces corresponding to delays $\tau = 0.2$ and 0.5 are shown (blue lines). The blue star represents the central continuum source of the AGN. The labeled filled circles represent clouds at various locations in the near side (A, B, C, etc.) and far side ($\rm{A}^\prime$}, $\rm{B}^\prime$, $\rm{C}^\prime$, etc.) of the torus, with respect to a distant observer.}
\end{figure*}

As an example, consider  a torus with angular width $\sigma=45^\circ$, inclined at $i=45^\circ$, as shown in Figure~\ref{torus_isodelay}. The isodelay surfaces corresponding to delays $\tau = 0.2$ and 0.5 are shown. For simplicity, we consider only clouds residing in the plane containing the torus axis and the LOS, which have $\phi = 0$ (near side) or $\phi = \pi$ (far side). In this case,

\begin{equation}\label{eq:cosalpha2}
 \begin{array}{ll}
\cos\alpha  & = \pm\cos\beta\sin i + \sin\beta\cos i \\
 & = \frac{1}{\sqrt 2}(\pm\cos\beta + \sin\beta).
\end{array}
\end{equation}

Recall that $\beta$ is the cloud's elevation angle, measured from the torus mid-plane. Therefore, clouds located in the near-side ``upper'' surface of the torus along the line A--D have $\beta = \sigma = 45^{\circ}$ and, thus,  $\cos\alpha = 1$. These clouds are observed to respond with zero delay with respect to the continuum pulse  ($\tau = 0$), as expected since they are located along the LOS. On the other hand, for a cloud located at $\rm{A}^\prime$ (at the sublimation radius on the far-side upper surface), $\cos\alpha = 0$, and since $\gamma = 1/Y$, it responds with a delay $\tau = 1/2Y$. Similarly, a cloud located at $\rm{D}^\prime$ ($\gamma = 1$) responds after a delay $\tau = 1/2$. As the isodelay surface is symmetric about the LOS, clouds located at the corresponding positions on the near side (i.e., C and F) have the same delays. For the clouds located at $\rm{C}^\prime$ and 
$\rm{D}^\prime$, along the LOS but on the far side of the torus, the delays are, respectively, $\tau = 1/Y$ and 1, i.e., the light-crossing times of the torus and the inner cavity, i.e., the spheres defined by $R_{\rm d, iso}$ and the torus outer radius, $R_o$. 
For convenience, the delays corresponding to the positions labeled in Figure~\ref{torus_isodelay} are listed in Table~\ref{clddelaytab}. 

Note that, in general, for inclinations $i < 90^{\circ} - \sigma$, the direct LOS to the central continuum source does not intercept any part of the torus, and therefore the onset of the torus response is delayed by 
\begin{equation}\label{eq:resbegin}
\tau=\frac{1}{2Y}\left(1-\sin(\sigma + i)\right).
\end{equation}
This is the delay corresponding to a cloud located at position A in Figure~\ref{torus_isodelay}. For larger inclinations, the response begins immediately, as in the situation depicted in Figure~\ref{torus_isodelay}.

\begin{table*}[h]
\caption{Normalized Delay Times}
\begin{center}
\label{clddelaytab}

\scriptsize{
\begin{tabular}{cc | lc | lc}
\hline
 \hline
 \multicolumn{2}{c}{Coordinates} &  \multicolumn{2}{c}{Near Side ($\phi = 0$)} & \multicolumn{2}{c}{Far Side ($\phi = 180^{\circ}$)}  \\
\hline
 $\gamma$ & $\beta$ & Position & $\tau$ &  Position & $\tau$  \\
\hline

$1/Y$ & $45^{\circ}$ & A &  0  & $\rm{A}^\prime$ &  $\frac{1}{2Y}$ \\
          & $0^{\circ}$   & B & $\frac{1}{2Y}(1-\frac{1}{\sqrt 2})$ & $\rm{B}^\prime$ & $\frac{1}{2Y}(1+\frac{1}{\sqrt 2})$ \\
          & $-45^{\circ}$ & C & $\frac{1}{2Y}$ & $\rm{C}^\prime$ & $\frac{1}{Y}$ \\
\\
\hline

$1$ & $45^{\circ}$ & D &  0  & $\rm{D}^\prime$ &  $\frac{1}{2}$ \\
          & $0^{\circ}$   & E & $\frac{1}{2}(1-\frac{1}{\sqrt 2})$ & $\rm{E}^\prime$ & $\frac{1}{2}(1+\frac{1}{\sqrt 2})$ \\
          & $-45^{\circ}$ & F & $\frac{1}{2}$ & $\rm{F}^\prime$ & 1 \\

\\
\hline
\end{tabular}}

\end{center}
\end{table*}

\onecolumngrid

\section{Additional Figures}\label{app:figs}
\begin{figure*}[!h]
\begin{center}
	\includegraphics[scale=0.6]{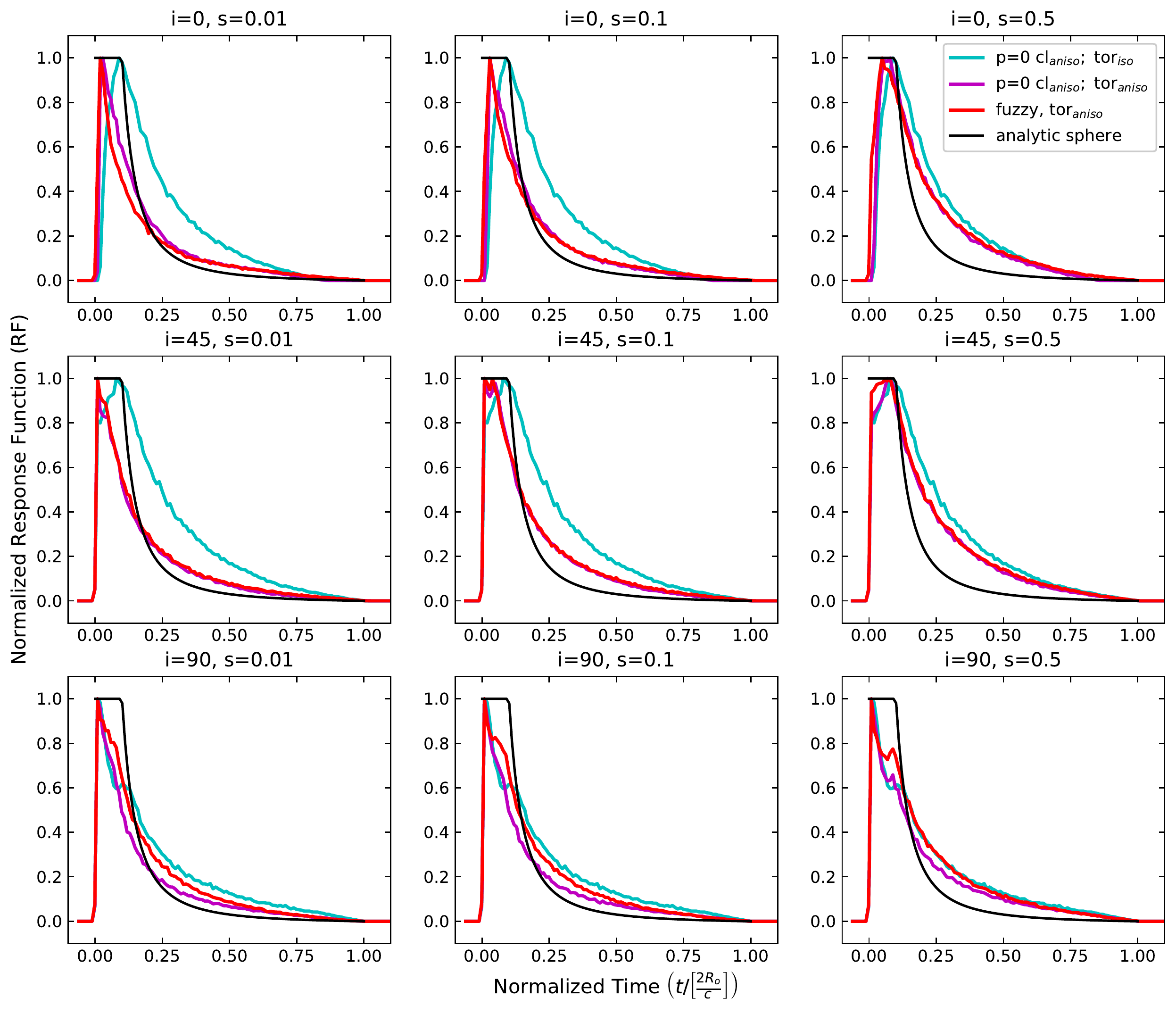}
\end{center}
\caption{\label{clorientvsani_sp30} \textbf{Globally optically thin torus varying $s$ at 30 $\mu$m:} Response functions for an isotropically or anisotropically illuminated torus with $p=0$, $\sigma=45^{\circ}$, and $Y=10$ at 30 $\mu$m, where each row represents a different inclination from top ($i=0^{\circ}$) to bottom ($i=90^{\circ}$). Each column represents a different $s$ value increasing from left ($s=0.01$) to right ($s=0.5$). The blue lines represent a sharp-edged isotropically illuminated torus. The purple lines represent a sharp-edged torus, and the red lines represent a fuzzy-edged torus, both of which are anisotropically illuminated. The black line is the analytic spherical transfer function for RWD/LWR.}
\end{figure*}

\begin{figure*}[!h]
\begin{center}
	\includegraphics[scale=0.49]{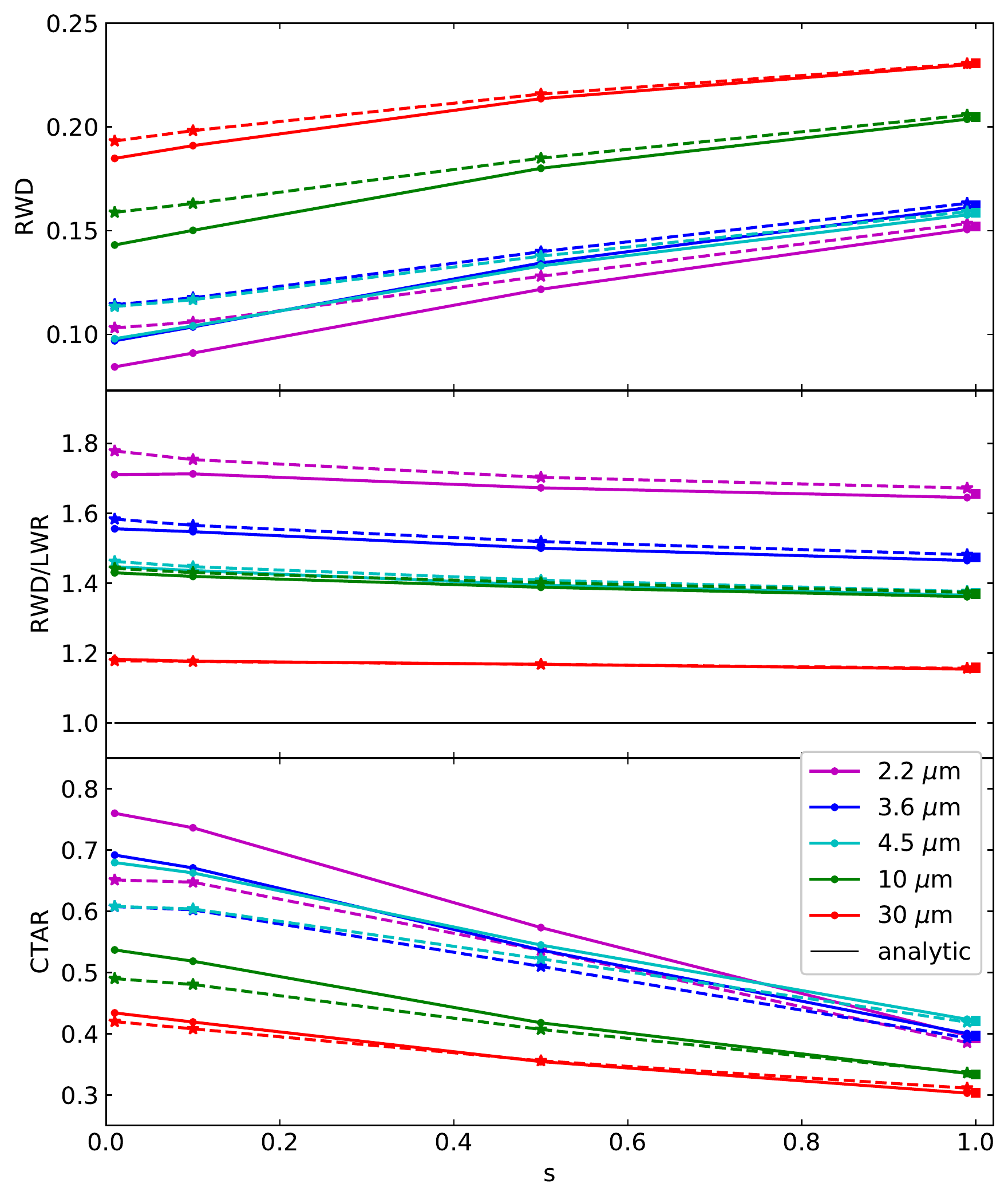}
\end{center}
\caption{\textbf{Globally optically thin torus with either a sharp- or fuzzy-edge varying $s$:} Response function descriptive quantities for an anisotropically illuminated torus with $p$=0, $\sigma=$45$^{\circ}$, $i$=45$^{\circ},$ and $Y$=10 with respect to different values of $s$. The circles and solid lines represent a sharp-edged torus, and the stars and dashed lines represent a fuzzy-edged torus. The isotropic illuminated torus cases are plotted as colored squares at $s=1$. The colors represent the different wavelengths (2.2, 3.6, 4.5, 10, and 30 $\mu$m). }
\label{rfdpdiffspvsfuzi45} 
\end{figure*}

\begin{figure*}[h]
\begin{center}
	\includegraphics[scale=0.6]{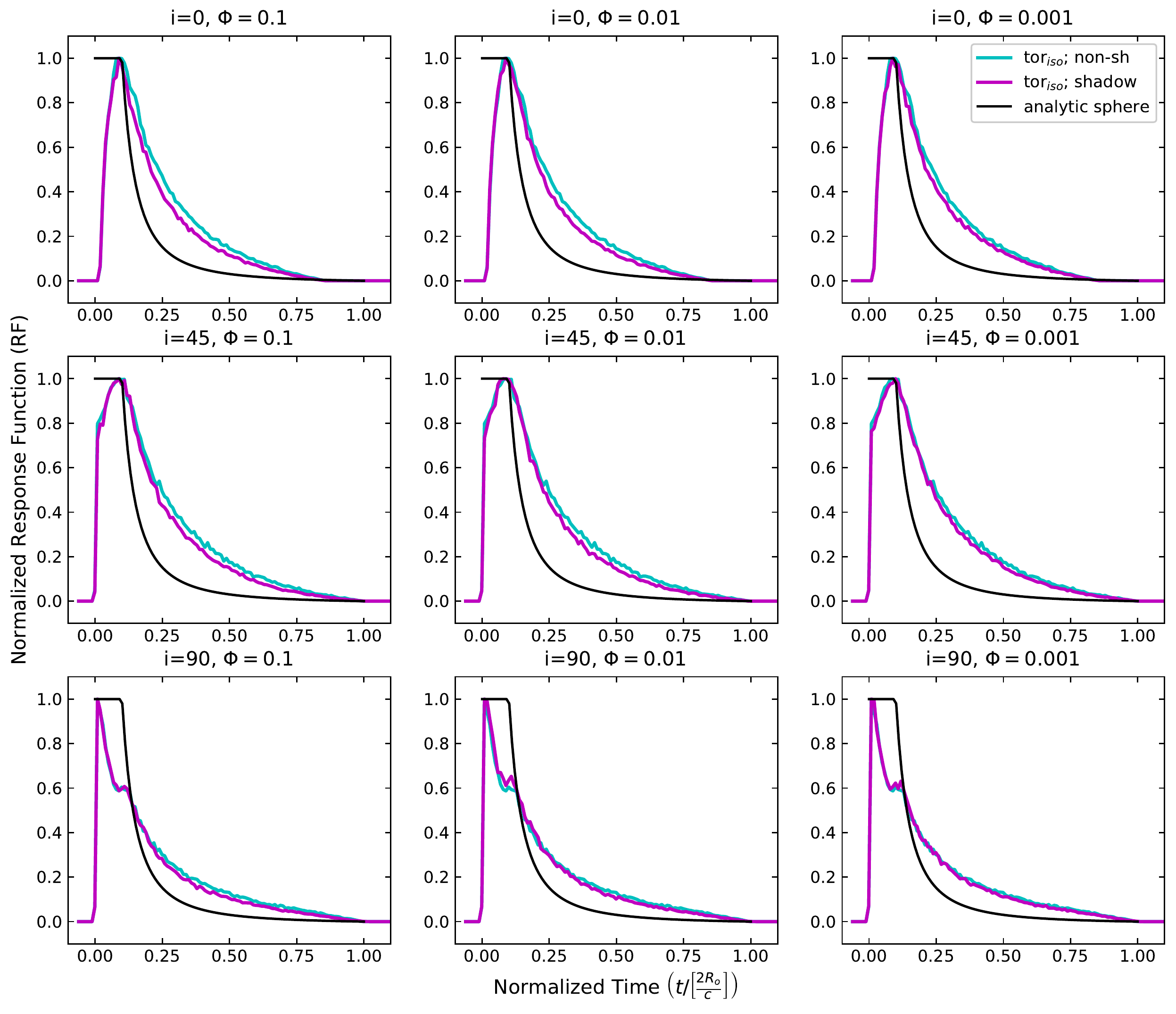}
\end{center}
\caption{\label{clshadow_vff30} \textbf{Torus with cloud shadowing varying $\Phi$ at 30 $\mu$m:} Response functions for an isotropically illuminated torus with $\sigma=45^{\circ}$; $Y=10$; $p=0$; for $i=0,$ 45, 90$^{\circ}$ at 30 $\mu$m. The blue lines represent simulations without cloud shadowing, the purple lines represent the simulations with cloud shadowing for values of the average volume filling factor ranging from $\Phi$=0.001 (right) to 0.1 (left), and the black line is the analytic spherical shell transfer function. } 
\end{figure*}

\begin{figure}
\begin{center}
	\includegraphics[scale=0.47]{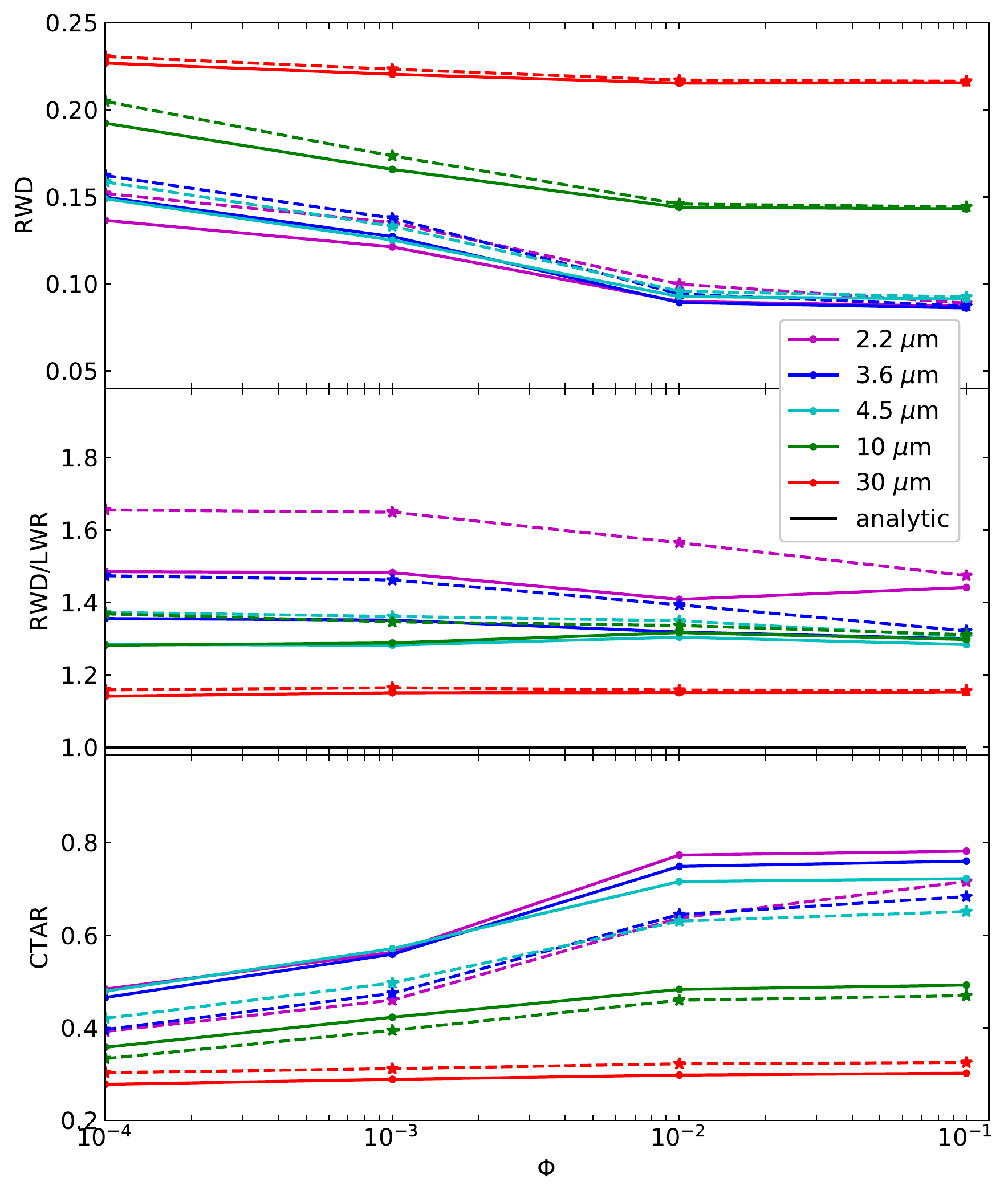}
	\caption{\label{rfdpdiffvffvsi045p0}  \textbf{Torus with cloud shadowing varying $\Phi$ at $i=0^{\circ}$ and $i=45^{\circ}$:} Response function descriptive quantities for isotropically illuminated tori with $p$=0, $\sigma=45^{\circ}$, and $Y$=10. The circles and solid lines represent the torus models with $i=0^{\circ}$, and the stars and dashed lines represent the torus models with $i=45^{\circ}$. Both tori have a sharp surface boundary and include cloud orientation and cloud shadowing. The colors represent the different wavelengths (2.2, 3.6, 4.5, 10, and 30 $\mu$m). The analytic spherical solution for RWD/LWR is plotted as a solid black line for comparison.} 
\end{center}
%
\begin{center}
	
	\includegraphics[scale=0.47]{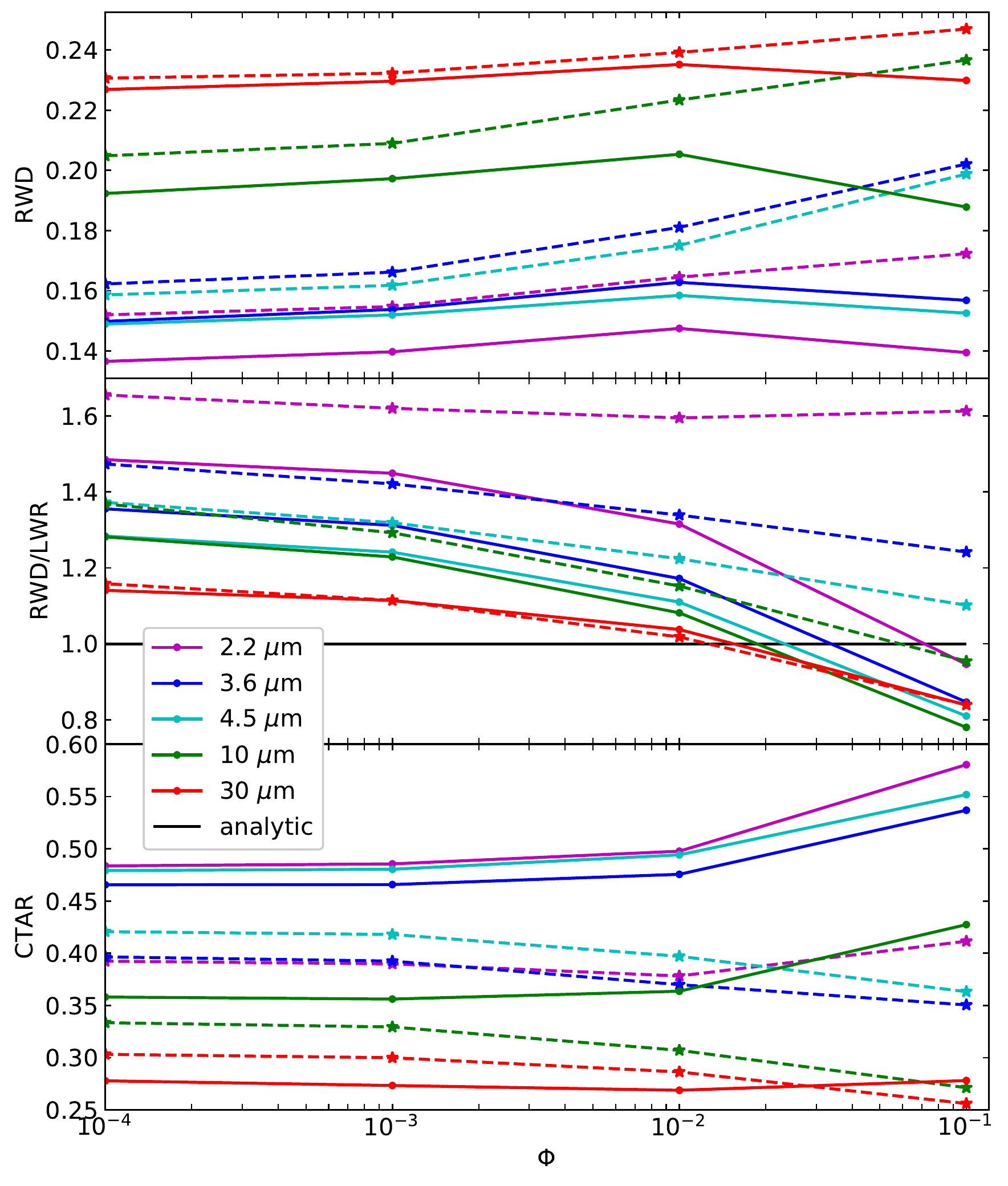}
	\caption{\label{rfdpdiffvffvsi0p0clocc} \textbf{Torus with cloud occultation varying $\Phi$ at $i=0^{\circ}$ and $i=45^{\circ}$:} Response function descriptive quantities for isotropically illuminated tori with $p=0$, $\sigma=45^{\circ},$ and $Y=10$. The circles and solid lines represent the torus models with $i=0^{\circ}$, and the stars and dashed lines represent the torus models with $i=45^{\circ}$. All models include cloud orientation and cloud occultation. The colors represent the different wavelengths (2.2, 3.6, 4.5, 10, and 30 $\mu$m). The analytic spherical solution for RWD/LWR is plotted as a solid black line for comparison. }
\end{center}
\end{figure}

\begin{figure}
 \begin{center}
	\includegraphics[scale=0.47]{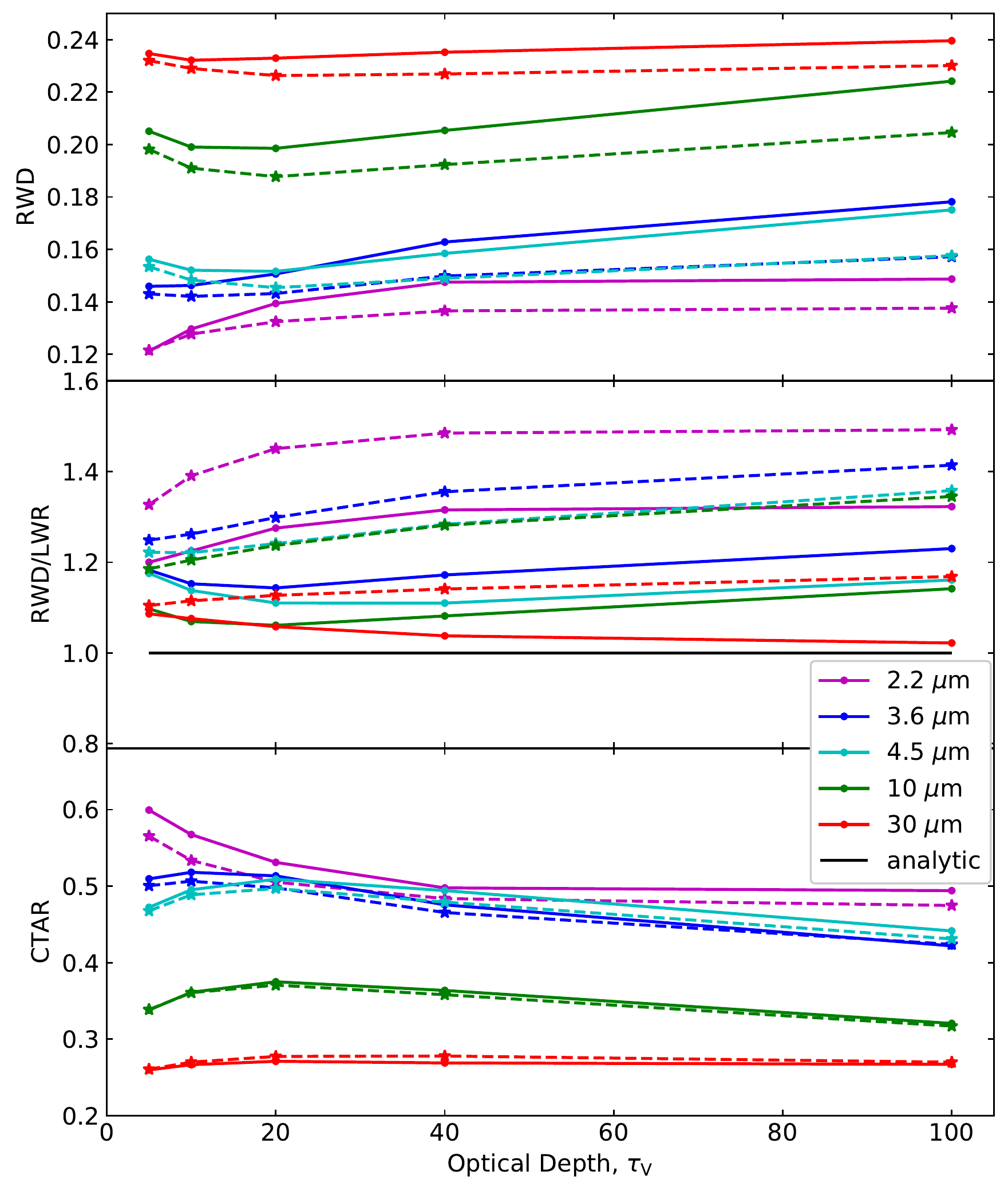}
\end{center}
\caption{\label{rfdpvstaui0clocc} \textbf{Torus with cloud occultation varying $\tau_V$:} Response function descriptive quantities for various simulation runs for an isotropically illuminated torus with $p=0$, $i=0^{\circ}$, $\sigma=45^{\circ}$, $\Phi$=0.01, and $Y=10$ with cloud occultation (circles and solid lines) or without cloud occultation (stars and dashed lines) with respect to different values of $\tau_V$ at select wavelengths. The colors represent the different wavelengths (2.2, 3.6, 4.5, 10, and 30 $\mu$m).  The analytic spherical solution for RWD/LWR is plotted as a solid black line for comparison.}

\begin{center}
	\includegraphics[scale=0.46]{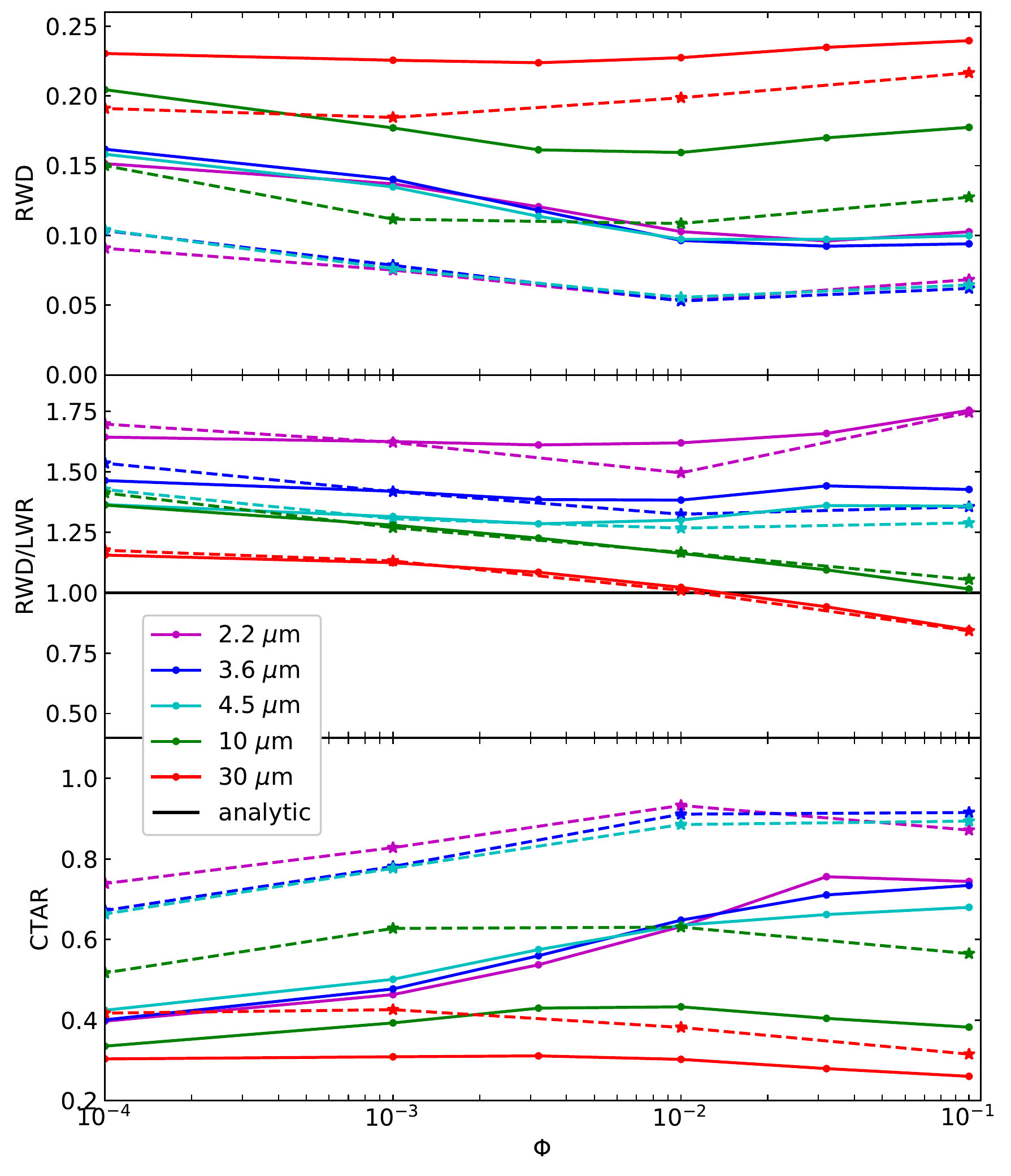}
\end{center}
\caption{\label{rfdpdiffvffvsi45p0_combo} \textbf{GOT torus varying $\Phi$ at $i$=45$^{\circ}$:} Response function descriptive quantities for a globally optically thick torus with $p$=0, $i$=45$^{\circ},$ $\sigma$=45$^{\circ},$ and $Y$=10 with respect to different values of $\Phi$ at select wavelengths. The circles and solid lines represent the models where the torus is illuminated isotropically, and the stars and dashed lines represent the models where the torus is illuminated anisotropically. The limiting globally optically thin case is represented at a nominal value of $\Phi = 0.0001$. Both tori have a sharp surface boundary. The colors represent the different wavelengths (2.2, 3.6, 4.5, 10, and 30 $\mu$m). The black line represents the analytic solution for RWD/LWR for a spherical shell with blackbody clouds.}
\end{figure}

 Additional figures referenced in Sections~\ref{sec:DustRF} and \ref{sec:combo}. 
 Figures~\ref{clorientvsani_sp30} and \ref{rfdpdiffspvsfuzi45} show RFs and descriptive quantities for the globally optically thin torus when the degree of anisotropy, $s$, is varied as discussed in Section~\ref{sec:aniscloud}. 
 Figures~\ref{clshadow_vff30} and \ref{rfdpdiffvffvsi045p0} show RFs and descriptive quantities for a torus with cloud orientation and cloud shadowing included when the volume filling factor, $\Phi$, is varied (Section~\ref{sec:shadowing}). 
 Figures~\ref{rfdpdiffvffvsi0p0clocc} and \ref{rfdpvstaui0clocc} show descriptive quantities for a torus with cloud orientation and cloud occultation included when the volume filling factor, $\Phi$, and cloud optical depth, $\tau_{\rm V}$, are varied, respectively (Section~\ref{sec:cloccrf}). 

\begin{figure}[h]
\begin{center}
	\includegraphics[scale=0.46]{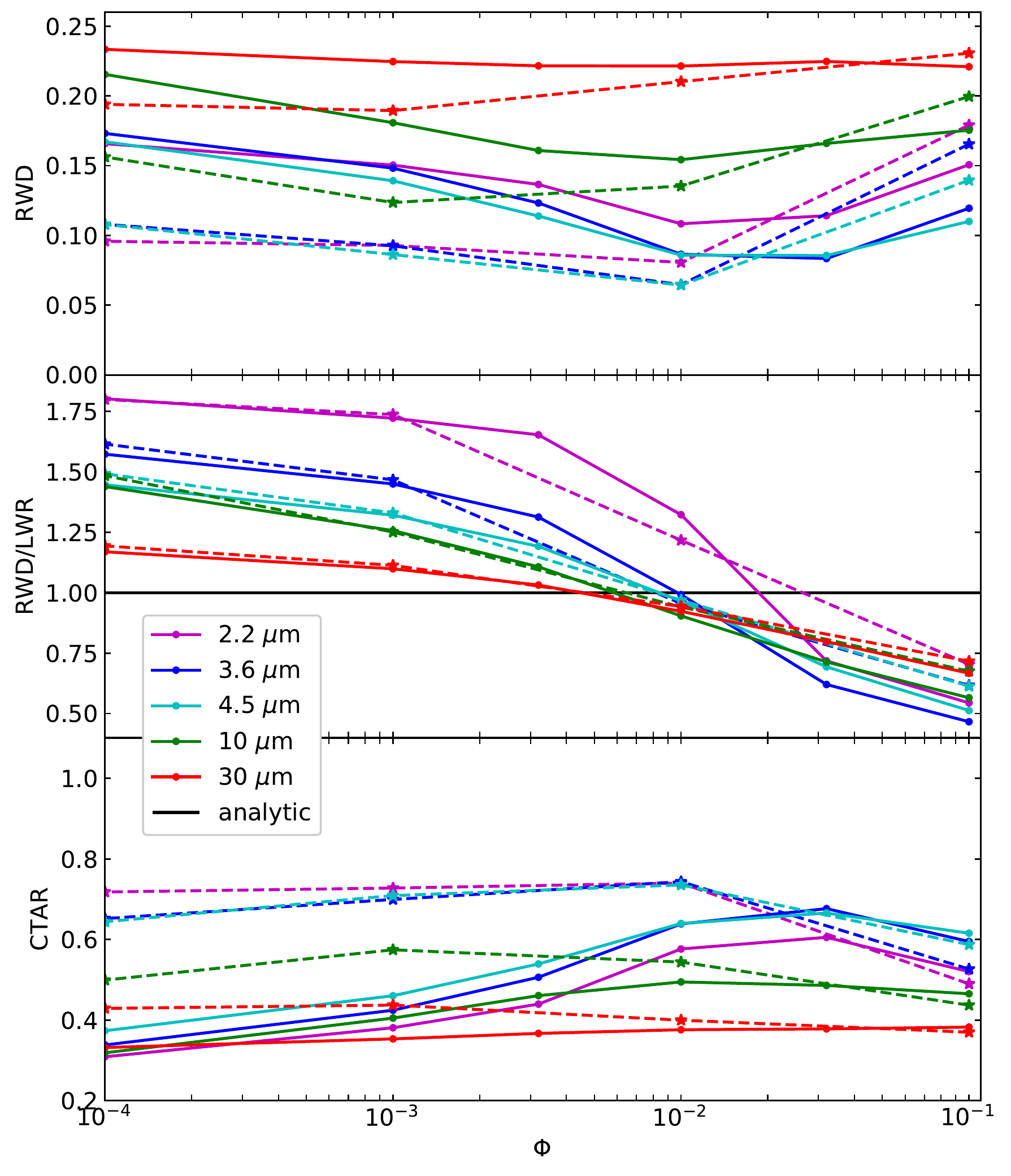}
\end{center}
\caption{\label{rfdpdiffvffvsi90p0_combo}\textbf{GOT torus varying $\Phi$ at $i$=90$^{\circ}$:}  Response function descriptive quantities for a globally optically thick torus with $p$=0, $i$=90$^{\circ},$ $\sigma$=45$^{\circ},$ and $Y$=10 with respect to different values of $\Phi$ at select wavelengths. The circles and solid lines represent the models where the torus is illuminated isotropically, and the stars and dashed lines represent the models where the torus is illuminated anisotropically. The limiting globally optically thin case is represented at a nominal value of $\Phi = 0.0001$. Both tori have a sharp surface boundary. The colors represent the different wavelengths (2.2, 3.6, 4.5, 10, and 30 $\mu$m). The black line represents the analytic solution for RWD/LWR for a spherical shell with blackbody clouds.}
\end{figure}

\begin{figure}[h]
\begin{center}
	\includegraphics[scale=0.46]{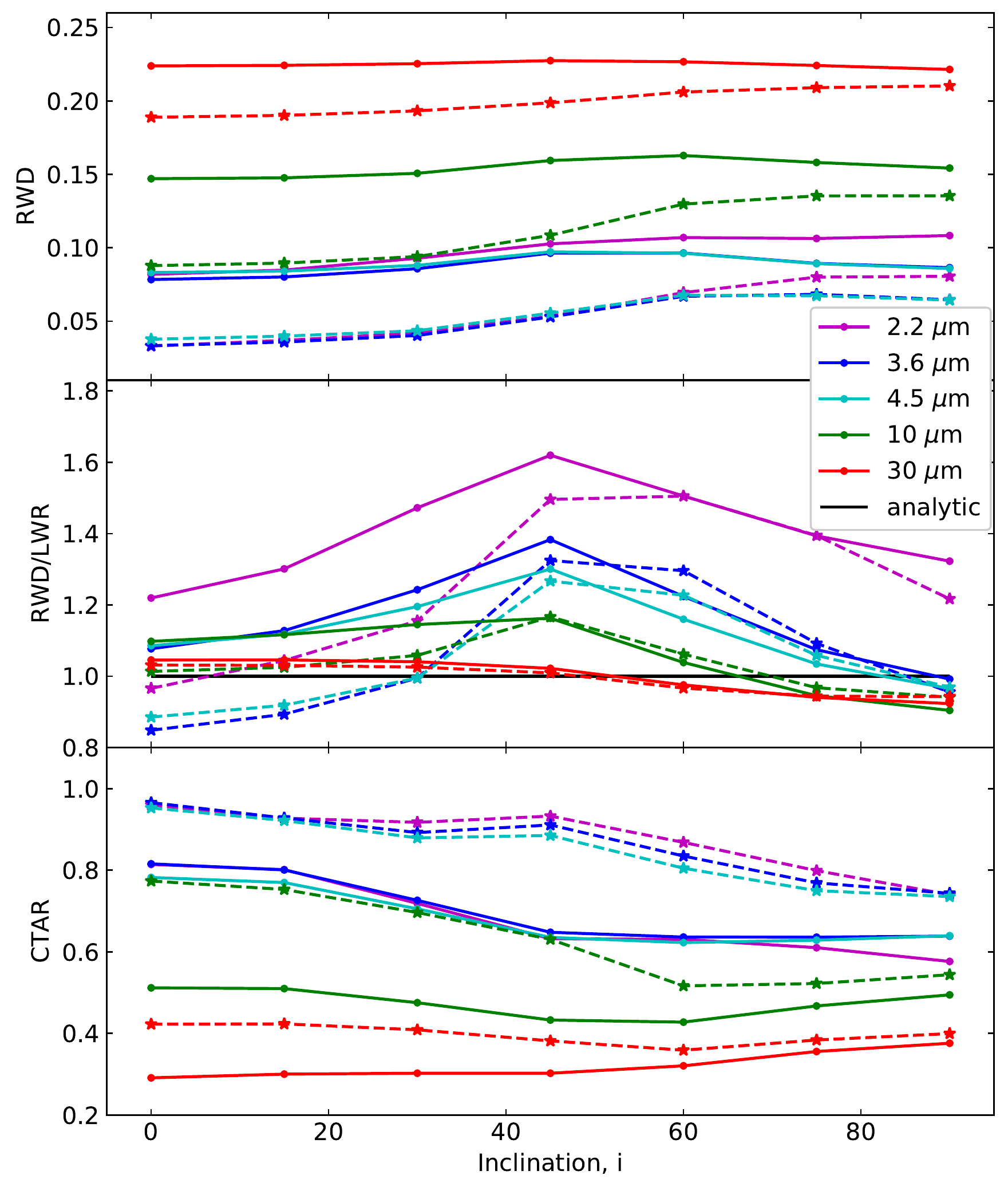}
\end{center}
\caption[DPs for tori with the ``complete" radiative transfer treatment that is illuminated either isotropically or anisotropically varying $i$ for $Y=10$ at $\Phi=0.01$, $\sigma=45^{\circ},$ and $p=0$]{\label{rfdpvsip0vsani_combo}\textbf{GOT torus varying $i$:}  Response function descriptive quantities for isotropically and anisotropically illuminated tori with $p=0$, $\Phi=0.01,$ $\sigma=45^{\circ},$ and $Y=10$. The circles and solid lines represent an isotropically illuminated torus, and the stars and dashed lines represent an anisotropically illuminated torus. Both tori have a sharp surface boundary and include all of the radiative transfer treatments. The colors represent the different wavelengths (2.2, 3.6, 4.5, 10, and 30 $\mu$m). The black line represents the analytic solution of RWD/LWR for a spherical shell with blackbody clouds.}
\end{figure}

\begin{figure*}[b]
\begin{center}
	\includegraphics[scale=0.47]{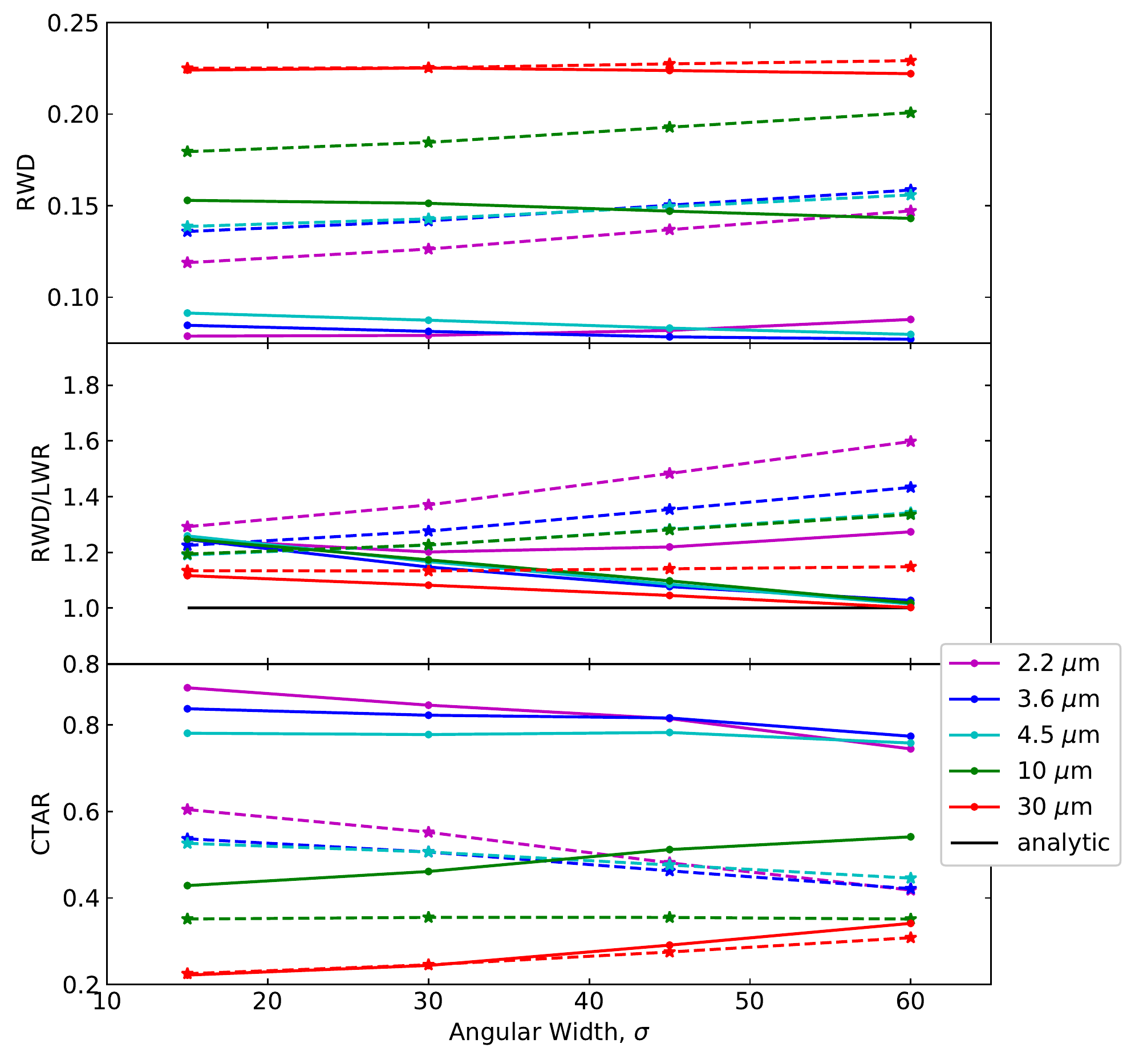}
\end{center}
\caption{\label{rfdpdiffsigcombo0} \textbf{GOT torus varying $\sigma$:} Response function descriptive quantities for an isotropically illuminated torus with $p = 0$, $\Phi=0.01$, $i=0^{\circ}$, and $Y$=10 with respect to different values of $\sigma$ at select wavelengths. The circles and solid lines represent the globally optically thick torus models, and the stars and dashed lines represent the torus models with only cloud orientation. The colors represent the different wavelengths (2.2, 3.6, 4.5, 10, and 30 $\mu$m). The analytic spherical shell solution is plotted as a solid black line for comparison.}
\end{figure*}

\begin{figure}[ht]
\begin{center}
	\includegraphics[scale=0.49]{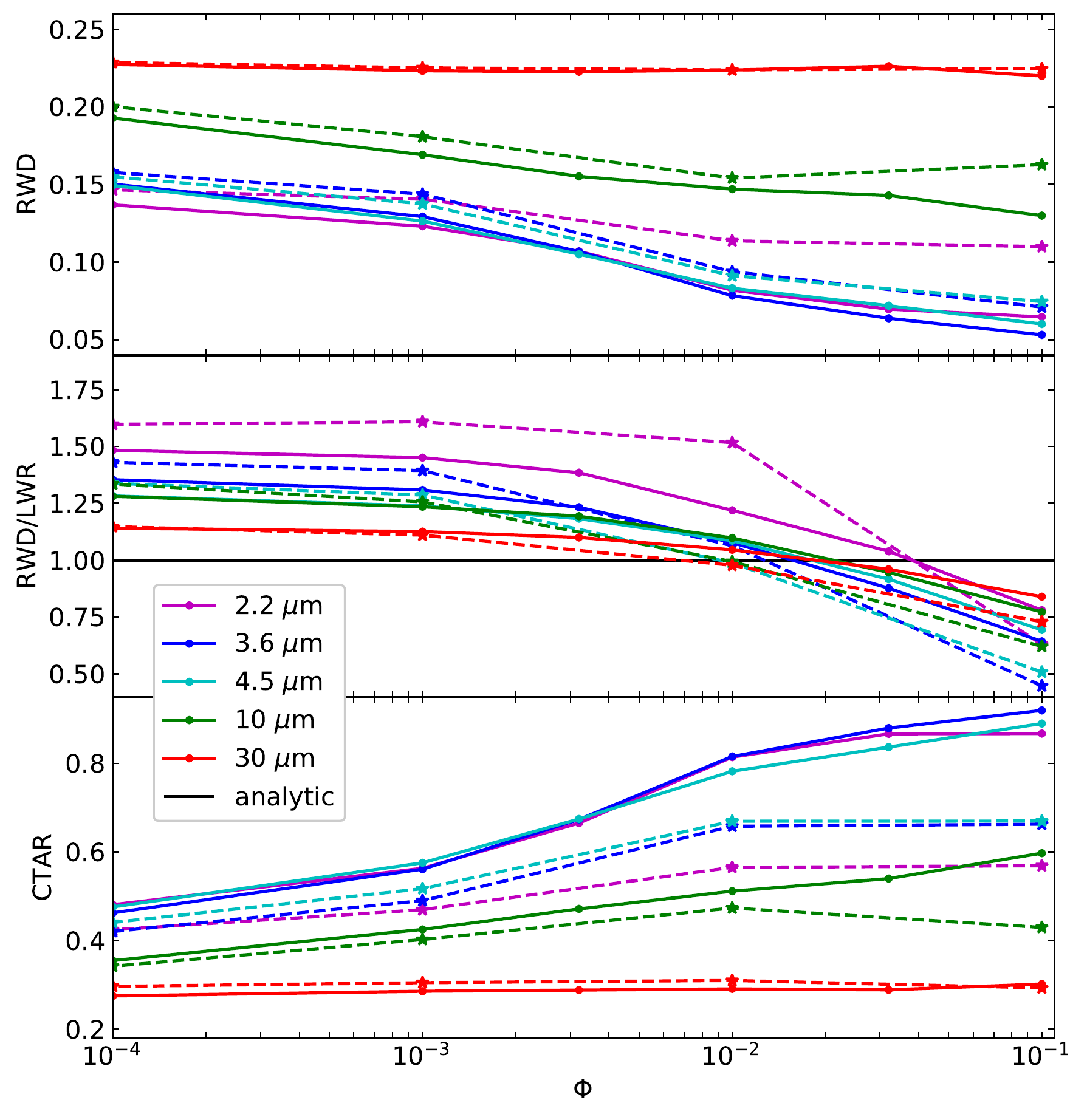}
\end{center}
\caption{\label{rfdpfuzzycombo0} \textbf{GOT torus with a sharp- or fuzzy-edge varying $\Phi$:} Response function descriptive quantities for an isotropically illuminated torus that has a sharp- or fuzzy-edge with $\sigma = 45^{\circ}$, $p=0,$ $i$=0$^{\circ},$ and $Y=$10 at select wavelengths. The circles and solid lines represent torus models with a sharp surface boundary, and the stars and dashed lines represent the torus models with a fuzzy surface boundary. The limiting globally optically thin case is represented at a nominal value of $\Phi = 0.0001$. Both tori have a sharp surface boundary. The colors represent the different wavelengths (2.2, 3.6, 4.5, 10, and 30 $\mu$m). The analytic spherical shell solution for RWD/LWR is plotted as a solid black line for comparison.}
\end{figure}

\begin{figure}[h]
\begin{center}
	\includegraphics[scale=0.49]{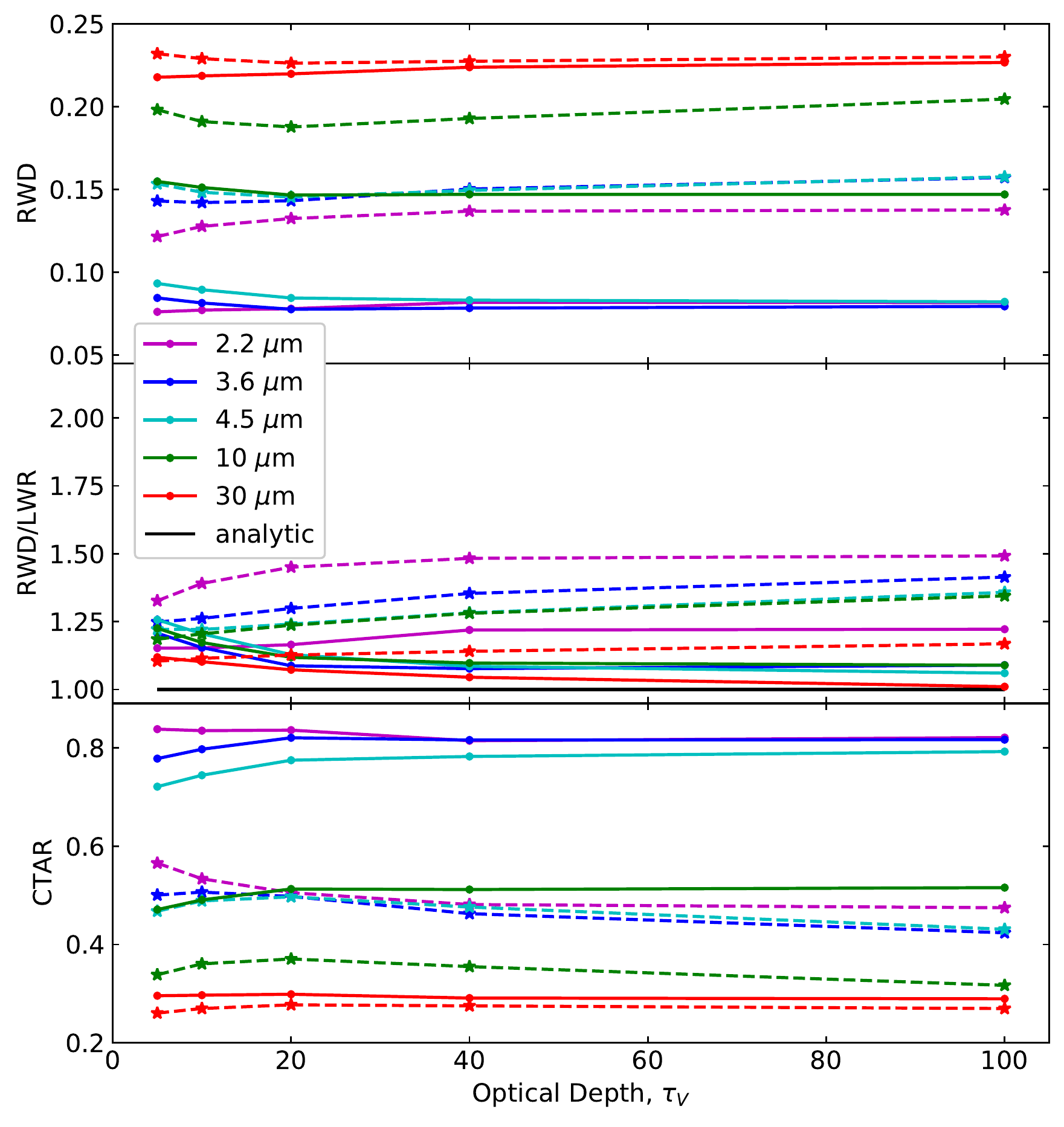}
\end{center}
\caption{\label{rfdpdifftaucombo0} \textbf{GOT torus varying $\tau_{V}$:} Response function descriptive quantities for an isotropically illuminated torus with $\sigma = 45^{\circ}$, $p=0,$ $\Phi=0.01$, $i$=0$^{\circ},$ and $Y=$10 with respect to different values of $\tau_{V}$ at select wavelengths. The circles and solid lines represent the globally optically thick (GOT) torus models, and the stars and dashed lines represent the torus models with only cloud orientation. The colors represent the different wavelengths (2.2, 3.6, 4.5, 10, and 30 $\mu$m). The analytic spherical shell solution is plotted as a solid black line for comparison.}
\end{figure}

Figures~\ref{rfdpdiffvffvsi45p0_combo}$-$\ref{rfdpvsip0vsani_combo} show descriptive quantities for the GOT torus when the volume filling factor, $\Phi$,  and inclination, $i$, are varied (Section~\ref{sec:gotvff}). 
Figures~\ref{rfdpdiffsigcombo0}$-$\ref{rfdpdifftaucombo0} show descriptive quantities for the GOT torus when the angular width, $\sigma$, torus surface (sharp or fuzzy), and cloud optical depth,  $\tau_{\rm V}$, are varied (Section~\ref{sec:gottor}).

\onecolumngrid

\end{document}